\documentclass[twocolumn,tighten,times]{aastex701}

\usepackage{color}
\usepackage{CJK}

\newcommand{\msun}{\ensuremath{M_\sun}}
\newcommand{\msuns}{\ensuremath{M_\sun\,{\rm s}^{-1}}}
\newcommand{\tpb}{\ensuremath{t_{\rm pb}}}
\newcommand{\tadv}{\ensuremath{\tau_{\rm adv}}}
\newcommand{\theat}{\ensuremath{\tau_{\rm heat}}}

\newcommand{\nue}{\ensuremath{\nu_{e}}}
\newcommand{\nuebar}{\ensuremath{\bar \nu_e}}
\newcommand{\numt}{\ensuremath{\nu_{\mu\tau}}}
\newcommand{\numtbar}{\ensuremath{\bar \nu_{\mu\tau}}}

\newcommand{\Ye}{\ensuremath{Y_{\rm e}}}

\newcommand{\Ediag}{\ensuremath{E^{+}}}

\newcommand{\Ediagov}{\ensuremath{E^{+}_{\rm ov}}}

\newcommand{\Mshock}{\ensuremath{M_{\rm sh}}}

\newcommand{\kms}{\ensuremath{\rm km \; s^{-1}}}

\newcommand{\kbbar}{\ensuremath{\rm k_B \; baryon^{-1}}}

\newcommand{\heateff}{\ensuremath{\eta_{\rm heat}}}
\newcommand{\accgain}{\ensuremath{\dot{M}_{\rm gain}}}

\newcommand{\isotope}[2]{\ensuremath{\mathrm {^{#2}#1}}}
\newcommand{\gcc}{\ensuremath{{\mbox{g~cm}}^{-3}}}
\newcommand{\ergg}{\ensuremath{{\mbox{erg~g}}^{-1}}}

\newcommand{\kmps}{\ensuremath{\mbox{km~s}^{-1}}}
\newcommand{\cmps}{\ensuremath{\mbox{cm~s}^{-1}}}
\newcommand{\Bethes}{\ensuremath{{\mbox{B~s}}^{-1}}}

\newcommand{\ctheta}{\ensuremath{\langle v_\theta^2}\rangle^{1/2}}

\newcommand{\coconut}{{\sc CoCoNuT-FMT}}

\newcommand{\fornax}{{\sc Fornax}}
\newcommand{\chimera}{{\sc Chimera}}
\newcommand{\prometheusvertex}{{\sc Prometheus-Vertex}}
\newcommand{\polaris}{{\sc Polaris}}

\newcommand{\dotQnu}{\ensuremath{\dot{Q}_{\nu}}}

\newcommand{\ee}[2]{\ensuremath{#1 \times 10^{#2}}}
\newcommand{\ZoverR}{\ensuremath{\langle Z \rangle / \langle R \rangle}}

\newcommand{\UTphys}{Department of Physics and Astronomy, University of Tennessee, Knoxville, TN 37996-1200, USA}
\newcommand{\ORNLphys}{Physics Division, Oak Ridge National Laboratory, P.O. Box 2008, Oak Ridge, TN 37831-6354, USA}
\newcommand{\NCCS}{National Center for Computational Sciences, Oak Ridge National Laboratory, P.O. Box 2008, Oak Ridge, TN 37831-6164, USA}
\newcommand{\NCSU}{Department of Physics and Astronomy, North Carolina State University, Raleigh, NC 27695-8202, USA}
\newcommand{\FAU}{Department of Physics, Florida Atlantic University, 777 Glades Road, Boca Raton, FL 33431-0991, USA}

\shorttitle{2D Progenitor impact on CCSNe}
\shortauthors{Chen et al.}	

\begin{document}

\title{Impacts of Multidimensional Progenitor Perturbations on Core-Collapse Supernova Explosions}

\author[0000-0001-7441-3113]{Chien-Hui Chen \begin{CJK*}{UTF8}{bsmi}(陳芊卉)\end{CJK*}}
\affiliation{\NCSU}
\affiliation{\UTphys}
\email[show]{cchen55@ncsu.edu}

\author[0000-0002-5231-0532]{Eric J. Lentz}
\affiliation{\UTphys}
\affiliation{\ORNLphys}
\email{elentz@utk.edu}

\author[0000-0002-9481-9126]{W. Raphael Hix}
\affiliation{\ORNLphys}
\affiliation{\UTphys}
\email{raph@ornl.gov}

\author[0000-0003-3023-7140]{J. Austin Harris}
\affiliation{\NCCS}
\email{harrisja@ornl.gov}

\author[0000-0001-5501-7625]{Chloe Keeling Sandoval}
\affiliation{\UTphys}
\email{chloe.sandoval91@gmail.com}

\author[0000-0003-0999-5297]{Stephen W. Bruenn}
\affiliation{\FAU}
\email{bruenn@fau.edu}

\correspondingauthor{Chien-Hui Chen}

\begin{abstract}
Numerical studies of core-collapse supernovae have demonstrated the importance of non-radial motions in pre-collapse progenitors on the explosion outcome. We use the \chimera\ neutrino radiation hydrodynamics code running seven two-dimensional simulations of 15 \msun\ progenitors with different progenitor structures introduced by different one and two-dimensional pre-collapse stellar evolution environments to examine the impacts of stellar structure and non-spherical motion in the pre-collapse progenitor on the development of explosions. We compare the explosion evolution of these models in terms of shock dynamics, diagnostic energy, neutrino heating, accretion, explosion geometry, nuclear abundances, and turbulent convection. We also analyze how stochastic variation impacts our simulations. Contrary to results reported in prior studies examining the impacts of multi-dimensional progenitors, we observe similar shock revival times and explosion development in our simulations despite differences in initial compositions and structures. We find no discernible impact from the accretion of non-radial perturbations from a multi-D progenitor onto the stalled shock in the revival and strength of explosion, as fully developed neutrino-driven convection behind the stalled shock is similar for all our models. For models with physically sourced noise in the iron core, a strong oscillation of the shock occurs after bounce and deflects infall laterally, and accelerates the saturation of the lateral turbulent kinetic energy. An examination of model stochasticity shows that any prior expected impacts on explosive outcome due to convection-related perturbations lie below the detectable threshold of numerical variation.
\end{abstract}

\keywords{Stellar convective shells (300); Core-collapse supernovae (304); Massive stars (732); Hydrodynamical simulations (767); Late stellar evolution (911); Silicon burning (1457) Supernova dynamics (1664)}

\section{Introduction}
\label{sec:intro}

As one of the most energetic phenomena in our Universe, core-collapse supernovae (CCSNe) not only represent the death of massive stars ($M_{\rm ZAMS} \gtrsim 8 \msun$), they also spread heavy elements synthesized by the progenitor stars into interstellar spaces. 
Understanding CCSNe is therefore essential for elucidating both their contribution to chemical enrichment and the physical mechanisms that power stellar explosions.

It is well-recognized that the neutrino heating mechanism is the primary engine for CCSN explosions. It increases the post-shock thermal pressure, breaks the balance with the pre-shock ram pressure, and thus pushes the stalled shock outward, which is known as shock revival. The shock eventually breaks through the stellar envelope and explodes the star. However, CCSNe are inherently complex multi-physics phenomena, and it has been proven in 1D studies that neutrinos are insufficient to solely power CCSNe due to stratification structures in the post-shock region imposed by spherical symmetry \citep[see, e.g.,][and references therein]{HiLeEn14, BrLeHi16}. In multi-dimensional (multi-D) studies, additional dimensionalities allow fluid motion development, thus breaking the adiabatically stratified structure, and setting the stage for fluid instabilities.  More importantly, multi-D allows for the co-existence of continuously accreting downstream flows and explosive outflows, with the former continuously converting gravitational potential energy into thermal energy and turbulent pressure, thereby enhancing neutrino heating and fueling the explosion. 

In multi-D studies, two fluid instabilities: neutrino-driven convection and the standing accretion shock instability (SASI), develop and take over the dynamics behind the shock during shock stagnation. It is the general consensus \citep[e.g.][]{Mezz05, Jank12, BrLeHi16, BrSiLe23} that one or both of these fluid instabilities are essential to shock revival and thus the neutrino-driven explosion. The SASI is the tendency for the standing shock to oscillate in non-radial low-order spherical harmonic modes, without convective instability. The SASI increases the mean shock radius and the advection time of matter as passing through the heating layer \citep{BlMeDe03, BlMe06, Jank12, BrLeHi16}. Neutrino-driven convection is tied to the entropy gradient imposed by the neutrino heating mechanism and tends to mix the post-shock fluid towards isentropy. In addition to exerting convective pressure to the shock, the convective flows enhance the critical conditions to revive the shock by increasing the effective neutrino heating efficiency and the dwell time of material in the heating layer \citep{Jank12, BrLeHi16, MuMeHe17, BrSiLe23}. 
The dwell time is commonly characterized by the advection timescale \tadv, which must be sufficiently long compared to the heating timescale \theat\ to allow neutrino heating to effectively energize the post-shock material before it leaves the heating region. 

Shock asphericity is augmented by large-scale convective plumes behind the shock forming lobular structures. When the accretion flow passes the shock, its velocity component normal to the shock surface is decreased subject to the shock jump conditions, while tangential velocity component remains unchanged and forms a post-shock boundary layer. The lobular structure of the shock focuses the accretion into streams that flow from the shock to the PNS surface.  The accretion stream accelerates radially inside the shock and may be supersonic and cause a secondary shock when hitting the PNS surface. As a result, a large amount of thermal energy is deposited onto the proto-neutron star (PNS) surface and increases the neutrino luminosity. 

\subsection{Multidimensional Perturbations and Explosion}

Because the tangential component of velocity survives passage through the shock, \citet{CoOt13} and \citet{CoChAr15} demonstrated that the non-spherical structure of the multi-D progenitors introduce fluctuations which enhance the post-shock turbulence. This provides a non-negligible turbulent pressure, pushing the shock outward and aiding in a successful explosion \citep{MuDoBu13, CoOt15}. This concept, that asphericity in the progenitor accelerates the development of asphericity during the explosion, motivates the work we present here and has been previously shown with progenitors with pre-collapse convection by \citet{CoChAr15,MuMeHe17, BoYaKr21} and \citet{VaCoBu22}.

\cite{CoChAr15} used a 15-\msun\ progenitor, evolving in 1D until the iron core mass is around 1.3~\msun\ during quasi-hydrostatic silicon shell burning, and then mapping to a 3D Cartesian octant mesh to build a 3D progenitor model that collapses, and compared it to a 1D version angle-averaged at the onset of collapse. They found that, prior to shock revival, the turbulent kinetic energy, measured by decomposition on a spherical harmonics basis, in the pre-shock region for the 3D progenitor model was an order of magnitude greater than for the angle-averaged progenitor model at all angular scales, $\ell$. This translates into greater turbulent energy in the gain region, the post-shock region outside the PNS with net neutrino heating, for spherical harmonics with $\ell=6$--10, where it is most effective at aiding shock expansion \citep{HaMaMu12, CoOc14, CoOt15}. 

In addition, \citet{CoChAr15} found that the neutrino heating efficiency, neutrino luminosity, and total neutrino heating rate in the gain region are basically identical, which narrows the cause of divergence between their full 3D model and its angle-averaged version to the turbulence introduced by asphericity in the 3D progenitor. They also observed a higher turbulent kinetic energy in the gain region of their 3D progenitor model between the shock reaching the silicon shell and revival, which also corresponds to the timing when the mean shock radius of the 3D progenitor model deviates from its angle-averaged counterpart.  Their results demonstrate the multi-D progenitor structure favoring the explosion and the importance of multi-D simulations to capture the lateral velocities and non-spherically symmetry structure.

\cite{MuMeHe17} computed the collapse and post-bounce evolution of a 1D progenitor model, s18-1D, and two 3D progenitor models, s18-3D and s18-3Dr, with an 18-\msun\ progenitor using the \coconut\ code \citep{MuJa15}. Their 3D progenitor models used the same 1D model, but computed the final 5 minutes of oxygen shell burning prior to the onset of collapse in 3D using the radial history of a mass zone in the silicon shell as an inner boundary before reinserting the 1D iron core and silicon shell at collapse. For s18-3Dr, the pre-collapse convective velocities were artificially reduced compared to s18-3D by lowering the nuclear burn rate. 
Although they did not find an essential turbulent kinetic energy difference between s18-3D and s18-3Dr before shock revival (see Section \ref{sec:structure} for more discussions) as \cite{CoChAr15}, they reported that, depending on the perturbation level, the shock revival occurred at $\tpb=300$ ms for s18-3D, $\tpb=500$~ms for s18-3Dr, but did not occur for s18-1D. Hence, \cite{MuMeHe17} concluded that initial perturbations from convective shell burning have a significant and qualitative impact on the fate of these progenitors.

\cite{BoYaKr21} evolved an 18.88-\msun\ progenitor in 1D until 7 minutes before core collapse, and then map it to 3D to the onset of collapse. They also compared a 1D progenitor run to the onset of collapse to this 3D progenitor model. From the onset of collapse, they varied angular grid resolution, equation of state (EoS), and muon formation in the PNS, building a series of 3D models based on 1D and 3D progenitors evolved with the \prometheusvertex\ hydrodynamics code until $\approx$7~s after bounce. The simulation results showed that all their 1D progenitor-based models do not explode, while all 3D progenitor-based models have successful explosions. In these 3D progenitor-based models, they find the infall of the large-amplitude density and velocity perturbations in the oxygen-burning shell fosters vigorous convection as the dominant hydrodynamic instability in the gain region, which is in line with the conclusions of \cite{CoChAr15} and \cite{MuMeHe17}. 

\cite{VaCoBu22} evolved two 3D progenitors from other groups and built their 1D counterparts to isolate the role of multi-D effects in the progenitor. The 1D progenitor models are mapped to 3D 10 ms after core bounce, so after core bounce all models are evolved in 3D with the \fornax\ code \citep{SkDoBu19}. 
The 12.5-\msun\ 3D progenitor model M12.5-3D3D from \cite{MuTaHe19} has non-radial perturbations from active oxygen shell burning, and the 15-\msun\ 3D progenitor model FC15-3D3D from \cite{FiCo20} has perturbations from silicon and oxygen-burning shells, while their 1D counterparts, M12.5-1D3D and FC15-1D3D, are evolved in spherical symmetry to collapse and thus have no perturbations. 
\cite{VaCoBu22} found that the 3D progenitor models preferentially explode. Their FC15-3D3D model explodes about 100 ms earlier than FC15-1D3D. The perturbed M12.5-3D3D explodes while unperturbed M12.5-1D3D does not. They concluded that the asphericity in the 3D progenitor models, around and interior to the Si/oxygen shell interface, seeds turbulence and is amplified during collapse, which aids the turbulence in the post-shock region and thus aids shock revival. 

\subsection{This Study}

\begin{figure}
    \centering
    \includegraphics[width=\columnwidth,clip]{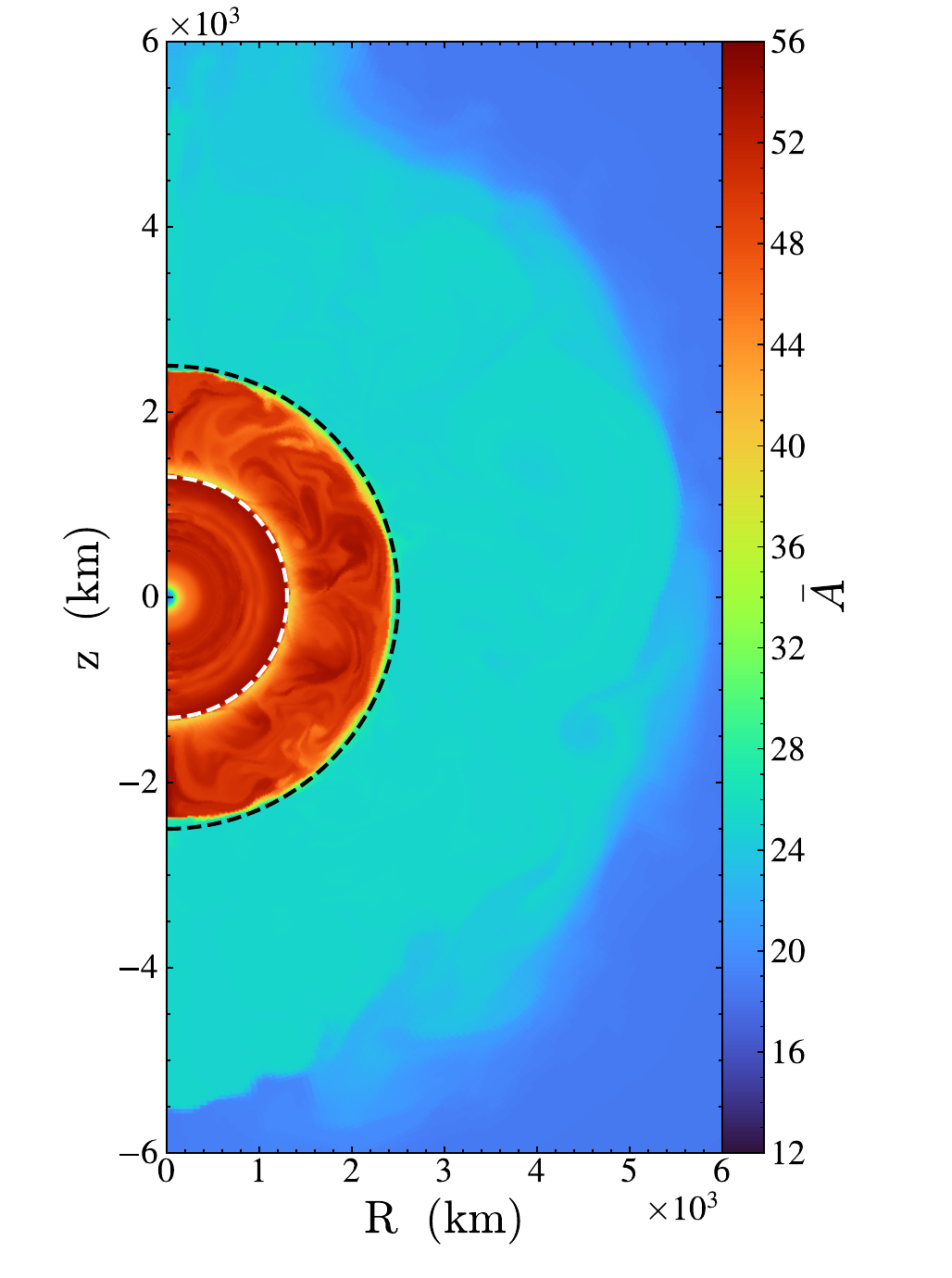}
    \caption{Mean atomic mass, $\bar{A}$, of G15-SiSB-M3 at the onset of collapse. The silicon-burning shell ranges from about 1300--2500 km (enclosed by two dashed contours) and perturbations can be seen within it, the convective oxygen-burning shell is the aqua region, and the non-convective oxygen shell is the blue region at the edge. }
    \label{fig:m3-abar}
\end{figure}

In the work described above, the sources of non-radial perturbations are from the silicon-burning shell in \cite{CoChAr15}, the oxygen-burning shell in \cite{MuMeHe17} and \cite{BoYaKr21}, and both shells in \cite{VaCoBu22}. In this paper we present a similar study for an axisymmetric progenitor where the asphericity in the multi-D progenitor structure comes from the silicon-burning shell and the convective oxygen-burning shell as shown in Figure~\ref{fig:m3-abar}. The earlier studies each compared the CCSN simulations started from multi-D progenitor to either the standard 1D stellar evolution model evolved to collapse or to models started from spherical averages of their multi-D progenitor. We make both comparisons. 

Section \ref{sec:methods} describes the methodology for our simulations and initial conditions of the progenitors. In Section \ref{sec:RD} we detail simulation results and differences caused by the density structure changes induced by multi-D pre-collapse evolution. In Section \ref{sec:structure} we isolate the impact of non-radial structure in the progenitor and discuss the impact of numerical noise on the supernova models.
We make a limited examination of model stochasticity in Section \ref{sec:stochasticity}, summarize our analysis in Section \ref{sec:summary}, and mark the major findings in Section~\ref{sec:conclusions}.

\section{Numerical methods and Model Set-up}
\label{sec:methods}

\subsection{Simulation code}

The supernova simulations that follow were computed using the \chimera\ neutrino radiation hydrodynamics code \citep{BrBlHi20} and are part of the \chimera\ `G-series' of simulations.
\chimera\ uses PPMLR hydrodynamics, multipole self gravitation with a monopole general relativistic (GR) potential correction, ray-by-ray 4-species neutrino transport by the flux-limited diffusion method, a dense nuclear equation of state at high temperatures where nuclear statistical equilibrium (NSE) applies and a nuclear network where it doesn't.
To resolve fluid flows as close to the center as possible without excessive time step restrictions we utilize a cell merging scheme (E.~J. Lentz et al., in prep.) in the lateral, $\theta$, hydrodynamic sweep.
Cells are merged during the lateral sweep to keep the radial and lateral dimensions approximately equal utilizing the closest integer divisor of merged cells that meets that goal.
A single, 1-km zone remains spherical at the center surrounded by shells inside the PNS approximately 100--200~m thick.

The neutrino opacities used are those used since the F-series simulations \citep{BrSiLe23} with the addition of neutrino pair conversion \citep[$\nue\nuebar \rightarrow \numt\numtbar$,][]{BuJaKe03} at for densities greater than $10^{12}$~\gcc.

The NSE EoS is the SFHo EoS of \citet{StHeFi13} incorporated into \chimera\ \citep{Land18}  using the WeakLib\footnote{\url{https://github.com/starkiller-astro/weaklib}} framework via the CompOSE\footnote{\url{https://compose.obspm.fr}} database.
New for Series-G runs, we use the \cite{TiSw00} `Helmholtz' electron EoS throughout.

In regions out of NSE, we use XNet \citep{HiTh99b} with the {\tt sn160} network designed to capture most isotopes and dominant energy-generating reaction flows relevant to CCSN conditions. 
The isotopes included are \isotope{n}{}, \isotope{H}{1\textrm{--}2}, \isotope{He}{3\textrm{--}4}, \isotope{Li}{6\textrm{--}7}, \isotope{Be}{7,9}, \isotope{B}{8,10,11}, \isotope{C}{12\textrm{--}14}, \isotope{N}{13\textrm{--}15}, \isotope{O}{14\textrm{--}18}, \isotope{F}{17\textrm{--}19}, \isotope{Ne}{18\textrm{--}22}, \isotope{Na}{21\textrm{--}23}, \isotope{Mg}{23\textrm{--}26}, \isotope{Al}{25\textrm{--}27}, \isotope{Si}{28\textrm{--}32}, \isotope{P}{29\textrm{--}33}, \isotope{S}{32\textrm{--}36}, \isotope{Cl}{33\textrm{--}37}, \isotope{Ar}{36\textrm{--}40}, \isotope{K}{37\textrm{--}41}, \isotope{Ca}{40\textrm{--}48}, \isotope{Sc}{43\textrm{--}49}, \isotope{Ti}{44\textrm{--}51},  \isotope{V}{46\textrm{--}52}, \isotope{Cr}{48\textrm{--}54}, \isotope{Mn}{50\textrm{--}55}, \isotope{Fe}{52\textrm{--}58}, \isotope{Co}{53\textrm{--}59}, \isotope{Ni}{56\textrm{--}64}, \isotope{Cu}{57\textrm{--}65}, \isotope{Zn}{59\textrm{--}66}, \isotope{Ga}{62\textrm{--}64}, and \isotope{Ge}{63\textrm{--}64}.
This is a slightly expanded version of the {\tt sn150} network \citep{Cher12,ChMeHi12,HaHiCh17} extended to include Ga and Ge and better flow into and out of \isotope{Ca}{48}, the most neutron-rich isotope in the network.
Reaction rates are taken from the REACLIB\footnote{\url{https://groups.nscl.msu.edu/jina/reaclib/db/}} compilation (V2.0) \citep{CyAmFe10} and supplemented/supplanted with $\beta$-decay rates and electron capture rates on heavy nuclei \citep{FuFoNe85,OdHiMu94,LaMa00}.

\subsection{Model Set-up}

The progenitors for this paper will be described in detail in a forthcoming \citep{LeKeHi26} study of late silicon shell burning in 2D.
All are started from a 15 $M_{\odot}$ stellar model computed with Modules for Experiments in Stellar Astrophysics (MESA) \citep{MESA11,MESA13,MESA15} stellar evolution code, from an unpublished 2016 study by J.~A. Harris using version r7624. A 204 species nuclear reaction network, \texttt{mesa\_204.net}, was used to sufficiently capture late stage nuclear burning and ensure an accurate electron fraction in the core at collapse by including neutron rich species. In general, parameters were set using the \text{inlist\_massive\_default} provided by MESA and the parameters from \citet{FaFiPe16}. 

All of our \chimera\ models include the inner 45,000~km, or 2.88~\msun, with 720 radial zones roughly geometric spacing and 240 uniform angular zones.
The first model in our set, G15-SiSB-M1, is started in the typical \chimera\ fashion from the 1D MESA progenitor at the onset of core collapse, which is defined as the progenitor reaches a negative radial velocity of 1000 \kms\ in the core. This progenitor is labeled `MESA' in Table~\ref{tab:Modelinfo}.

The rest of our models were started from the same MESA progenitor in the final silicon-shell phase, starting from 271~s before the MESA model collapses and computed to collapse using the \polaris\ stellar modeling code.
\polaris\ is derived from \chimera\ without the neutrino transport and with additional tuning for non-CCSN stellar evolution simulations.
The rest of the physics in \polaris\ are treated with the same methods as in \chimera\ including hydrodynamics, gravity, EoS, and nuclear network.
The inner 1.0~\msun\ of the 15-\msun\ MESA progenitor was treated in NSE with electron capture on nuclei computed using the LMSH \citep{LaMaSa03,HiMeMe03} formulation assuming instantaneous escape. The rest of the star used the \texttt{sn160} network in XNet.
Neutrino cooling for the whole model was implemented in \polaris\ using the formulation of \citet{Itoh96}.
The are two \polaris\ models used here, a 2D model (\polaris\ model SiSB-2-2D) matching the 720-radial, 240-lateral zone resolution of our CCSN models labeled `\polaris-2D', and a 1D version (\polaris\ model SiSB-2-1D) labeled `\polaris-1D' in Table~\ref{tab:Modelinfo}.
The \polaris-1D progenitor is the least physically realistic of all three progenitors as it includes only 1D hydrodynamics that suppresses all convection as there is no mixing length theory (MLT) in it. This model was created to help sort out differences due to dimensionality and those due to code physics and implementation.

The second model, G15-SiSB-M2, was started with the \polaris-1D progenitor.
A `stochastic partner' called G15-SiSB-M2b was created by restarting G15-SiSB-M2 at core bounce with a small code change (see below).
The remaining \chimera\ models derive from the \polaris-2D progenitor.
G15-SiSB-M3 utilizes the full 2D progenitor as the initial state, while the remaining three models have been spherically averaged prior to their \chimera\ runs.
In G15-SiSB-M4, the lateral velocities, $v_\theta$, were averaged, but not zeroed, providing a second model which collapses with non-zero $v_\theta$.
In the final two models, the lateral velocities were set to zero prior to the \chimera\ run.
G15-SiSB-M5 was collapsed from the spherically averaged state in 2D as is the typical \chimera\ practice, but G15-SiSB-M6 was collapsed in spherical symmetry with \chimera\ before being remapped to 2D at core bounce without added perturbations.
For brevity, the models are labeled, M1, M2, etc. in plots and the presentation of results below.
One small code change made between models M1 and M2 and the rest was an improvement in angular momentum conservation in the radial hydrodynamic sweep update of $v_\theta$ \citep[Equation (117) in][]{BrBlHi20}. It was updated from a thin-shell approximation for the moment of inertia to a thick, uniform shell approximation. This modest change reduces $v_\theta$ `noise' amplification during collapse in test problems and was intended for unrelated future work with rotating progenitors. 

\section{Simulation results}
\label{sec:RD}

\begin{deluxetable*}{@{\extracolsep{4pt}}lcccccccc}
	\tablewidth{0pt}
	\tablecolumns{9}
	\tablecaption{Model Overview \label{tab:Modelinfo}}
	\tablehead{
		 \colhead{} & \colhead{}  & \multicolumn{3}{c}{Progenitor} & \colhead{Shock\tablenotemark{a}} & \multicolumn{3}{c}{Shock Arrival\tablenotemark{a}}  \\
         \cline{3-5} 
         \cline{7-9}
         \colhead{Model}	& \colhead{Duration\tablenotemark{a}} & \colhead{Source\tablenotemark{b}}	& \colhead{Radial}	& \colhead{$v_\theta$} & \colhead{Revival}    & \colhead{ at 1.3 \msun}    & \colhead{ at 1.5 \msun}    & \colhead{ at 1.7 \msun} \\
		 \colhead{} 		& \colhead{[ms]}	            & \colhead{}	                    & \colhead{structure}	                & \colhead{}                    & \colhead{[ms]}             & \colhead{[ms]}                        & \colhead{[ms]}                        & \colhead{[ms]}
	}
	\startdata
	\hline
	G15-SiSB-M1				& 1128	        & MESA	        & 1D	    & 0   & 238      & 34.4     & 244   & 697	\\
	G15-SiSB-M2 			& 1097			& \polaris-1D	& 1D		& 0	 & 203      & 32.2     & 205   & 618    \\
	G15-SiSB-M2b 			& 600			    & \polaris-1D	& 1D		& 0   & 205      & 32.2         & 205         & \nodata     \\
	G15-SiSB-M3				& 1341	        & \polaris-2D	& 2D		& 2D	 & 240      & 37.6     & 263   & 715    \\
	G15-SiSB-M4			    & 1159			& \polaris-2D	& averaged		& averaged	 & 222 & 37.2     & 258   & 657    \\
	G15-SiSB-M5				& 1000			& \polaris-2D	& averaged		& 0	 & 225      & 37.6     & 255   & 674    \\
	G15-SiSB-M6				& 1000			    & \polaris-2D	& averaged		& 0	 & 230      & 37.8     & 256   & 663    \\
	\enddata
	\tablenotetext{a}{Times given relative to core bounce.}
        \tablenotetext{b}{See text.}
\end{deluxetable*}

\subsection{Initial Conditions and Density Structure}
\label{sec:inicondi}

In this subsection, we will discuss the differences in internal structure from the treatment of silicon shell burning (between 1.3 \msun\ and 1.5 \msun) and how they are translated after core bounce. We compare M1 for the MESA progenitor, M2 for the \polaris-1D progenitor, and M3 for the \polaris-2D progenitor. The differences in their precollapse density structures and nuclear compositions lead to different behaviors in the later explosions.

\begin{figure}
    \fig{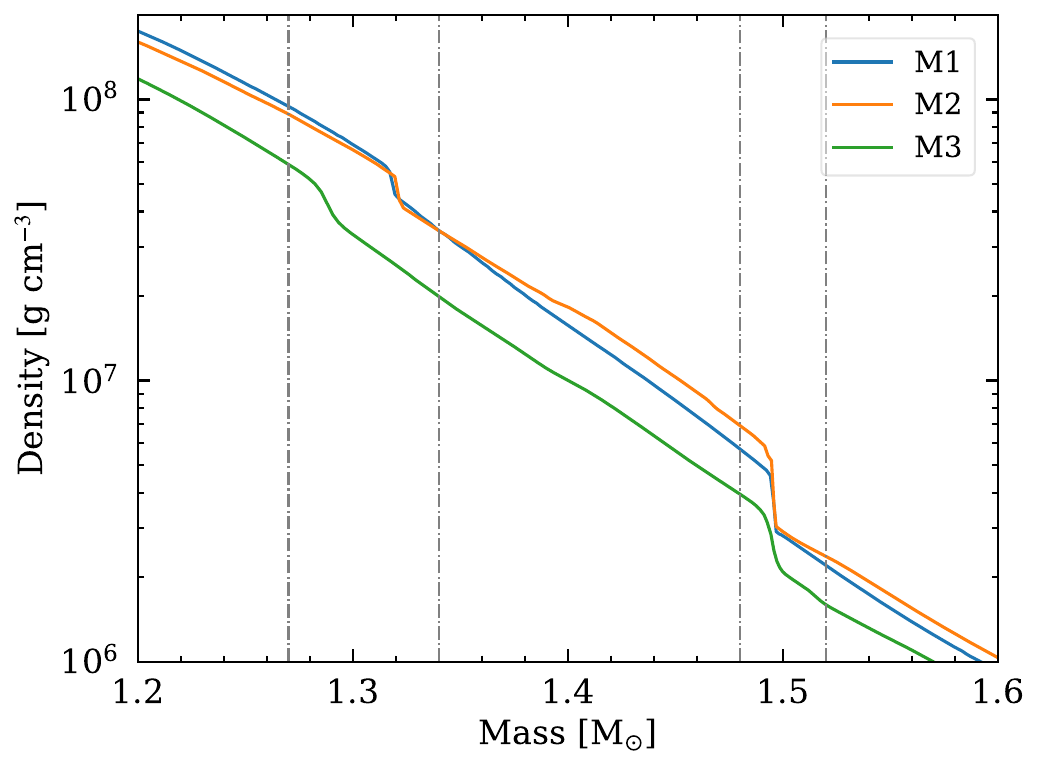}{\columnwidth}{(a)}
    \fig{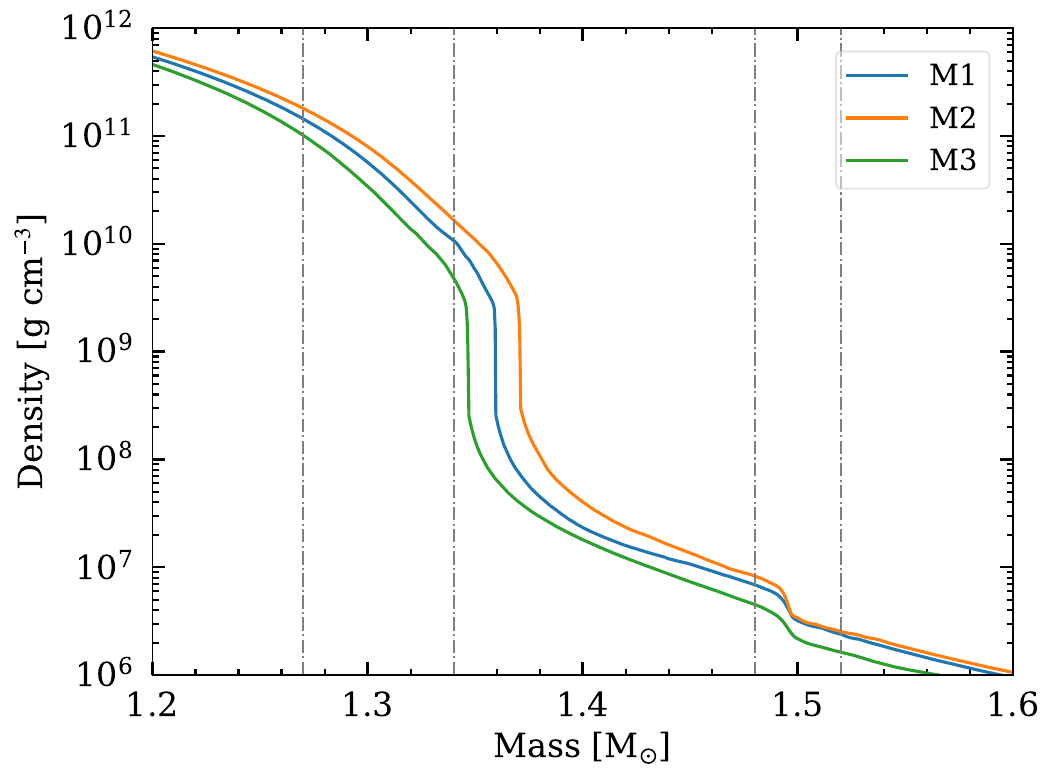}{\columnwidth}{(b)}
    \caption{Angle-averaged density profiles of models from different pre-collapse progenitors (a) in the collapsing phase when the central density reaches \ee{1.5}{10} \gcc, and (b) at 60 ms post-bounce. Vertical dashed lines indicate mass coordinates of 1.27, 1.34, 1.48, and 1.52 \msun. }
    \label{fig:dens-diff}
\end{figure}

Figure \ref{fig:dens-diff}(a) shows the angle-averaged density profile  for M1, M2, and M3 early in core collapse when the central density reaches $1.5\times10^{10}\ \gcc$. The steep negative density gradients near 1.3 \msun\ and 1.5 \msun\ correspond to Fe/Si and Si/oxygen shell interfaces, respectively. Similar sharp decline in density at the Si/O interfaces enhance the possibility of explosion in many simulations \citep{VaLaRe21, VaCoBu22, BrSiLe23}.

The \polaris-2D progenitor (green line in Figure \ref{fig:dens-diff}) is less compact in overall structure, therefore lower density at collapse, and thus takes longer to collapse. Its lower density persists, despite the longer collapse, throughout shock passage through the silicon-burning shell. 
At $\tpb=60$ ms, the mass coordinate 1.34 \msun\ is post-shock while 1.48 \msun\ is pre-shock. Both the densities and density differences between the models are amplified during the collapsing phase and by the shock wave propagating through. 

\begin{deluxetable}{lccc}
	\tablewidth{0pt}
	\tablecolumns{4}
	\tablecaption{Densities near silicon shell \label{tab:density}}
	\tablehead{
		 \colhead{G15-SiSB-}	& \colhead{-M1} 	& \colhead{-M2}	& \colhead{-M3}	    
	}
    \startdata
     \hline
        \sidehead{onset of collapse, inner edge}
        \hline
      	$\rho_{1.27}$ [\gcc] & \ee{9.46}{7}  & \ee{8.88}{7}  & \ee{5.88}{7}  \\
		relative to M3               & 1.61  & 1.51  & 1 \\
		$\rho_{1.34}$ [\gcc] & \ee{3.42}{7}  & \ee{3.42}{7}  & \ee{1.99}{7}\\
		relative to M3               &  1.72 & 1.72  & 1 \\
		$\rho_{1.34}/\rho_{1.27}$     &  0.36 & 0.38  & 0.34 \\
        \hline
        \sidehead{onset of collapse, outer edge}
        \hline
		$\rho_{1.48}$  [\gcc] & \ee{5.71}{6}   & \ee{6.93}{6}  & \ee{3.95}{6} \\
		relative to M3               & 1.45  & 1.76  & 1 \\
		$\rho_{1.52}$  [\gcc] & \ee{2.20}{6}  & \ee{2.37}{6}   & \ee{1.59}{6}  \\
		relative to M3               & 1.38  & 1.49  & 1 \\	
		$\rho_{1.48}/\rho_{1.52}$     & 2.60  & 2.92  & 2.47\\
    \hline
    \sidehead{60 ms after bounce, inner edge}
    \hline
        $\rho_{1.27}$ [\gcc] & \ee{1.45}{11}  & \ee{1.81}{11}  & \ee{1.02}{11} \\
		relative to M3               & 1.42  & 1.78 & 1 \\
		$\rho_{1.34}$ [\gcc] & \ee{1.08}{10}  & \ee{1.66}{10}  & \ee{4.82}{9}\\
		relative to M3               & 2.23  & 3.45  & 1 \\
		$\rho_{1.34}/\rho_{1.27}$     &  0.074 &  0.092 & 0.048 \\
        \hline
        \sidehead{60 ms after bounce, outer edge}
        \hline
		$\rho_{1.48}$ [\gcc] &  \ee{6.88}{6}  & \ee{8.28}{6}  & \ee{4.51}{6} \\
		relative to M3               & 1.52  &  1.83 & 1 \\
		$\rho_{1.52}$ [\gcc] & \ee{2.39}{6}  & \ee{2.53}{6}  & \ee{1.63}{6} \\
		relative to M3               & 1.46  &  1.55 & 1 \\	
		$\rho_{1.48}/\rho_{1.52}$     &  2.88 & 3.27  &  2.76\\
    \enddata
    
\end{deluxetable}

Values for the density of the three models around the silicon shell are given in Table~\ref{tab:density}.
For all three models, the measured density contrast through both the upper and lower boundaries, $\rho_{1.34} / \rho_{1.27}$ and $\rho_{1.48} / \rho_{1.52}$, of the silicon shell is 0.3--0.4. 
The reference mass points are set to capture the shift of the 2D progenitor silicon shell inward due to convective boundary mixing dredging up the outer portion of the pre-collapse iron core and the less sharp nature of the interface in the 2D progenitor.
The 1D progenitors have densities between 1.4 and 1.8 times larger at the four control points than the 2D progenitor, with the larger ratios inside the silicon shell and the smaller differences beyond the silicon shell. 
By 60~ms, the shock has propagated into the silicon shell passing the two mass reference points of the inner silicon shell edge, the passage of the shock steepens the density contrast and leaves these three models with significantly different density slopes behind it. 
The pre-shock density contrast at the outer edge, $\rho_{1.48} / \rho_{1.52}$, also increases from the values at the onset of collapse. 
The 1D progenitors still have densities larger at the four control points than the 2D progenitor, with the larger ratios inside the silicon shell and the smaller differences beyond the silicon shell. 
We also note that the density ratios at two control points in the silicon shell, $\rho_{1.34}$ and $\rho_{1.48}$ compared to M3, are larger than the ratios at the onset of collapse. 
This buttresses the idea that the pre-bounce features will be preserved and amplified through collapse and core bounce, thus affecting the eventual explosion.

The multi-D nuclear burning in the progenitor in M3 introduces turbulence that is not found in 1D models.  This will be amplified during the collapsing phase. As the result, we will have pre-existing turbulence in models using the \polaris-2D progenitor at the moment of core bounce allowing comparison to other groups \citep{CoChAr15, MuMeHe17, BoYaKr21, VaCoBu22} that suggest these amplified perturbations revive the stalled shock earlier and induce more powerful explosions. 
The silicon shell in the \polaris-2D progenitor, 1.3--1.5~\msun, has convective velocities of $\approx$500~\kmps\ and $\ell\approx7$--8, while the oxygen burning shell has convective velocities of $\approx$200~\kmps\ and $\ell\approx4$--5.

\begin{figure}
    \fig{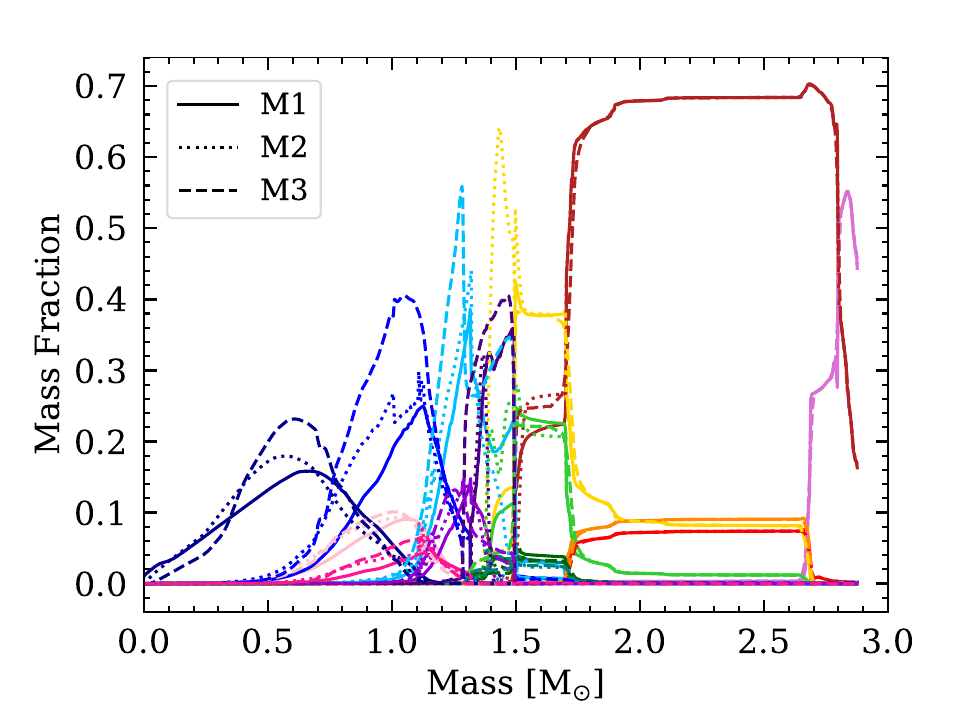}{\columnwidth}{(a)}
    \fig{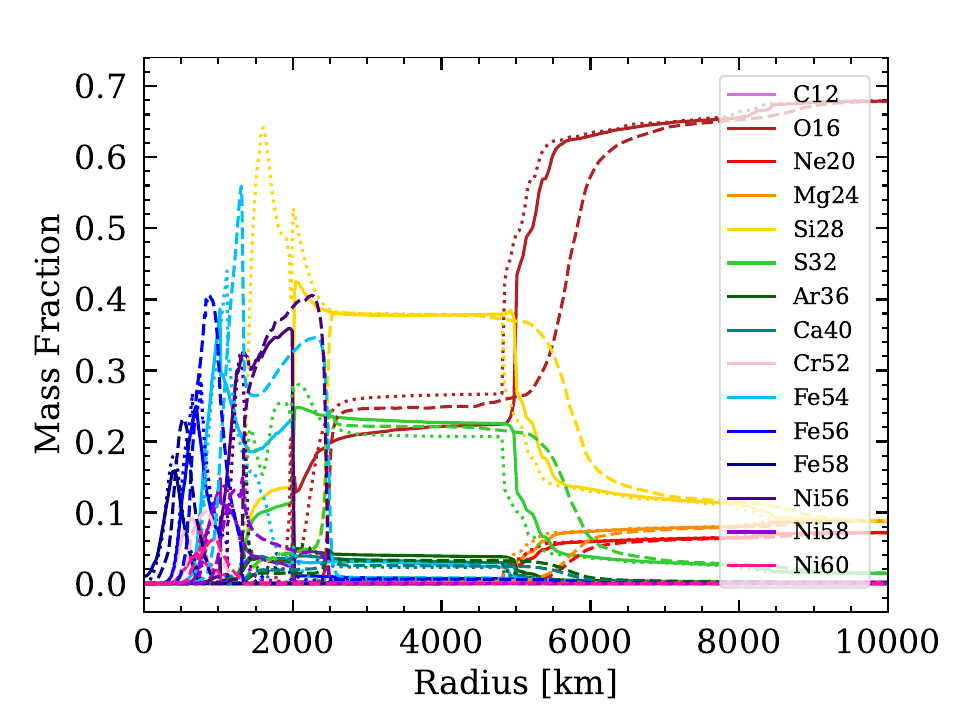}{\columnwidth}{(b)}
    \caption{Initial nuclear abundances of M1 (solid line), M2 (dotted line), and M3 (dashed line) as a function of (a) enclosed mass and of (b) radius from the center of the core to a radius of 10000 km at the onset of collapse.}
    \label{fig:abund-cc}
\end{figure}

Figure \ref{fig:abund-cc} shows the nuclear abundances of our models at the onset of core collapse. Differences in the progenitor structure are located within the convective oxygen shell (the inner part of the oxygen shell) enclosing 1.7~\msun. Between 1.5--1.7 \msun, the lower \isotope{O}{16} mass fraction indicates that oxygen burning is more active in the MESA model (M1) than in the \polaris-1D (M2) and \polaris-2D (M3) models.  They all have similar mass fractions of silicon, which indicates that the more vigorous oxygen burning in the MESA model (M1) generates more intermediate mass nuclei, such as \isotope{S}{32}, \isotope{Ar}{36}, and \isotope{Ca}{40}, shown in Figure \ref{fig:abund-cc}. In the silicon shell between 1.3--1.5 \msun, silicon burning significantly differs under the MESA, \polaris-1D (M2), and \polaris-2D environments. Under \polaris-2D (M3), \isotope{Si}{28} and \isotope{S}{32} are almost depleted throughout the shell with only mass fractions $X \approx 0.04$ left, while under MESA, there are still $X\approx 0.14$ of \isotope{Si}{28} and $X\approx 0.12$ of \isotope{S}{32} remaining, except at the inner edge of the shell. For \polaris-1D (M2), the \isotope{Si}{28} mass fraction remains high, up to 0.64 in the upper silicon-burning shell between 1.4--1.5~\msun. This \isotope{Si}{28} abundance peak corresponds to the concavity in \isotope{S}{32}, \isotope{Ar}{36}, and \isotope{Ca}{40}, which is an internal structure only seen in this progenitor. Little iron group material is synthesized in the upper silicon-burning shell in this model. Still, in the lower silicon-burning shell between 1.3--1.4 \msun, \isotope{Si}{28} depletes and \isotope{Ni}{56} accumulates, which peaks to $X \approx 0.32$. This comparison shows that without convection, treated in 2D in \polaris-2D (M3) or approximated with mixing length theory (MLT) in MESA (M1), silicon shell burning is highly localized at the inner edge, as seen for \polaris-1D (M2). 
We conclude that, in the silicon-burning shell, the \polaris-2D progenitor shows very efficient mixing and complete burning of silicon and sulfer to iron-group elements, while the MESA progenitor, with MLT, is less efficient at mixing silicon and sulfer into the burning layer and leaves more unburnt in the outer silicon-burning shell. The lack of mixing in the \polaris-1D model results in localized burning only sufficient to trigger core collapse. 
This highlights the need for some convective treatment for silicon shell burning and suggests that MLT may underestimate the extent or speed of mixing \citep{LeKeHi26}. 

\begin{figure}
    \fig{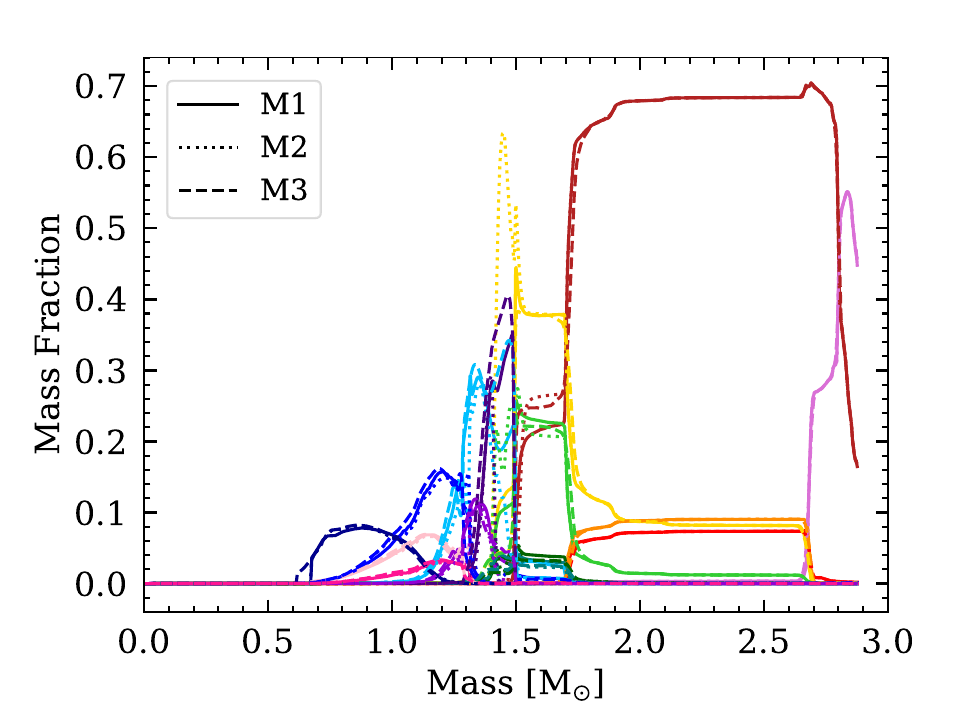}{\columnwidth}{(a)}
    \fig{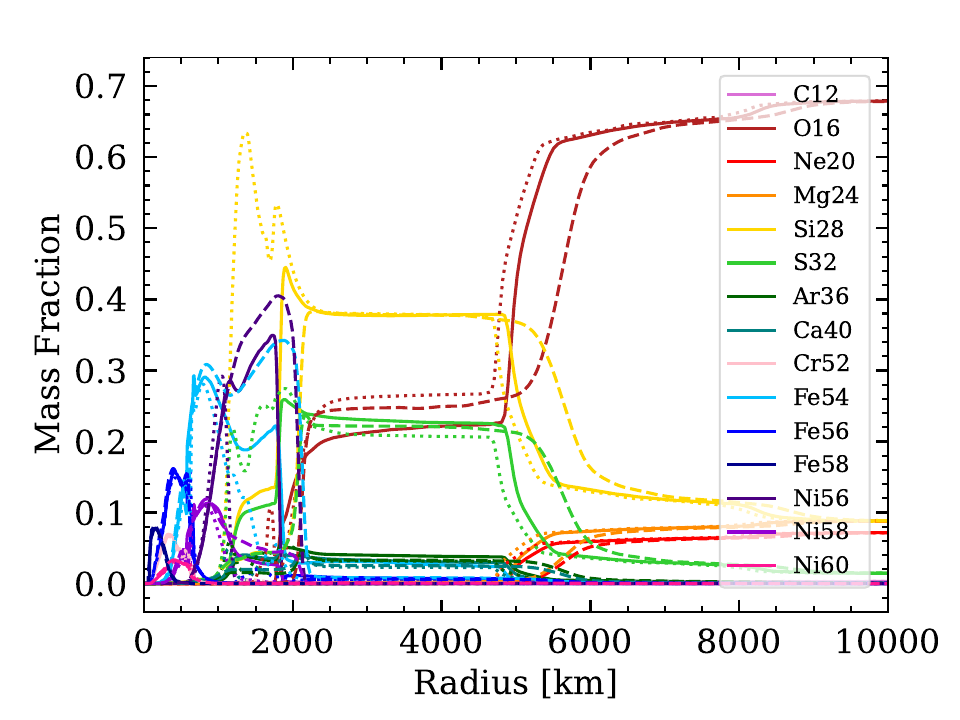}{\columnwidth}{(b)}
    \caption{Nuclear abundances  at core bounce of M1 (solid line), M2 (dotted line), and M3 (dashed line) as a function (a) of enclosed mass and (b) of radius from the center of the core to a radius of 10000 km. }
    \label{fig:abund-ms0}
\end{figure}

The nuclear abundances at core bounce above the iron core are largely unchanged from those at the onset of collapse as seen when we compare Figure~\ref{fig:abund-ms0}(a) to Figure~\ref{fig:abund-cc}(a).  Within the iron core, nuclei are transformed to more neutron-rich species by electron capture during collapse, hence the mass fractions of iron group nuclei are lower in Figure \ref{fig:abund-ms0} compared to Figure \ref{fig:abund-cc}, replaced by species not plotted. 

M1 and M2 both have 1D progenitors, so the comparison of their nuclear abundances at core bounce is dominated by the difference in pre-collapse burning in MESA and \polaris-1D. The high fraction of silicon in the silicon-burning shell between 1.4--1.5\ \msun\ in M2, due to reduced burning resulting from the lack of mixing, is still present. As a result, there is less \isotope{Fe}{54} and almost no \isotope{Ni}{56} but more \isotope{Si}{28} and \isotope{S}{32} between 1.4--1.5 \msun, which is shown in the upper panel of Figure \ref{fig:abund-ms0} and corresponds to the range 1000--2000 km in the lower panel. The previously discussed higher \isotope{O}{16} fraction between 1.5--1.7 \msun\ in M2 persists, as does the higher mass fraction of \isotope{S}{32}, \isotope{Ar}{36}, and \isotope{Ca}{40} in M1.  
The comparison between M1 and M3 in Figure \ref{fig:abund-ms0} is likewise dominated by differences between the 1D and 2D progenitors pre-collapse burning environments. Inside of 1.5 \msun\ enclosed, M3 has more \isotope{Ni}{56} and less \isotope{Si}{28} and \isotope{S}{32} than M1, indicating its silicon burning has been more active. In contrast, the convective oxygen-burning shell (about 1.5--1.7\ \msun), M3 has more \isotope{O}{16} and less intermediate nuclei: \isotope{S}{32}, \isotope{Ar}{36}, and \isotope{Ca}{40}, indicating its oxygen burning has been less active than in M1. 

The coincident shell interfaces between these models in Lagrangian space (Figure \ref{fig:abund-ms0}(a)), which have diverged significantly in Eulerian space (Figure \ref{fig:abund-ms0}(b)), highlight the differences prior to core collapse. 
While these differences have developed over the course of these stars' final evolution, the differences at core bounce are significant.  In M3, the 1.5 \msun\ interface lies at a radius 300 km ($\sim$20\%) larger than the other two models.   At the 1.7~\msun\ interface, M2 has a slightly ($< 200$~km) smaller radius than M1, whereas M3 has a much larger ($>600$ km) radius.  Since M2 and M3 have identical EoS, the implication is that turbulent pressure in M3 slowed the contraction of its shells. The combination of these differences with differences due to weaker silicon burning in the 1.3--1.5 \msun\ shell for M2 discussed above, results in the edge of the iron core is displaced by almost 1000 km in that model.

\begin{figure}
    \fig{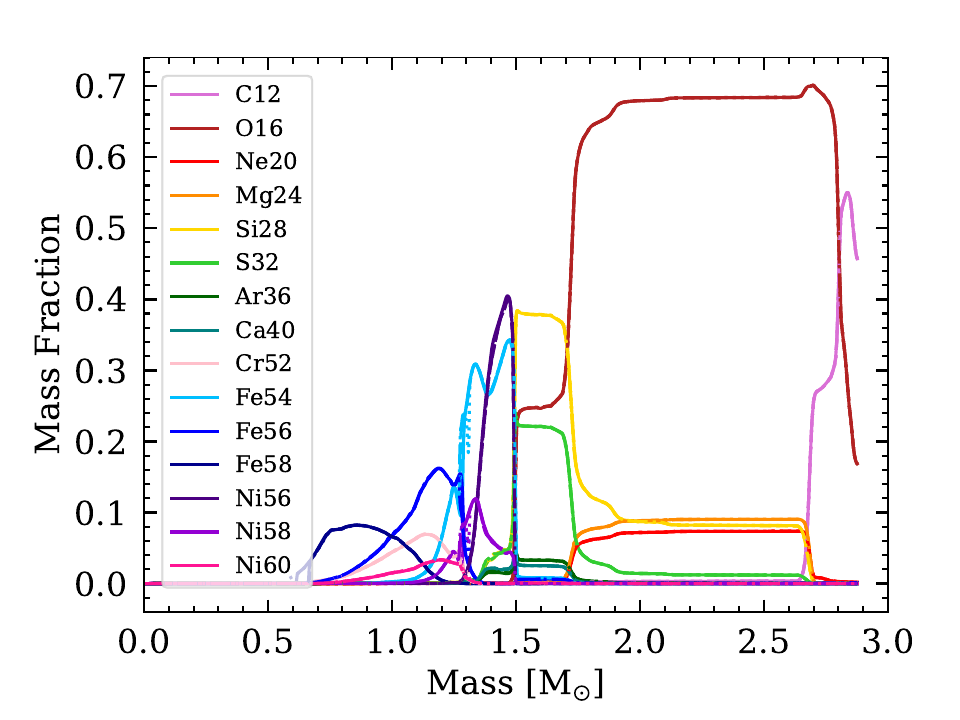}{\columnwidth}{(a)}
    \fig{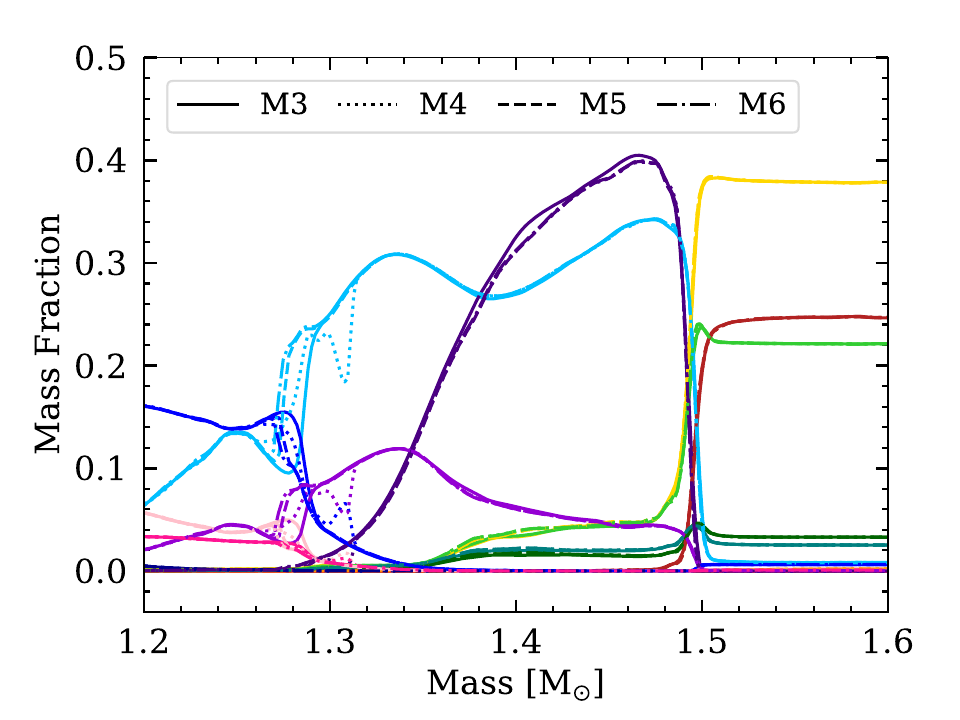}{\columnwidth}{(b)}
    \caption{Panel (a): Nuclear abundances at core bounce of M3 (solid line), M4 (dotted line), M5 (dashed line), and M6 (dashdot) as a function of enclosed mass. Panel (b): Zoomed in to 1.2--1.6 \msun, to illustrate details of silicon-burning shell including in the convective erosion of the iron core. }
    \label{fig:abund-ms0-pol}
\end{figure}

To isolate the effect of spherical symmetry on the models, we compare the nuclear abundances of M3, M4, M5, and M6 at bounce since they originate from the same 2D progenitor through the pre-collapse burning stage. As shown in Figure \ref{fig:abund-ms0-pol}, their nuclear abundances are nearly identical except for nuanced differences among iron group isotopes within the silicon-burning shell. This difference has a negligible impact on what follows.

\begin{figure*}[htbp]
    \gridline{
        \fig{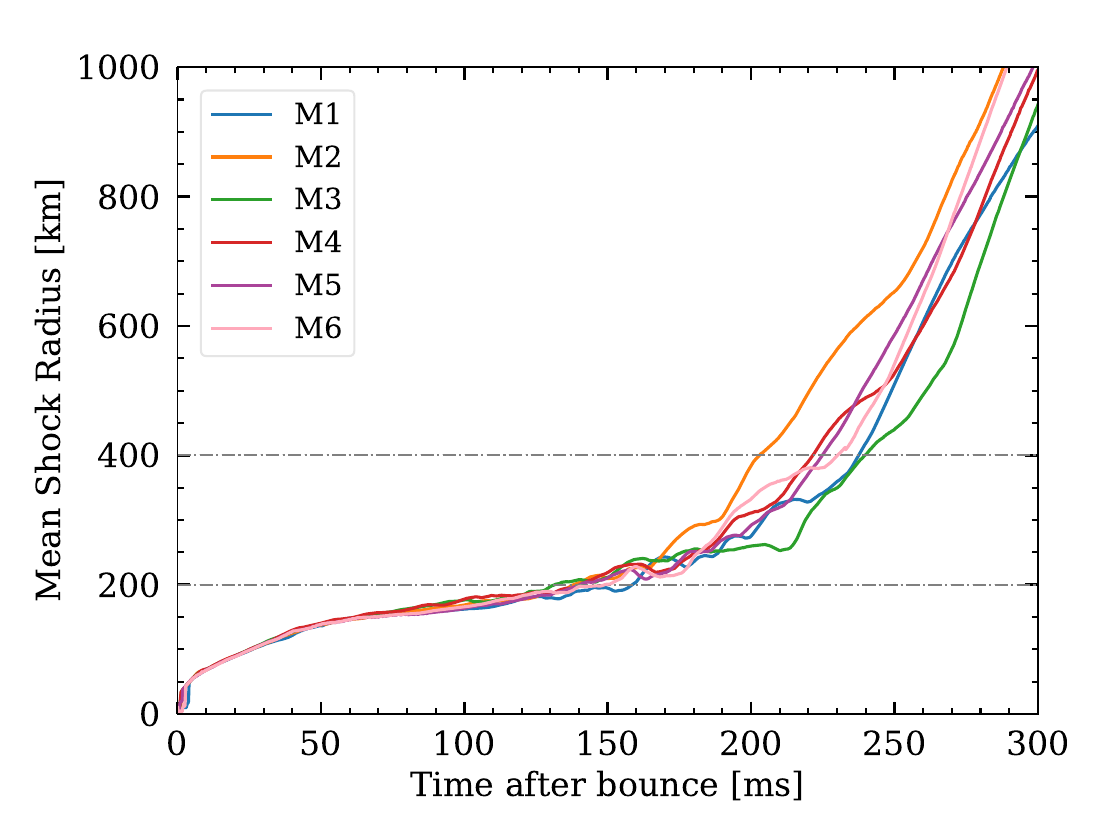}{\columnwidth}{(a)}
        \fig{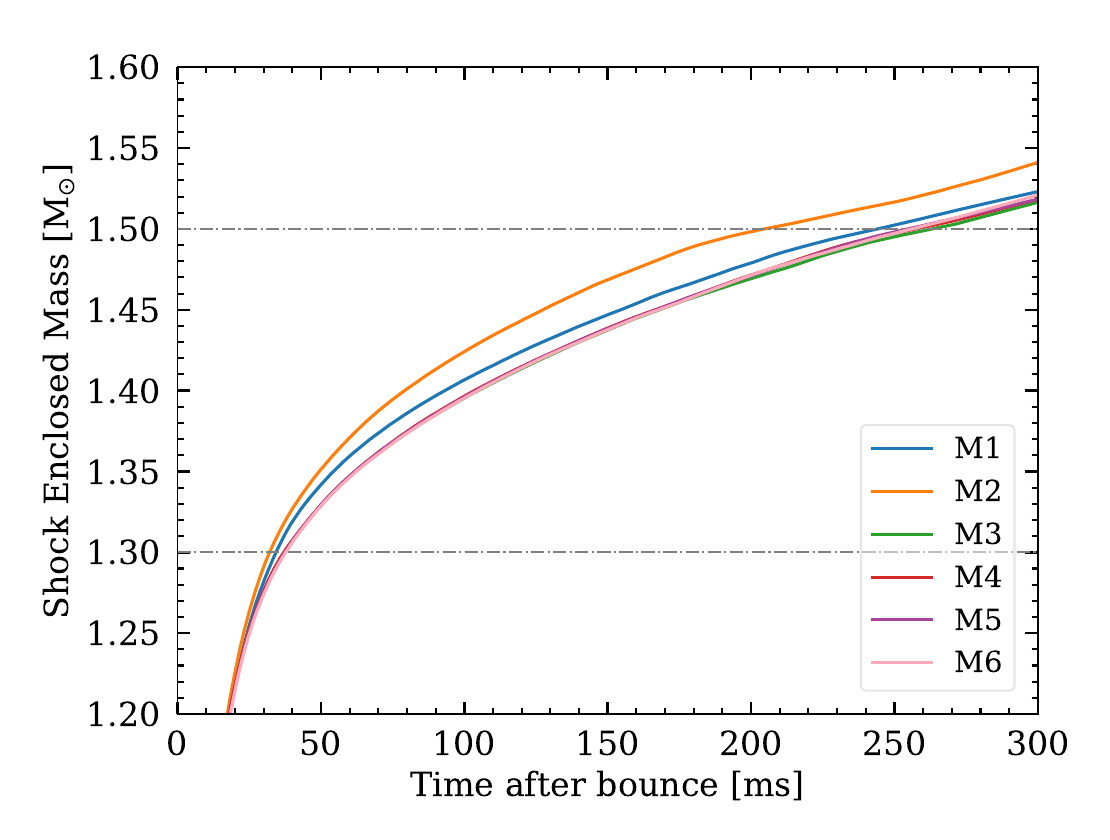}{\columnwidth}{(b)}
    }
    \gridline{
        \fig{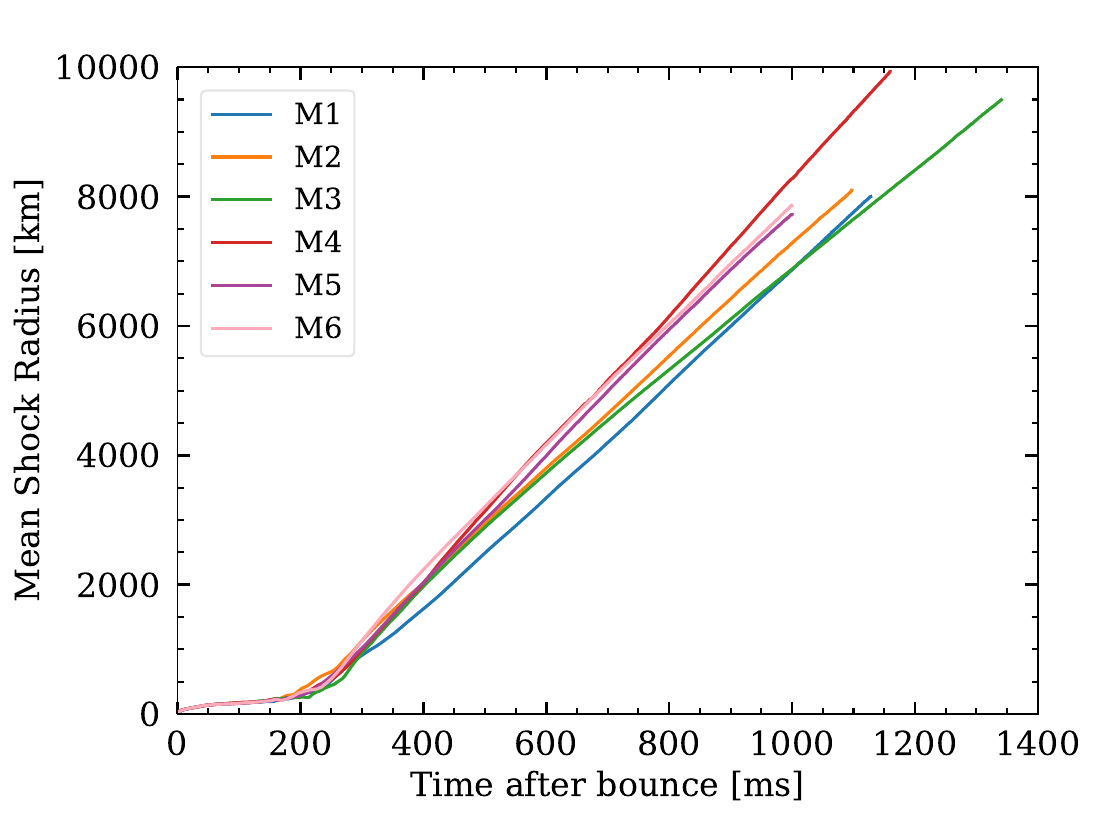}{\columnwidth}{(c)}
        \fig{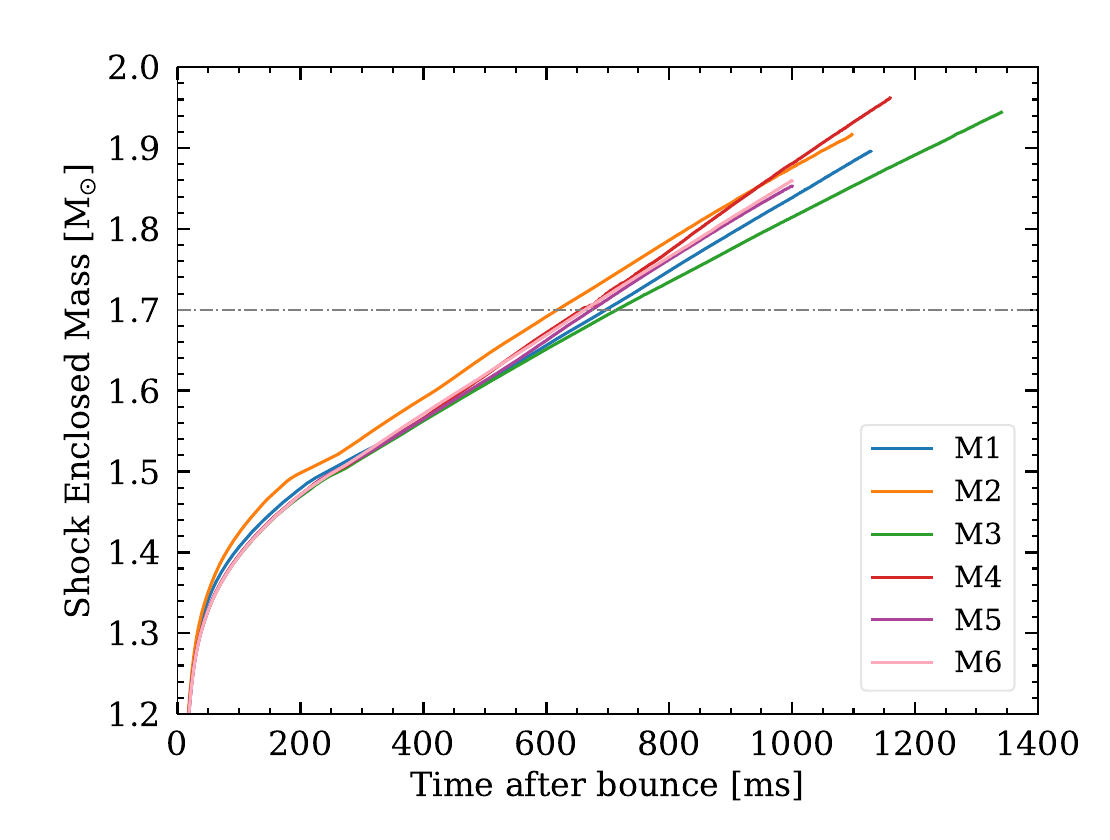}{\columnwidth}{(d)}
    }
    \caption{Left two panels, (a) and (c): mean shock radius; Right two panels, (b) and (d): mass enclosed by shock \Mshock. Upper panels, (a) and (b): cover first 300 ms after bounce; Lower panels, (c) and (d): full simulation. Radius of 200~km and 400~km in panel (a), Fe-core edge at 1.3 \msun\ and Si/O interface at 1.5 \msun\ in panel (b), and the top of convective O-shell at 1.7 \msun\ in panel (d) are marked with gray dash-dot lines.
    }
    \label{fig:mean-shock}
\end{figure*}

\subsection{Shock Development}
\label{sec:shock_dev}

Figure \ref{fig:mean-shock}(a) shows that the expansion of the mean (angle-averaged) shock slows significantly by $\tpb=50$ ms for all models, stagnating into an accretion shock. Shock expansion never completely halts or reverses, but it does slow down or 'pause'. 
The mean shock radii of six models start to deviate from each other around 130 ms after bounce with radius close to 200 km. 
Starting around $\tpb = 130$ ms, fluctuations in the shock radius prior to shock revival, marking the onset of explosion, become noticeable in all models. 
The shock revival happens at around 200 ms after bounce for all six models (Figure~\ref{fig:mean-shock}(a)). 
To quantify the timing of shock revival, we define it as the moment that the shock radius exceeds 400 km and does not retreat thereafter. 
As the shock radius expands beyond 400 km (Table \ref{tab:Modelinfo}), the models enter the rapid shock expansion phase with M2 exploding earlier than the other five models, which have a similar time for the onset of explosion, with M1 and M3 being the last. 
In the view of mean shock radius development, we do not see a large difference between these six models.

To understand the factors impacting the timing of shock revival, it is important to identify the shell in which the shock revives, as well as tracing its progression through the convective-burning shells. The nuclear abundance plot at core bounce (Figure \ref{fig:abund-ms0}(a)) shows the Fe/Si interface is 1.3 \msun, the Si/O interface is at 1.5 \msun, and the top of the convective oxygen-burning shell is at 1.7 \msun. Figure \ref{fig:mean-shock}(b) shows when the shocks breach the Fe/Si and Si/O interfaces, and Figure \ref{fig:mean-shock}(d) shows when the shocks exit the convective oxygen-burning shells. The shock waves of all six models breach their Fe/Si interfaces between $\tpb=30$--40~ms, before shock stagnation, but they breach the Si/O interfaces and enters the convective oxygen-burning shells at very different times. The shock of M2 breaches the Si/O interface and convective oxygen-burning shell the earliest, at $\tpb=205$ ms and $\tpb=618$ ms, respectively. The shock of M1 enters the convective oxygen-burning shell at $\tpb=244$ ms, and the shock of models with \polaris-2D progenitors cross the Si/O interface nearly simultaneously at $\tpb\approx 260$ ms. 
 
The shocks of M4 and M6 break through the convective oxygen-burning shells near $\tpb\approx 660$ ms, the shock of M5 enters its non-convective oxygen shell at $\tpb=674$ ms, the shock of M1 enters its non-convective oxygen shell later ($\tpb=697$ ms), and the shock propagation of M3 being the last, breaches the convective oxygen shell at $\tpb=715$ ms. At the end of these models runs, the enclosed mass of the shock \Mshock\ exceeds 1.7 \msun\ for all our models, which indicates the shock has crossed the Si/O interface and exited the convective oxygen-burning shell, propagating into the non-convecting part of the oxygen shell.

In all six simulations, shock stagnation and revival both happen in the silicon-burning shell. This means the stalled shock wave stays in the silicon-burning shell for a sufficient time that perturbations in this shell, if they are present in a multi-D progenitor, can affect shock revival. We will explore the role of these non-radial perturbations in Section \ref{sec:structure} through comparisons between M3, M5, and M6.

Aside from shock propagation itself, an indicative criterion for shock revival by neutrino heating is the timescale ratio $\tadv/\theat$. The advection timescale \tadv\ measures the time spent by a fluid element in the gain region, which is approximated by the ratio of mass in the gain region ($M_\mathrm{gain}$) over the mass accretion rate of this region (\accgain), assuming steady-state conditions, $\tadv \equiv M_\mathrm{gain} / \accgain$. The heating timescale \theat\ measures the time required to increase the total energy of a fluid element $e$-times by neutrino heating. This is calculated for the gain region by the ratio of the integrated total energy ($E_\mathrm{gain}$) to the net neutrino heating rate (\dotQnu), $\theat \equiv E_\mathrm{gain} / \dotQnu$ \citep{MuMeHe17, BrSiLe23}. For $\tadv/\theat>1$, fluid elements remain in the gain region long enough to significantly increase their total energy, enabling them to overcome the gravitational potential and contribute to a successful explosion. 

\cite{MuMeHe17} report the explosions in their models are triggered before reaching the criterion $\tadv/\theat = 1$. Specifically, the explosions in their models s18-3D and s18-3Dr occur roughly 22\% and 16\% below the critical value. They deduce the early explosions may originate from the shock reaching the Si/O interface and experiencing strong forcing by infalling perturbations. \cite{BrSiLe23} report explosions in two similar progenitor mass models, with their F15.78 model reviving roughly when \tadv/\theat\ exceeds unity (coinciding to a similar interface interaction), while F15.79 waits $\approx$50 ms after exceeding the threshold value before exploding. 

\begin{figure}
    \includegraphics[width=\columnwidth,clip]{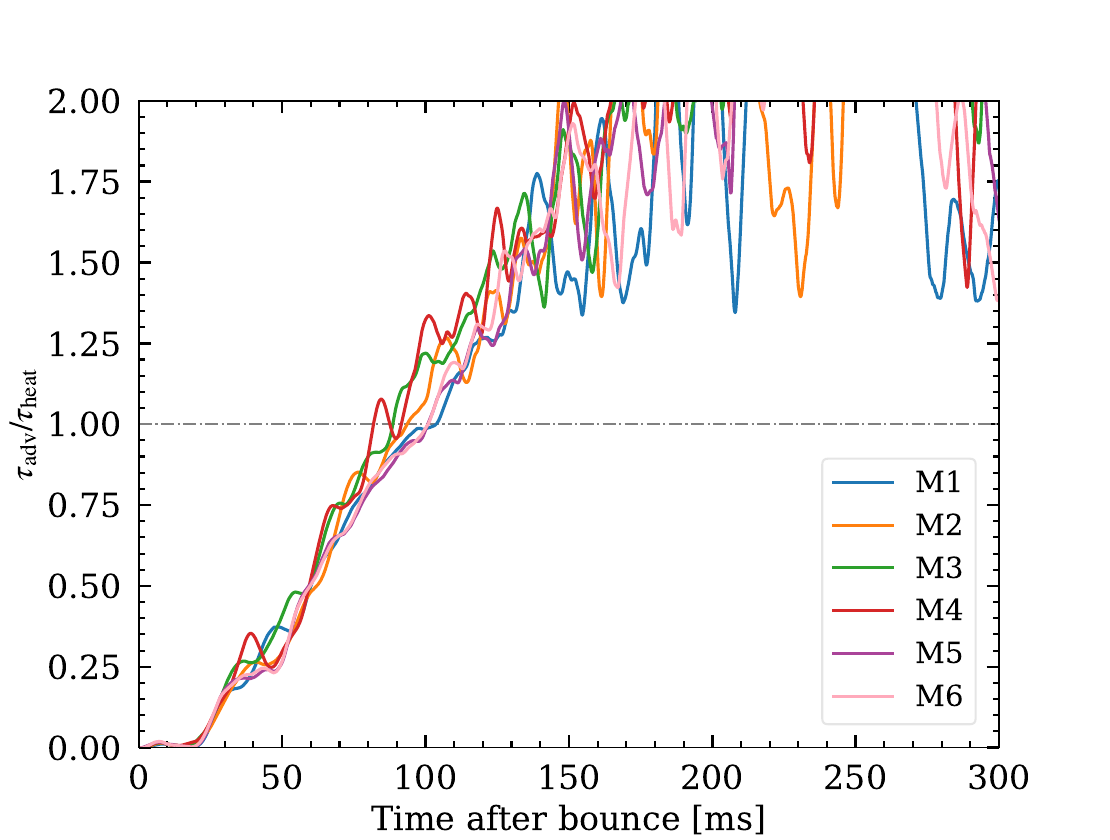}
    \caption{Timescale ratio \tadv/\theat\ to $\tpb=300$ ms for all six models. The shock revival threshold value $\tadv/\theat=1$ is marked by a gray dashdot line. This plot is smoothed over a 10 ms interval using the Savitzky-Golay filter with order of the polynomial $n=3$ to reduce the noise. The Savitzky-Golay filter with order of the polynomial $n=3$ is applied to all our smoothed plots but with different time intervals. }
    \label{fig:tau-ratio}
\end{figure}

Figure \ref{fig:tau-ratio} illustrates that the timescale ratio \tadv/\theat\ of all our models exceeds unity between $\tpb=85$--105~ms, with M1 being the latest reaching this criterion. Despite the oscillation of \tadv/\theat, the overall trend of \tadv/\theat\ stays increasing for all six models, indicating the success of explosions. However, the time when \tadv/\theat\ reaches the critical value is approximately 100 ms before actual shock revival. This means the explosions in our models are relatively delayed after reaching the $\tadv/\theat = 1$ criterion, similar to F15.79 of \cite{BrSiLe23}. 

In Figure \ref{fig:mean-shock}(b), we can see around 100 ms after bounce, the shock is still in the middle of silicon-burning shell ($\Mshock \approx 1.4\ \msun$). By 200 ms after bounce, when the shock revives, the shock enclosed mass reaches 1.46--1.50~\msun, approaching the 1.5-\msun\ Si/O interface. We conclude that the infalling perturbation to the shock as it crosses the Si/O interface does not trigger the onset of the explosion in these models, which is different from \cite{MuMeHe17}, but is typical of most \chimera\ models.

\begin{figure}
    \fig{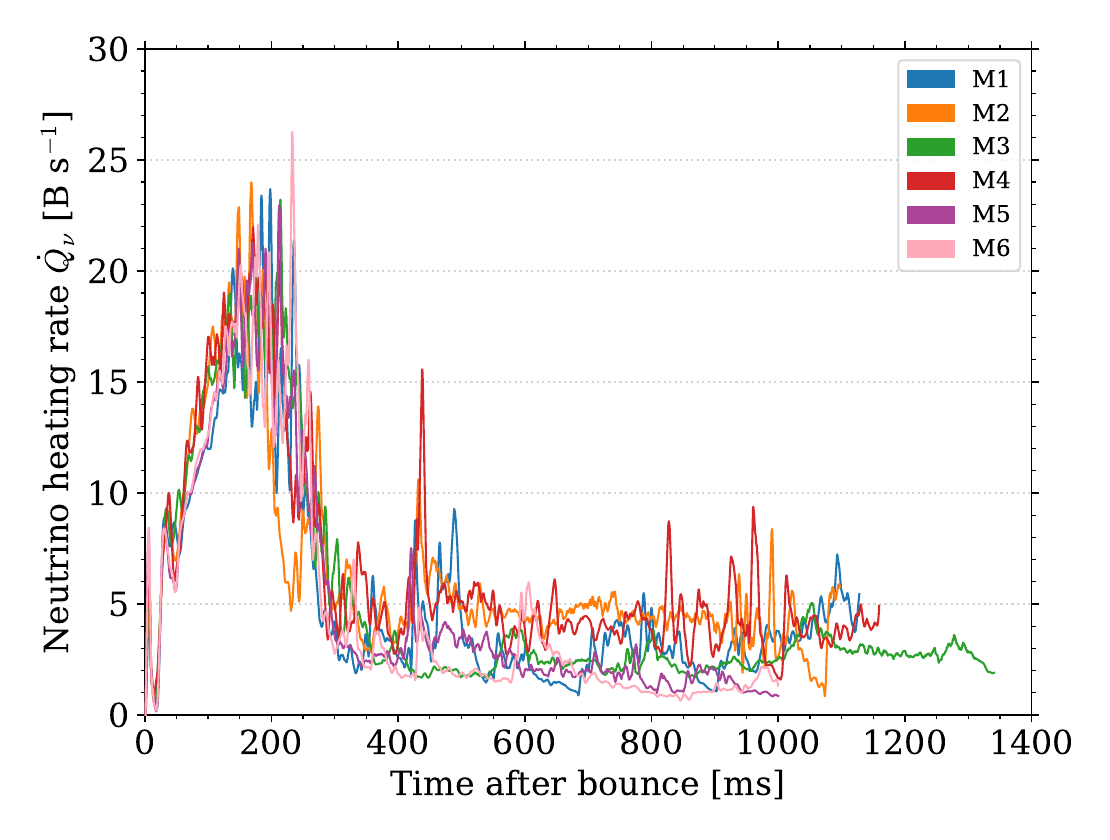}{\columnwidth}{(a)}
    \fig{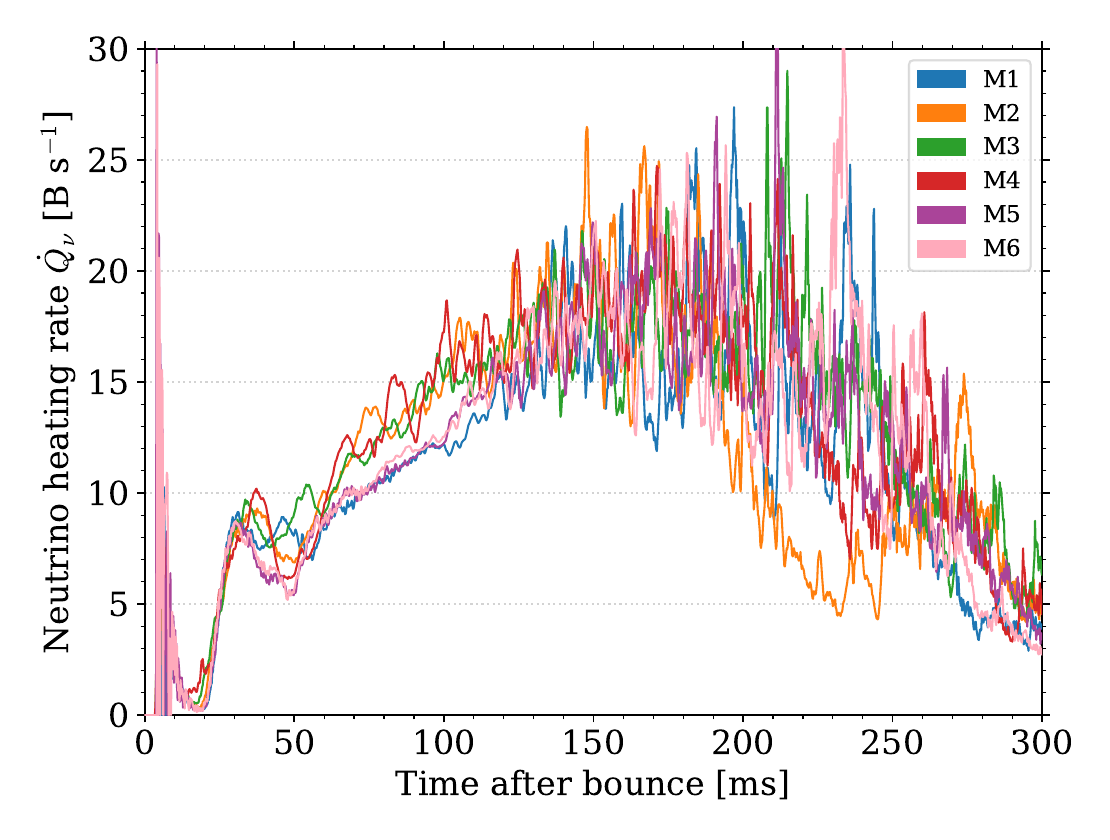}{\columnwidth}{(b)}
    \caption{Neutrino heating rate in the gain region (a) smoothed over a 10 ms interval and (b) unsmoothed in the first 300 ms post-bounce. }
    \label{fig:heat-gain}
\end{figure}

\subsection{Neutrino Heating}
\label{sec:neutrino_heating}

We use the net total neutrino heating rate \dotQnu\ in the gain region to examine how the energy is deposited into the gain region by the absorption of electron flavor neutrinos and anti-neutrinos.

In Figure \ref{fig:heat-gain}(b), the first peak of \dotQnu\ is associated with the breakout burst and ceases within 20 ms post-bounce in all our models. 
Starting from 30 ms after bounce, the escaping neutrinos gradually heat up the materials in the gain region, leading to an entropy gradient and the development of the neutrino-driven convection, which effectively deposits energy into the gain region, so \dotQnu\ increases over time and peaks near shock revival. In Figure \ref{fig:heat-gain}(b), we can observe multiple peaks over 20 \Bethes\ between $\tpb = 150$--250~ms, corresponding to the highest peaks in Figure \ref{fig:heat-gain}(a). 
One example of this is M6 which spikes at about 230 ms post-bounce, caused by a short accretion event. 

After shock revival, the drastic expansion of the post-shock region lowers the density of matter behind the shock and hence the average accretion rate onto the PNS, dropping \dotQnu\ below 5 \Bethes\ overall around $\tpb=300$ ms. The peaks thereafter are caused by brief accretion episodes onto the PNS, which heats the PNS surface, increases the neutrino luminosity, and thus temporarily enhances the neutrino heating. 

\begin{figure}
    \includegraphics[width=\columnwidth,clip]{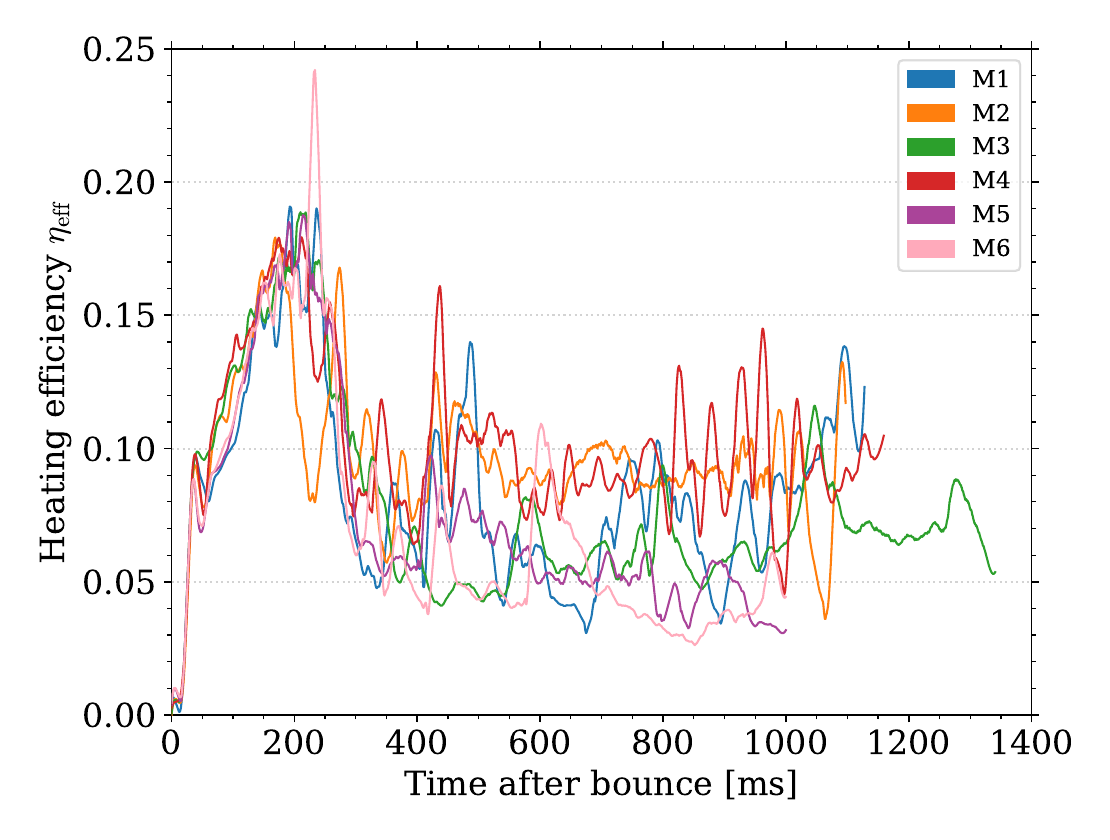}
    \caption{Neutrino heating efficiency in the gain region smoothed over 32.4 ms interval. }
    \label{fig:heat-eff}
\end{figure}

Another indicator is the neutrino heating efficiency \heateff\ in the gain region, which is defined as the ratio of the neutrino energy deposition rate in the heating layer to the electron flavor neutrino luminosity at the gain surface \citep{BrLeHi16, BrSiLe23},
\begin{equation}
    \heateff = \frac{\dotQnu}{L_{\nue} + L_{\nuebar}}. 
\end{equation}
Figure \ref{fig:heat-eff} shows \heateff\ significantly increases as the neutrino-driven convection develops, stably grows, peaks with \dotQnu\ around the onset of explosion ($\tpb = 150$--250~ms) and gradually drops as the explosion matures. 

\subsection{Accretion Stream}
\label{sec:accretion}

\begin{figure}
    \fig{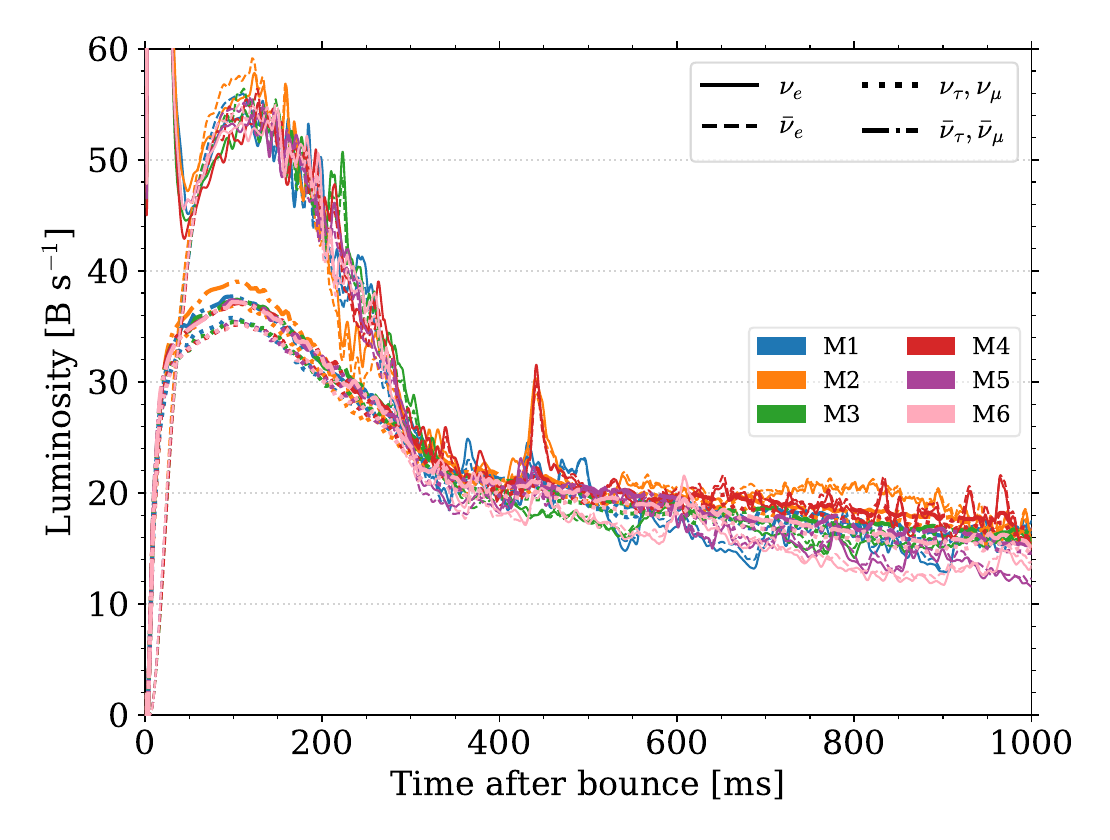}{0.95\columnwidth}{(a)}
    \fig{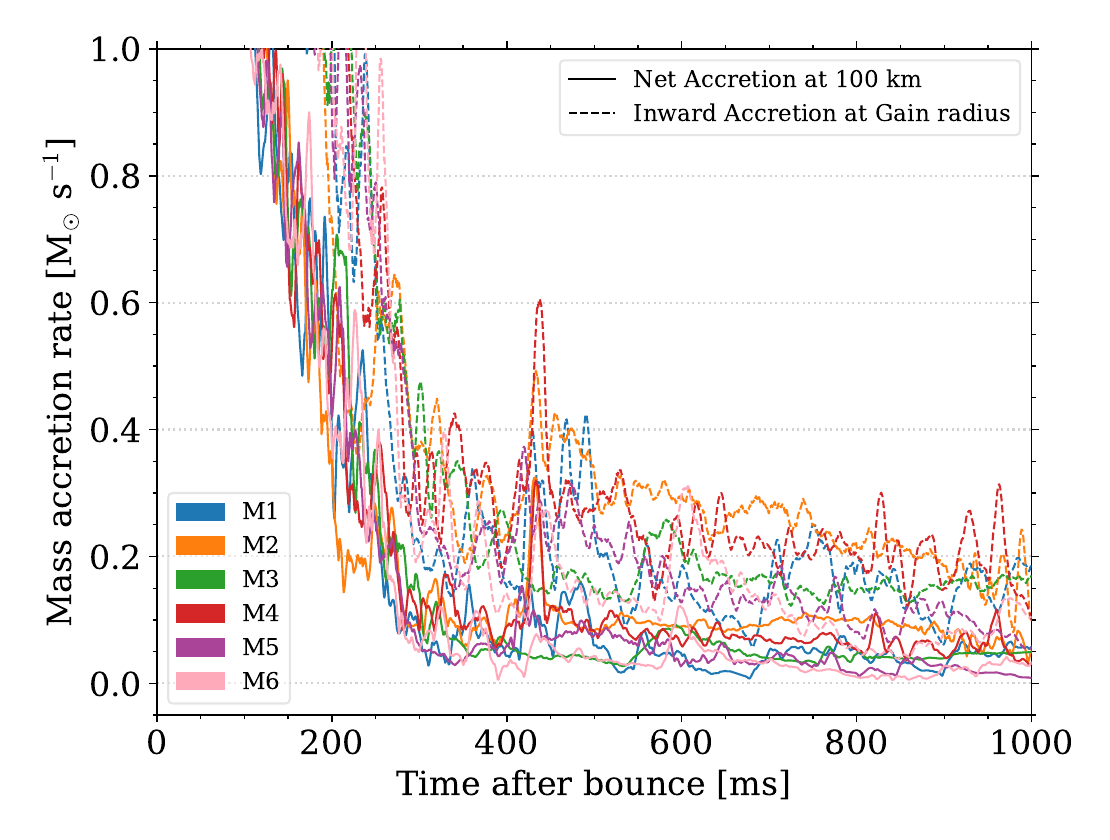}{0.95\columnwidth}{(b)}
    \fig{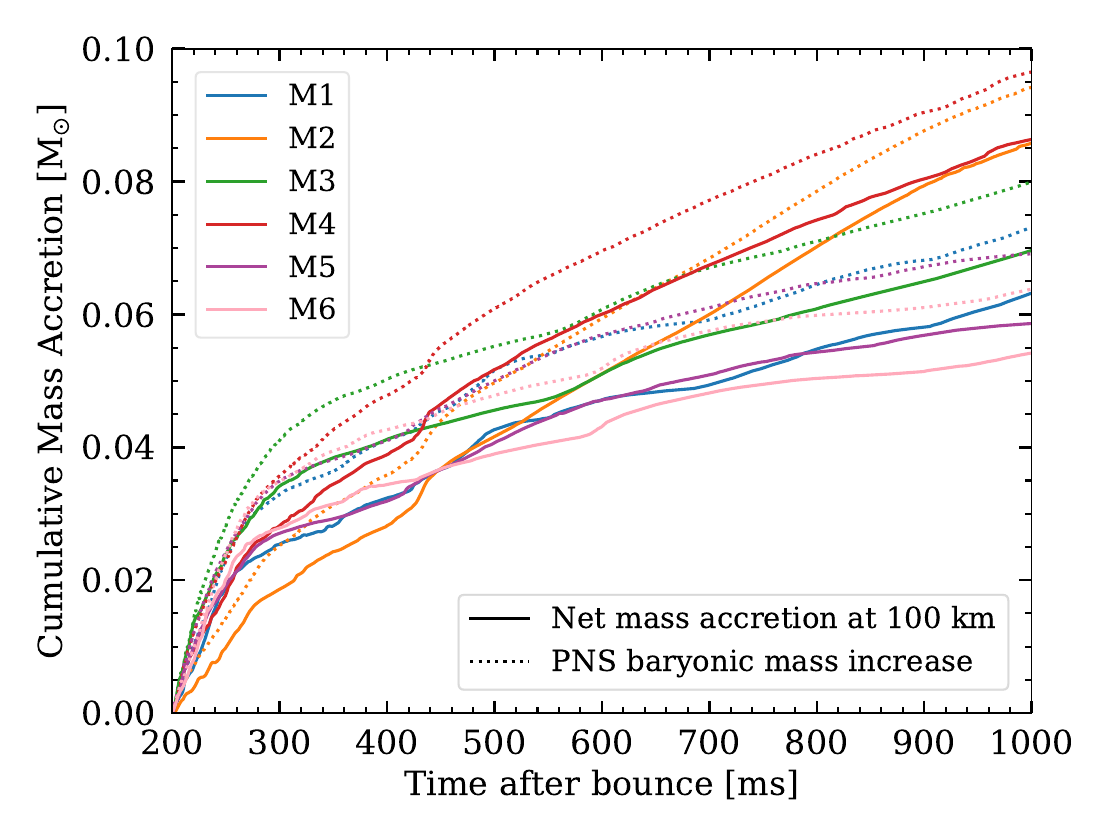}{0.95\columnwidth}{(c)}
    \caption{Panel (a): Neutrino luminosity on a linear scale.  Panel (b): Net accretion rate through 100 km radius (solid lines) and inward mass accretion rate through the gain surface (dashed lines). Panels (a) and (b) are smoothed over 10 and 20 ms intervals, respectively. Panel (c): Cumulative net mass accretion through 100 km radius  (solid lines) and PNS mass increase (dotted lines), relative to 200 ms after bounce. }
    \label{fig:lumin-accrete} 
\end{figure}

Accretion onto the PNS contributes to the neutrino luminosity and thus the neutrino heating. We can observe the trend of neutrino heating (Figure \ref{fig:heat-gain}(a)) roughly follows the trend of luminosity (Figure \ref{fig:lumin-accrete}(a)). The steep peak of neutrino luminosity shortly after the core bounce is due to the release of trapped \nue\ (``the breakout burst''). The post-bounce peak of neutrino luminosity occurs between $\tpb = 150-250$ ms, matching the highest \dotQnu\ in Figure \ref{fig:heat-gain}(a) and \heateff\ in Figure \ref{fig:heat-eff}.

Following prior practice with \chimera\ models, we treat the inward mass accretion rate through the gain surface as the accretion rate onto the PNS surface for matter with $v_{r} < 0$ \citep[e.g.][]{BrLeHi16, BrSiLe23}. 
As shown in Figure \ref{fig:lumin-accrete}(b), the mass accretion rates have the same trend as the neutrino luminosity in (a), which drop sharply immediately after core bounce. 
During shock stagnation, the rate drops more steadily, smooths, and eventually flattens as the explosion develops. 
At the end of our simulations, the inward mass accretion rates through the gain surface maintain approximately constant values around 0.2 \msuns\ for M1 to M4 and around 0.1 \msuns\ for M5 and M6. In contrast, the net accretion at a fixed radius of 100 km exhibits a similar behavior but stabilizes at a smaller value between 0 and 0.1 \msuns. 
Here, we notice that the inward mass accretion rates measured at the gain surface are consistently higher than the net (inward minus outward) accretion rates and show larger fluctuations.

In Figure \ref{fig:lumin-accrete}(c), we focus on the interval from 200--1000 ms after bounce, corresponding to the onset of explosions in all six models, and extending to the end of the simulations for M5 and M6, and compare the cumulative accreted mass through a radius of 100 km with the growth of the PNS mass (defined as the mass where $\rho > 10^{11} \gcc$). 
``Double counting'' due to convective vortices near the PNS surface cause the cumulative mass accretion measured at the gain radius to be overestimated by $\sim$50\%. 
This also explains why the inward mass accretion rate measured at gain radius is $\sim$50\% higher than the net accretion rate measured at 100 km. 
The gain radius is also subject to local variability. 
On the contrary, the net accretion at 100 km gives a strong correlation between the accretion and the PNS mass increase. 
The increases in PNS baryonic mass are about 0.01 \msun\ larger than the cumulative net accretion measured at 100 km for all six models, a difference that can be attributed to mass fallback onto the PNS surface.

Both accretion rates tend to stabilize after $\tpb=400$~ms except for M2 and M4, which happen to be the models with higher overall accretion rates. The narrow peaks in the orange and red lines in Figure \ref{fig:lumin-accrete}(b) indicate sudden accretion events for M2 and M4 near $\tpb=400$ ms, seen also in \dotQnu\ (Figure \ref{fig:heat-gain}(a)). Examination of the data, especially spatial entropy images, for  $\tpb=300$--500~ms, confirms major hydrodynamic accretion events causing \dotQnu\ to suddenly increase at these points. 

The sudden \dotQnu\ peak in M2 is caused by changes in accretion onto the PNS. M2 has continuous accretion primarily on to the PNS north pole due to its explosion morphology. At 300 ms, the accretion is limited to a cone extending $\sim 30^{\circ}$ from the pole. However, a secondary accretion stream from more equatorial regions reaches the PNS at 320 ms, excising part of the hot bubble into a minor lobe.  This lobe is pushed southward by the north pole accretion stream, inhibiting equatorial accretion.  By 360 ms, the equatorial accretion stream is reaching the PNS near the south pole.  This circuitous path is very inefficient, delaying equatorial accretion. As a result, at 370 ms, and again at 400 ms, additional equatorial streams subdivide the lobe, only to be pushed south by the influence of the north polar accretion stream on the remnants of the lobe.  In the time interval $\tpb=400$--433~ms, the minor lobe gets pushed back by the combined action of the accretion streams, forcing its matter to accrete onto the PNS. For $\approx$50~ms, accretion occurs onto the PNS from the entirety of the northern hemisphere, before gradually resuming a configuration similar to that at $\tpb=300$~ms. This event increases the accretion to the north pole suddenly, and causes the peak in \dotQnu\ near $\tpb=450$~ms  (orange line, Figure \ref{fig:heat-gain}(a)).

While the shock in M4 is initially spherical, by $\tpb=300$ ms, plumes along both poles have made the shock prolate, pushing accreting matter towards the equator. At 335 ms, the primary accretion stream hits the PNS at about 35$^{\circ}$ from the north pole, separated from the pole by an outflow. By 350 ms, this stream, though fed from the equator, has been pushed to the north pole. Again this circuitous path inhibits accretion, building up mass in the stream.  As a result, a series of small equatorial accretion streams try to penetrate the hot bubble. At $\tpb=426$ ms, one of these equatorial accretion streams briefly succeeds, hitting the PNS near the south pole.  This stream however gets blended with the south hemisphere ejecta as another equatorial accretion stream reaches the PNS near its equator at $\tpb=436$ ms.  With the polar stream still active, adding the second stream enhances the total accretion. The timing of these accretion events matches the interval when \dotQnu\ peaks (Figure \ref{fig:heat-gain}(a)). Unlike what we observe in M2, maintenance of an equatorial accretion stream consolidates the bipolar explosion morphology of M4. More details of the explosion morphology will be discussion later in Section \ref{sec:morph}.

\subsection{Explosion Energy}
\label{sec:exploE}

\begin{figure}
    \includegraphics[width=\columnwidth,clip]{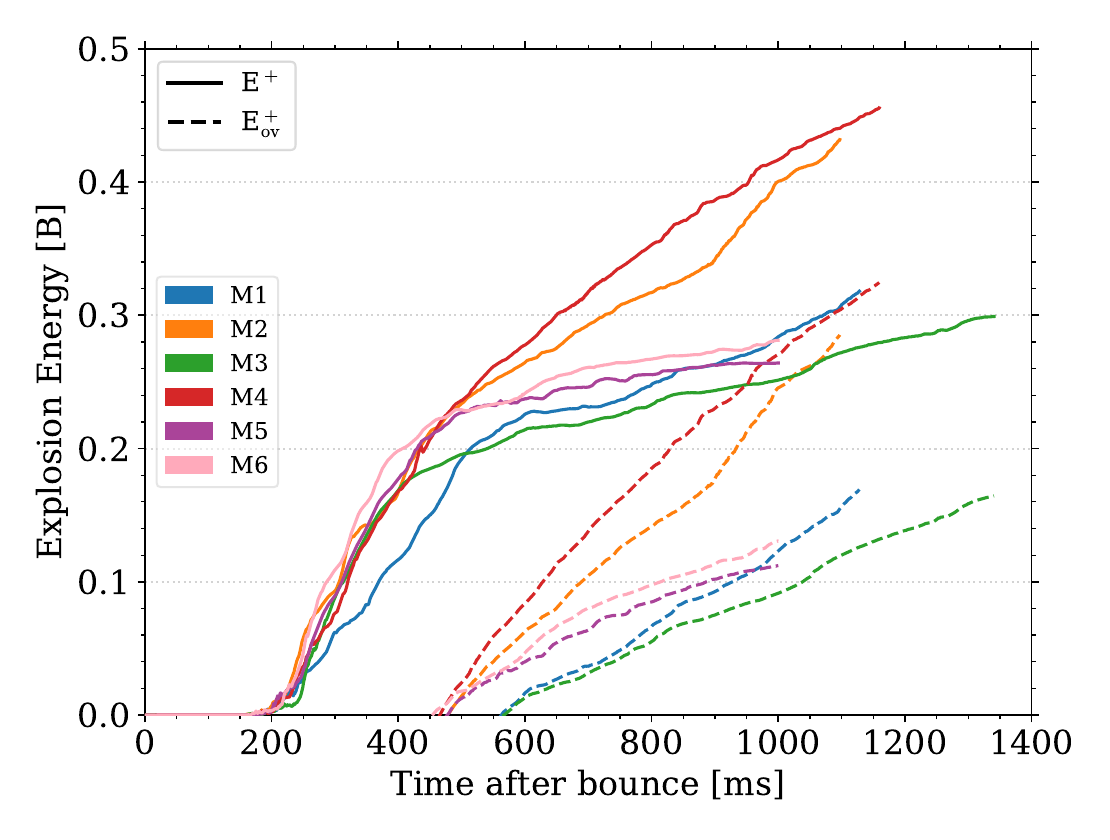}
    \caption{Growth of diagnostic energy \Ediag\ (solid lines) and overburden-corrected diagnostic energy \Ediagov\ (dashed lines). The gap between the solid line and the dashed line gives the overburden binding energy of each model. }
    \label{fig:explo-energy}
\end{figure}

\begin{figure*}
    \includegraphics[width=\textwidth]{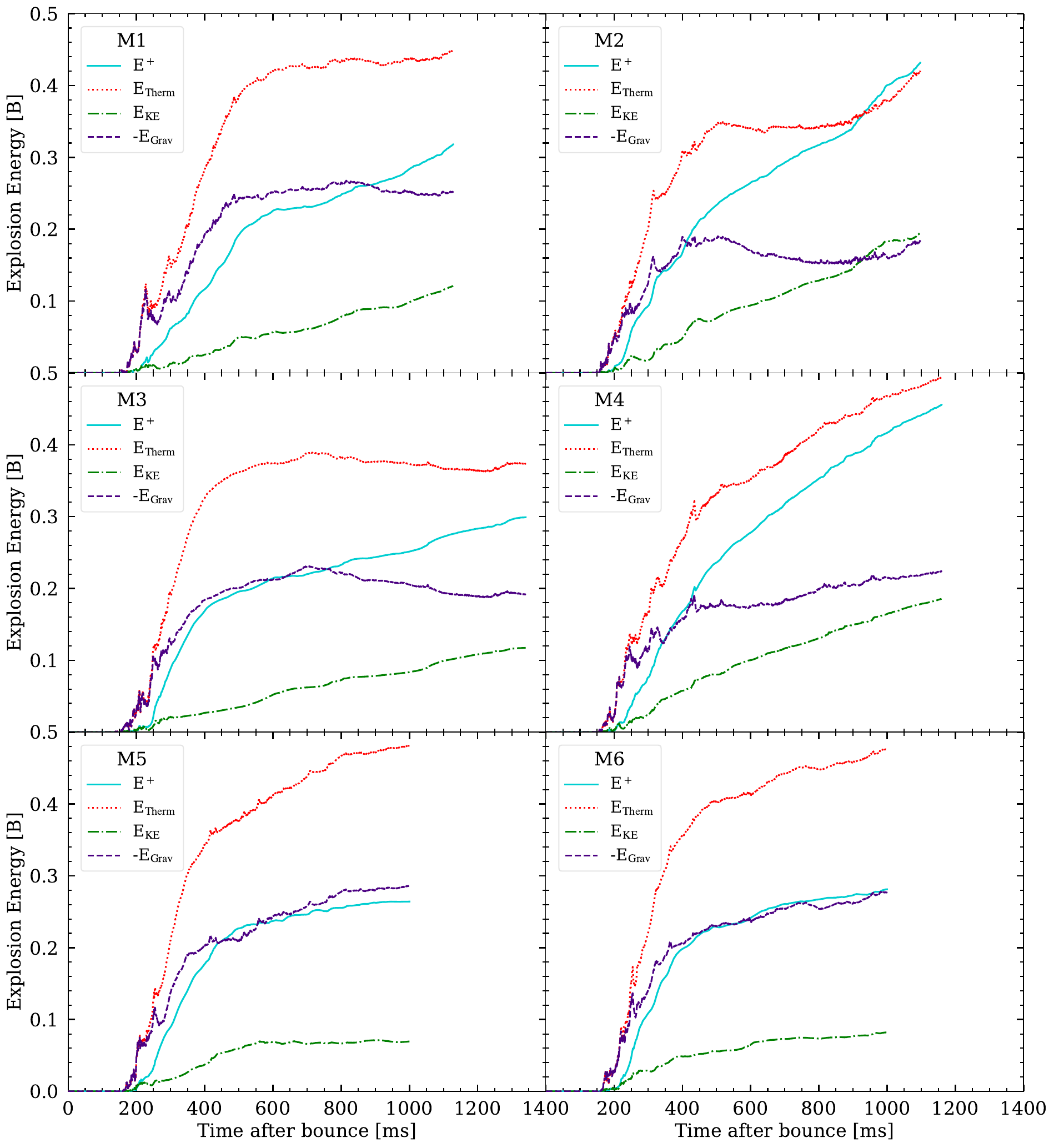}
    \caption{Diagnostic energy \Ediag\ (cyan lines) and components for all six models M1 to M6 in panels left-to-right, top-to-bottom. Red dashed lines represent the thermal energy component,  green dashed lines the kinetic energy component, and purple dashed lines the `negative' gravitational potential energy component. }
    \label{fig:energy-split}
\end{figure*}

Following \cite{BrLeHi16}, the zones between the shock and PNS are divided into positive energy zones or negative energy zones based on the sum of kinetic energy, thermal energy, and gravitational energy in each zone.  In each positive energy zone, this specific energy density is
\begin{equation}
    e_{\mathrm{tot}} = e_{k} + e_{\rm th} + e_{\mathrm{grav}} > 0 . 
    \label{eq:postive_energy_zone}
\end{equation}
These positive energy zones are defined as the unbound region.
As of yet unshocked material above the unbound region, both on- and off-grid, are defined as the overburden, whose energy contribution is negative due to the binding energy of the material in the overburden. Two quantities commonly used to track growth of the explosion energy are the diagnostic energy \Ediag\ and the overburden-corrected diagnostic energy \Ediagov. Diagnostic energy \Ediag\ is defined as the sum of $e_{\mathrm{tot}}$ over positive energy zones, and the overburden-corrected diagnostic energy \Ediagov\ is defined as the sum of the diagnostic energy and the overburden binding energy.  It is ultimately the overburden-corrected diagnostic energy that should be compared to observed explosion energies.

The growth of \Ediag\ starts around 160 ms after bounce for all six models, earlier than the actual shock revival. With the overburden energy $\sim 0.2$ B, the positivity of \Ediagov\ trails \Ediag\ by several hundred ms. We observe \Ediagov\ turns positive in the following order: M6 ($\tpb=453.8$ ms), M4 ($\tpb=465.2$ ms), M5 ($\tpb=477.4$ ms), M2 ($\tpb=478.0$ ms), M1 ($\tpb=562.8$ ms), and M3 ($\tpb=566.4$ ms), from which we find no correlation with the shock shell breaching order (see Table \ref{tab:Modelinfo} for the shell breach timing). 
We find that M2 and M4 have more aggressive \Ediag\ growth pattern than other models, which we will discuss later in Section~\ref{sec:morph&E}. 
As the shock unbinds the rest of the star, it eliminates the on-grid portion of the overburden binding energy as it expands. Hence, \Ediag\ and \Ediagov\ converge. As the explosion enters the final explosion state, \Ediag\ tends to saturate while \Ediagov\ continues to converge toward it.

The saturation of \Ediag\ and the convergence of \Ediag\ and \Ediagov\ are reported in previous simulation results (see, e.g., model B12-WH07 and B15-WH07 in \cite{BrLeHi16} and 18.88-\msun\ progenitor model in \cite{BoYaKr21}). However, the models reported here do not run to sufficient times for  \Ediag\ to saturate nor for the convergence of \Ediag\ and \Ediagov\ to complete. In Figure \ref{fig:explo-energy}, for each model, \Ediag\ and \Ediagov\ are still increasing and the gap between \Ediag\ and \Ediagov\ remains large. 
Figure \ref{fig:energy-split} breaks down \Ediag\ into components, which shows \Ediag\ still being dominated by the thermal energy.  

As a CCSN evolves into the late explosion phase, the post-shock density dilutes, and the neutrino mean free path correspondingly increases, rendering neutrino heating ineffective. 
At this point, the thermal component saturates, although it can subsequently grow somewhat through nuclear recombination and eventually nuclear decay. 
The thermal and gravitational components of \Ediag\ gradually diminish to zero on a timescale of hours, transferring to the kinetic component through buoyancy \citep[see, e.g.][]{HaPlOd14,SaHiMe21,NeSaHi26}. 
As shown in Figure \ref{fig:energy-split}, only M3 seems to  have reached the transition toward a later-stage explosion, with M1 possibly showing similar behavior. 
M2 also appears to exhibit this tendency before a turning point at 900 ms after bounce, when the thermal component begins to grow again and regains dominance. 
When the high entropy outflowing matter between the main south ejecta and a newly developed minor north convective plume, which becomes noticeable at $\tpb\approx$ 700 ms with a maximum radius between 50--100 km and gradually growing to 250 km on the z-axis, is forced to accrete back to PNS, it leads to an enhanced neutrino luminosity and maintains a higher accretion rate between 700--900 ms post-bounce. 
Starting from $\tpb\approx$ 900 ms, the change in accretion pattern and explosion morphology results in an \Ediag\ increase we observe.

\subsection{Explosion Morphology}
\label{sec:morph}

The explosion morphology is one of the factors that determines if neutrino heating in the explosion stage effectively and efficiently deposits energy in the ejecta and thus increases the explosion energy. This is closely related to how the accretion streams bring material from the shock to the PNS. The geometry of the accretion streams can also affect the geometry of the explosion if sufficiently strong (see Section \ref{sec:accretion}). 

\cite{BrLeHi16} reported that the spherical shock shape in their 20-\msun\ model B20-WH07 blocks the accretion streams to the PNS, reducing the accretion rate through the gain surface. This lowers the resulting neutrino luminosity and slows the growth of \Ediag, breaking the typical correlation between a higher progenitor mass and a higher \Ediag. Another case is presented in \cite{VaCoBu22}, where models with unipolar ejecta maintain higher explosion energies than those with bipolar ejecta, which they credit to the preferential neutrino heating in one hemisphere. 

\subsubsection{An Early SASI?}\label{sec:SASI}

\begin{figure}
    \fig{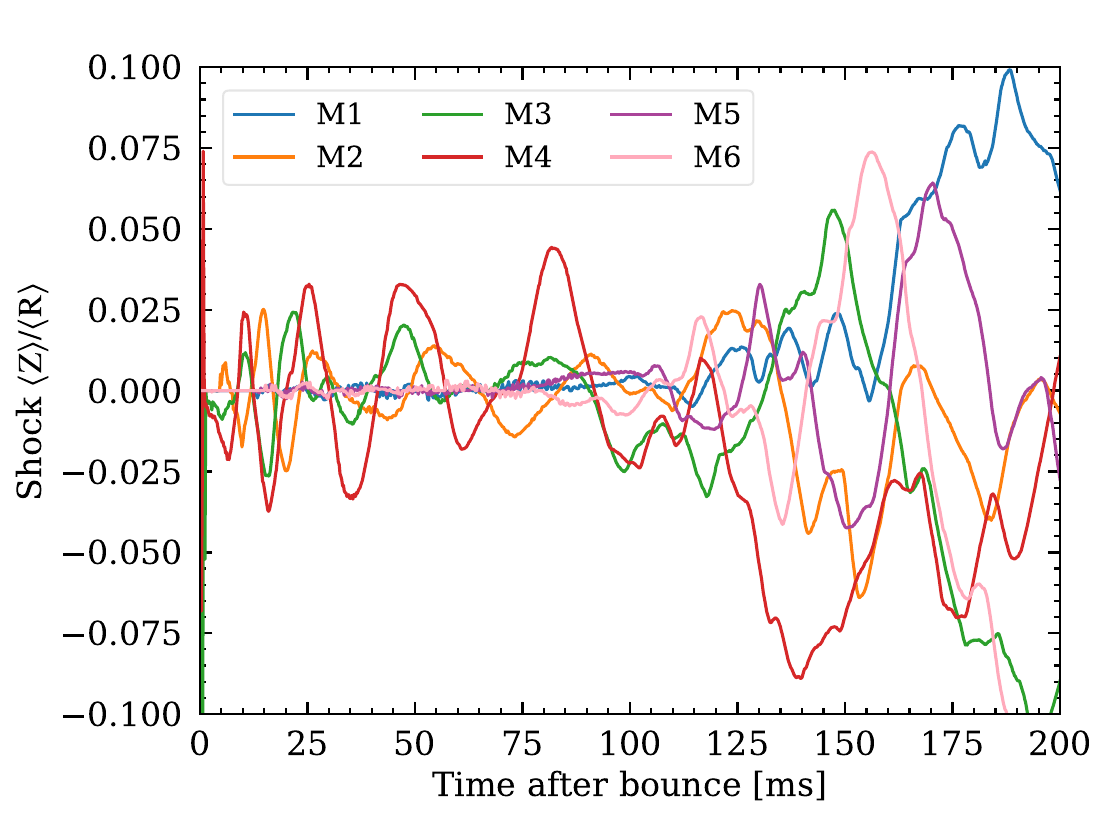}{\columnwidth}{(a)}
    \fig{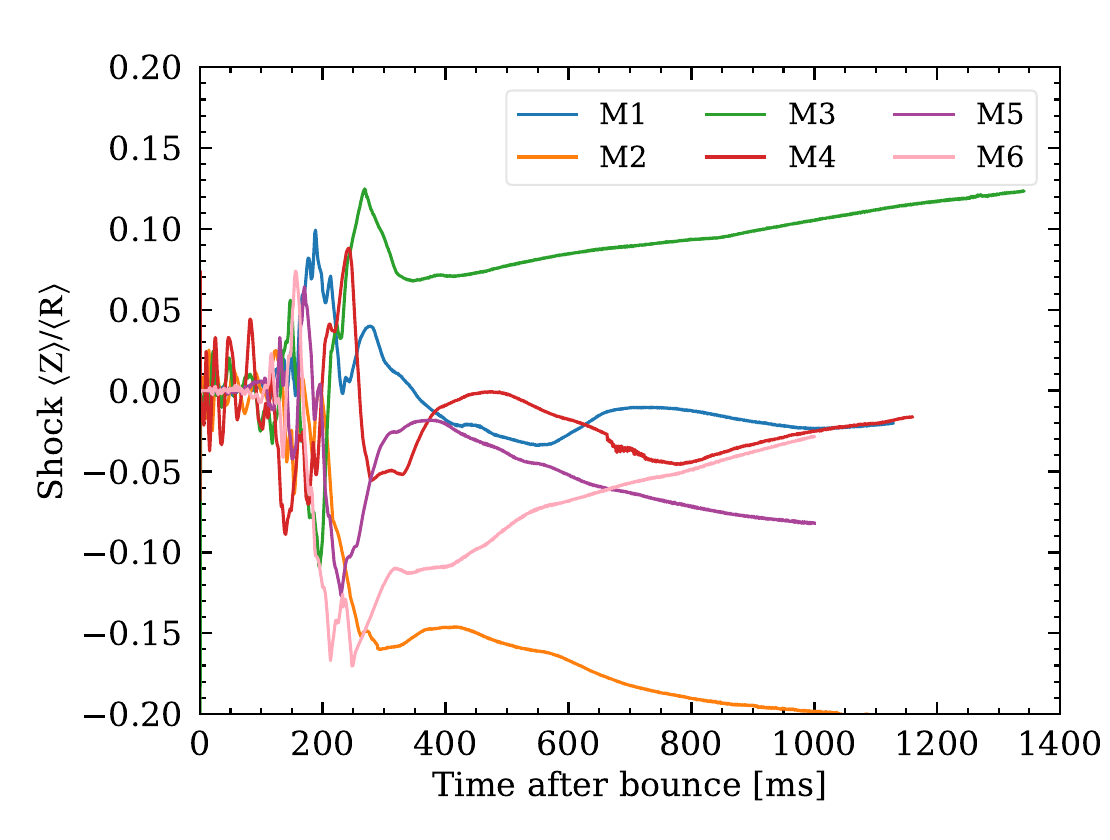}{\columnwidth}{(b)}
    \caption{Dipole deformation of the shock normalized by mean shock radius for (a) the first 200 ms after bounce and (b) the entire run. We refer to the positive $z$-direction as `north' and the negative $z$-direction as south.}
    \label{fig:dipole-time}
\end{figure}

Figure \ref{fig:dipole-time} shows the dipole deformation of the shock, which is measured as mean shock $z$-coordinate position normalized by mean shock radius, \ZoverR. Factors that affect the dipole include bulk fluid motions like progenitor asymmetries, prompt convection in the PNS, SASI, and neutrino-driven convection. M2, M3 and M4 show obvious deformation of the shock immediately after bounce. In contrast, the shock dipole in M1, M5, and M6 remain close to zero until 75--100 ms after bounce. 

M3 and M4 have lateral velocities in their progenitors at the onset of core-collapse, so it is not surprising that both models rapidly develop asphericity including a strong dipole. The other four models have spherical progenitors with zero lateral velocities, and the dipole \ZoverR\ is weak during prompt convection in three of four. M2, however, exhibits strong post-shock sloshing fluid motions and a strong dipole signal.
Visual inspection of the prompt convection phase in M2 to M4 shows a distinct dipole oscillation resembling the classical SASI, though shock is neither standing, nor accreting, but rather expanding through the collapsing outer Fe core. 

\begin{figure}
    \fig{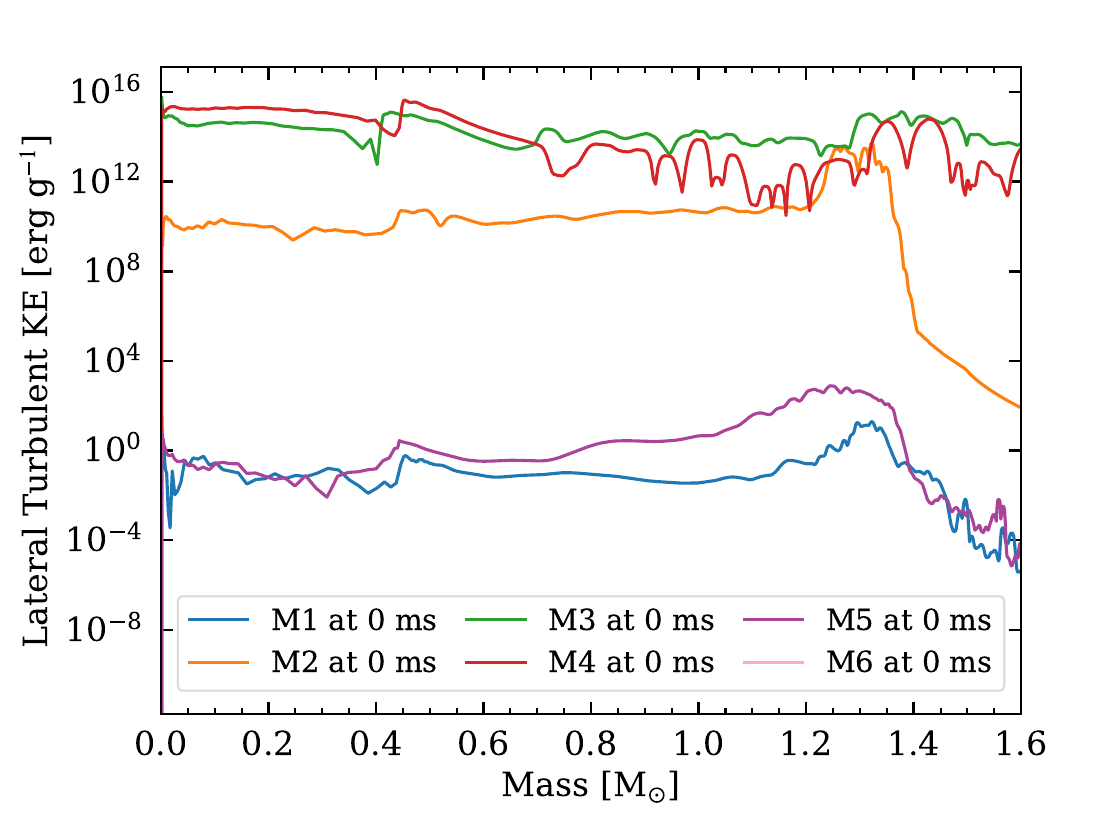}{\columnwidth}{(a)}
    \fig{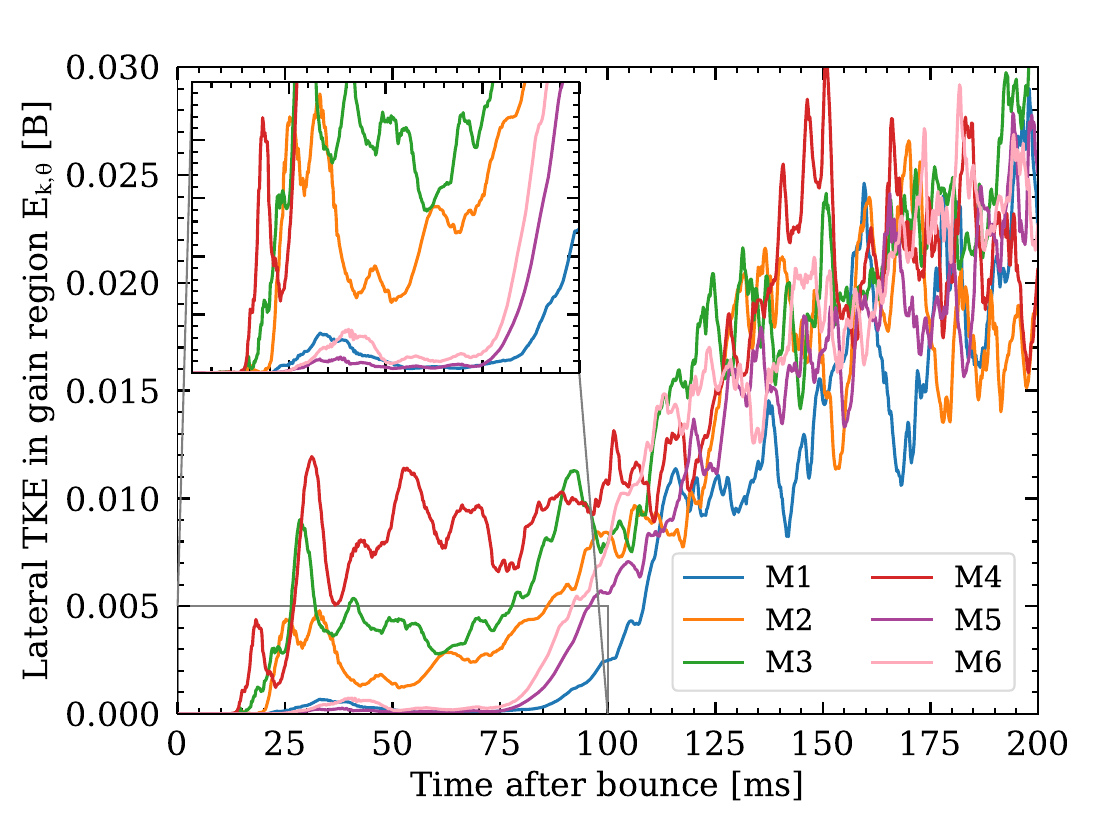}{\columnwidth}{(b)}
    \caption{Specific lateral TKE (a) averaged over shells between 0--1.6~\msun\ at bounce for all models except M6 since it is 1D at bounce; and (b) integrated over the gain region, $E_{k,\theta}$ for the first 200 ms after bounce, smoothed over a 10 ms interval.}
    \label{fig:ke-lat}
\end{figure}

In this analysis we use `noise' to refer to any small scale motion relative to bulk flow regardless of source, including numerical noise and small scale turbulent motions from convection.
The common, simplified measure of this is the aniosotropic, or turbulent, velocity, which is the local velocity with the mean flow subtracted. 
For our purposes, we consider only the lateral, $\theta$, component of the kinetic and assume that their is no background flow toward either pole, such that the local lateral turbulent kinetic energy (lateral TKE) is $\rho v_\theta^2/2$. 
We consider lateral TKE as the specific shell average value, integrated over a spherical shell and normalized by shell mass, as in Figure~\ref{fig:ke-lat}(a), or as the integrated over the gain region $E_{k,\theta}(t)$ as in Figure~\ref{fig:ke-lat}(b).

Though M2 does not \emph{start} with lateral velocities from the progenitor, if we examine the profile of lateral TKE) in the collapsed cores at bounce (Figure~\ref{fig:ke-lat}(a)) we see that it more closely resembles M3 and M4. The mean lateral speed $\ctheta \sim 1$~\kmps\ throughout the core is more similar to other `noisy' models M3 and M4 with $\ctheta \sim 100$~\kmps\ than to the three quiet models (M1, M5, and M6) with $\ctheta \lesssim 1$~\cmps. As will be discussed in Section~\ref{sec:structure}, the the lateral velocity `noise' in M2 grows rapidly to nearly the level seen in the models with initially non-zero $v_\theta$.

Prompt convection begins slightly earlier in the `noisy' models (3--5 ms after bounce) than in the `quiet' models (8--10 ms after bounce). The motions from the prompt convection does not seem to be sufficient to trigger the SASI-like oscillations of the concurrent shock propagation through the Fe core in the `quiet' models but the oscillations are present in the `noisy' models.

This seems to be the SASI, distorting the shock rapidly after core bounce, while the shocks in M1, M5, and M6 remain relatively spherical until neutrino-driven convection develops.  The non-zero small dipole signals in M1, M5, and M6 suggest the existence of a weak SASI filling the gap between the concession of prompt convection and development of neutrino-driven convection.

When strong enough, the traditional SASI acts to expand the mean shock radius, so theoretically the mean shock radii in M2, M3, and M4 should be larger than other three models. However, we do not find this tendency in our models. In fact, Figure \ref{fig:mean-shock}(a) shows the mean shock radii are very similar until the onset of explosion. 

Figure \ref{fig:ke-lat}(b) shows that between 20--50 ms after bounce, though small, there is a growth of the lateral TKE $E_{k,\theta}$ in the gain region for M1, M5, and M6. The early timing of this increase in lateral TKE makes it most likely due to the SASI instead of neutrino-driven convection.  The lateral TKE of the `quiet' models remain low, below 0.001~B, during the prompt convection phase and then decreases until about 75~ms after bounce when neutrino driven convection drives new non-radial motion. Lateral TKE in the three `noisy' models that show the early SASI-like behavior is likely dominated by that activity in the shocked cavity. Once neutrino-driven convection becomes fully developed (around 100~ms) the behavior of the $E_{k,\theta}$ and \ZoverR\ become similar for all models.

\subsubsection{Explosion morphology}
\label{sec:morphM1toM6}

\begin{figure*}[htbp]
  \gridline{
    \fig{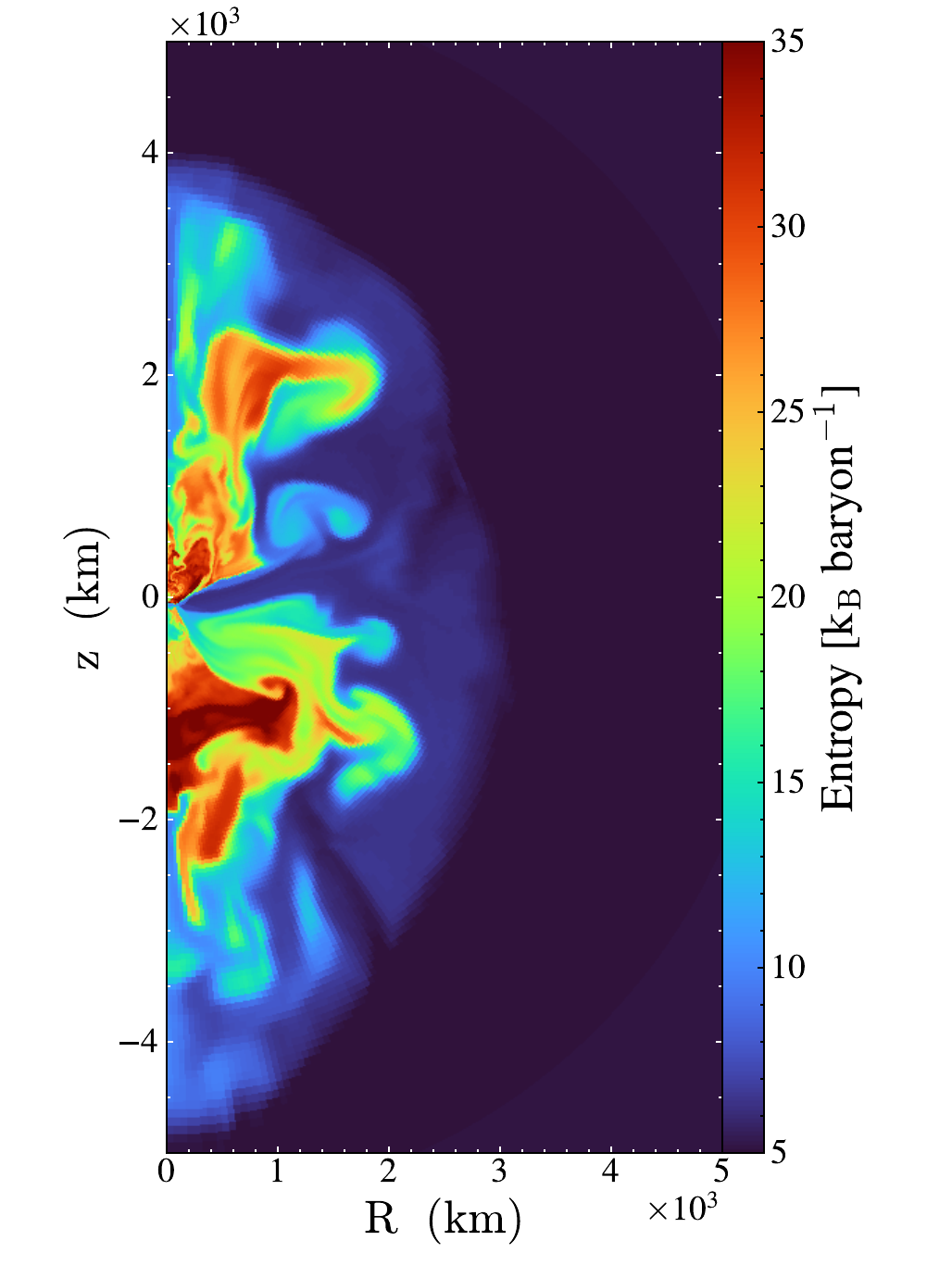}{0.32\textwidth}{(a) M1}
    \fig{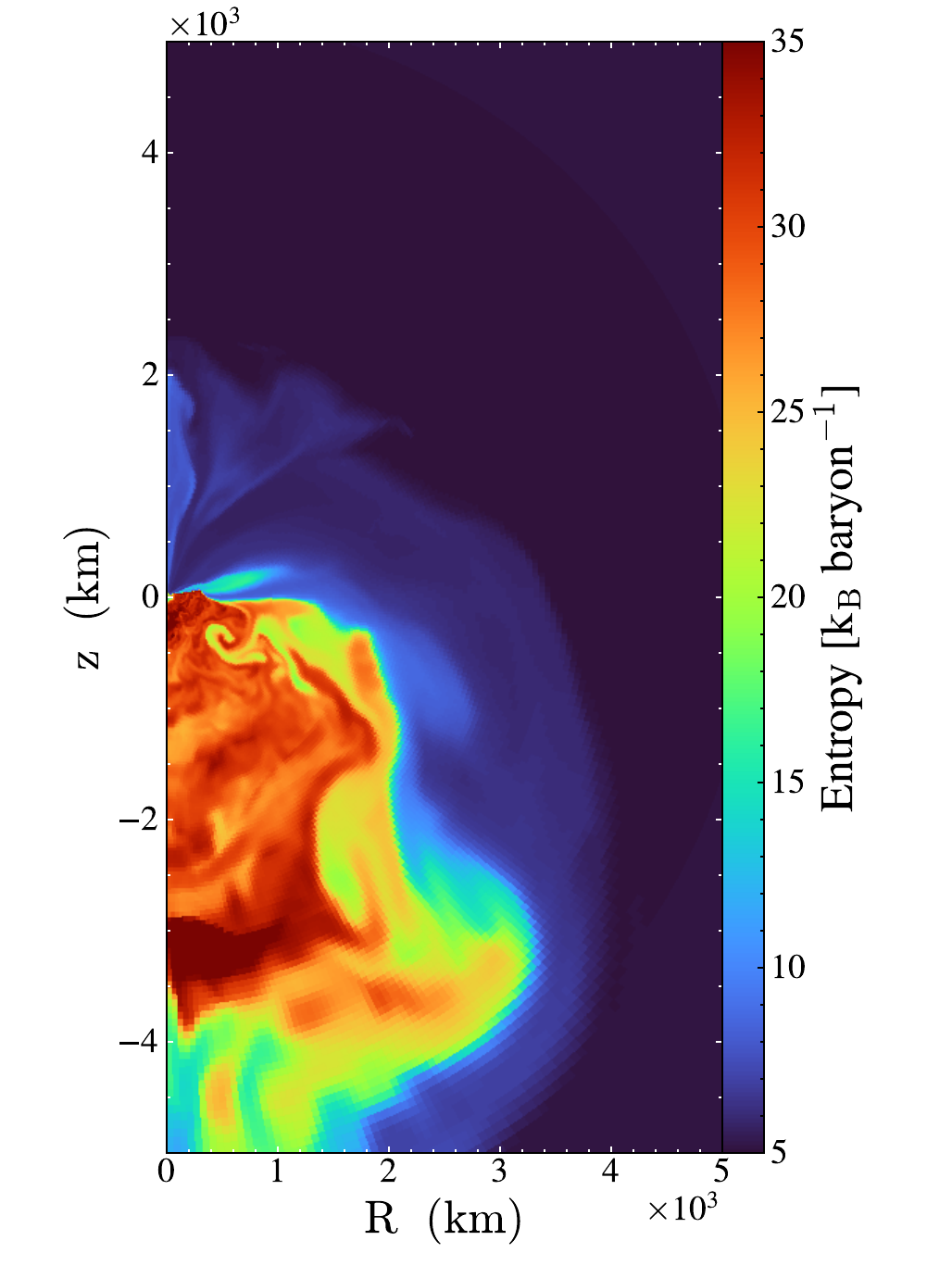}{0.32\textwidth}{(b) M2}
    \fig{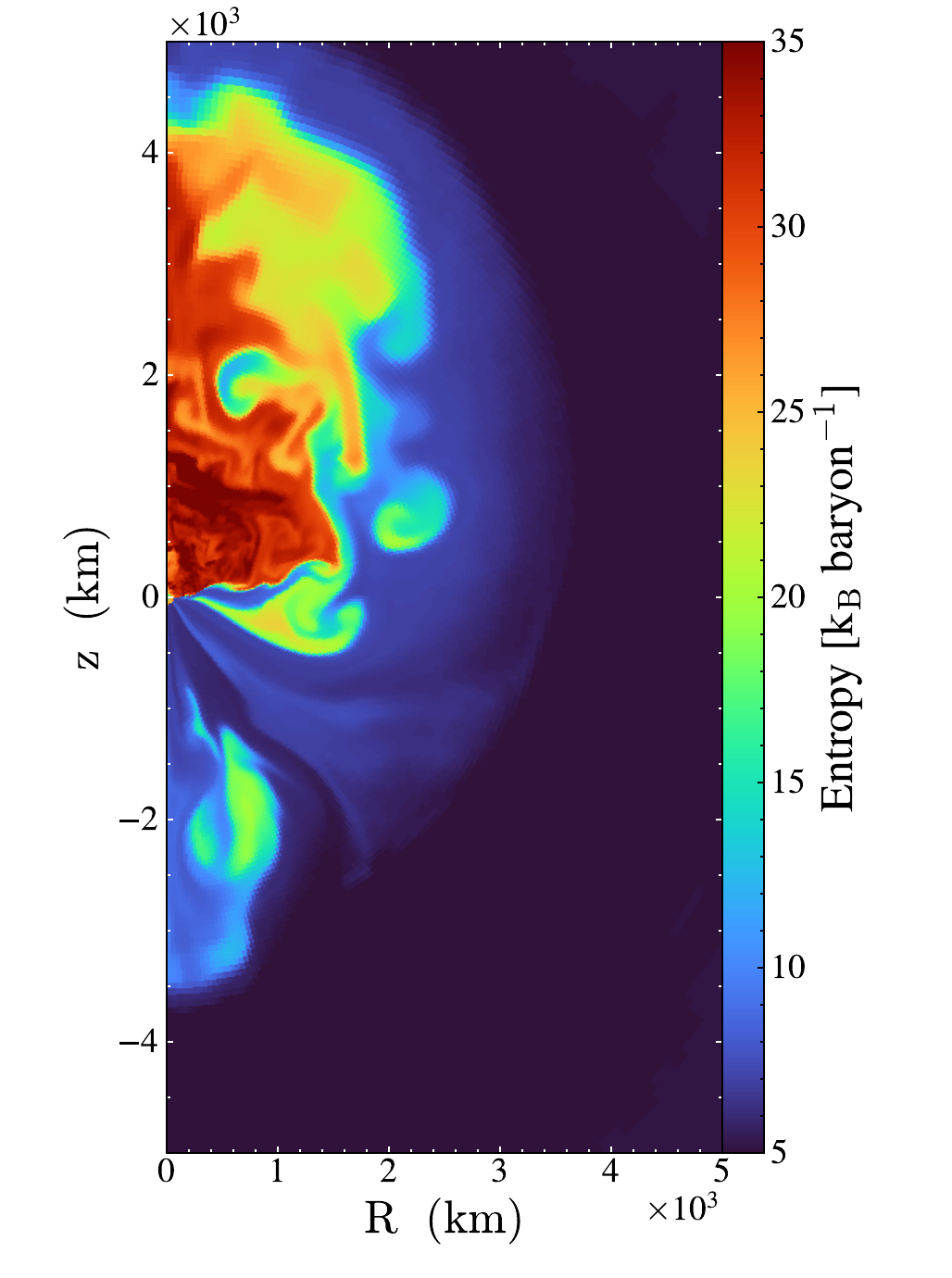}{0.32\textwidth}{(c) M3}
  }
  \gridline{
    \fig{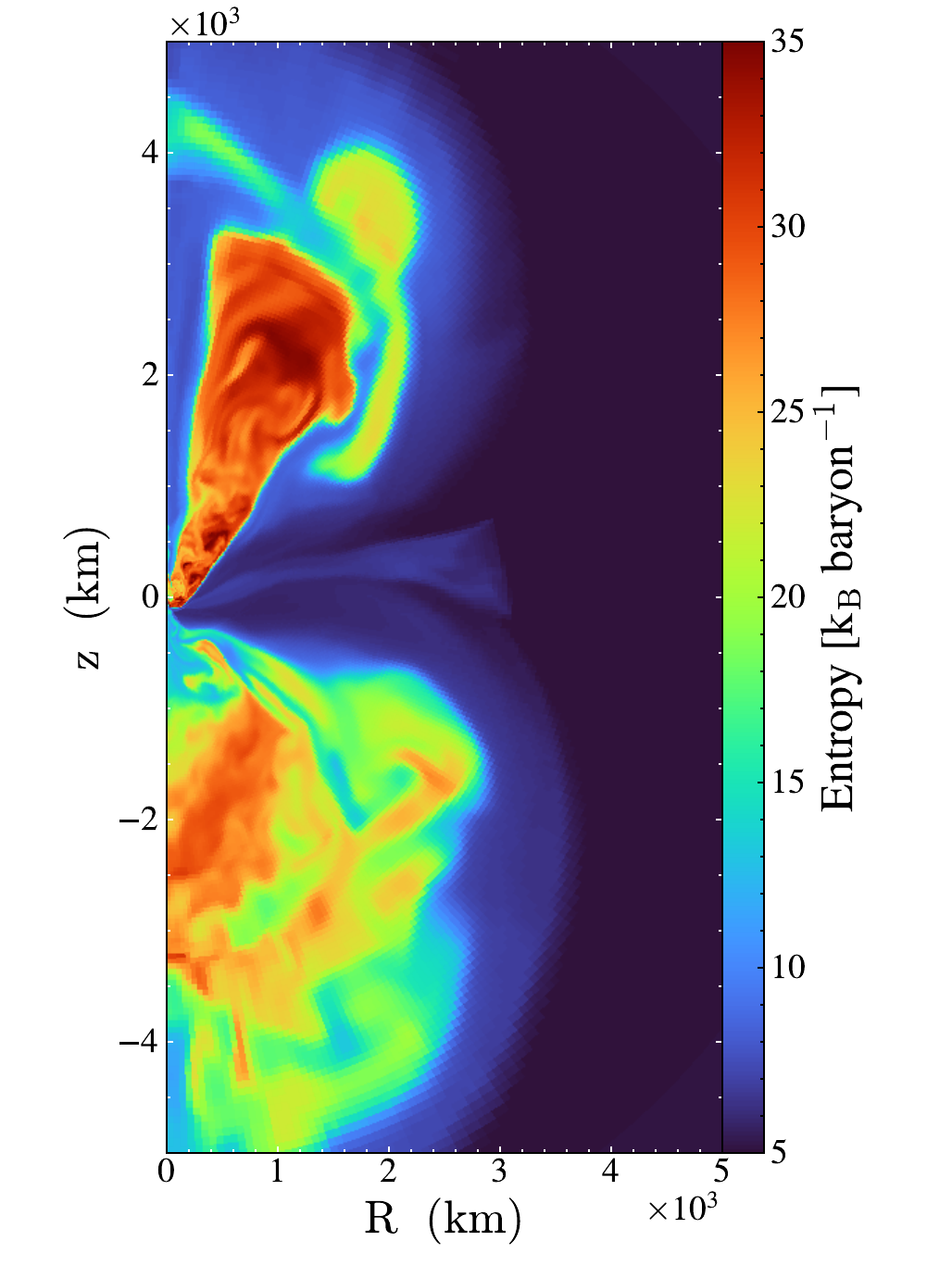}{0.32\textwidth}{(d) M4}
    \fig{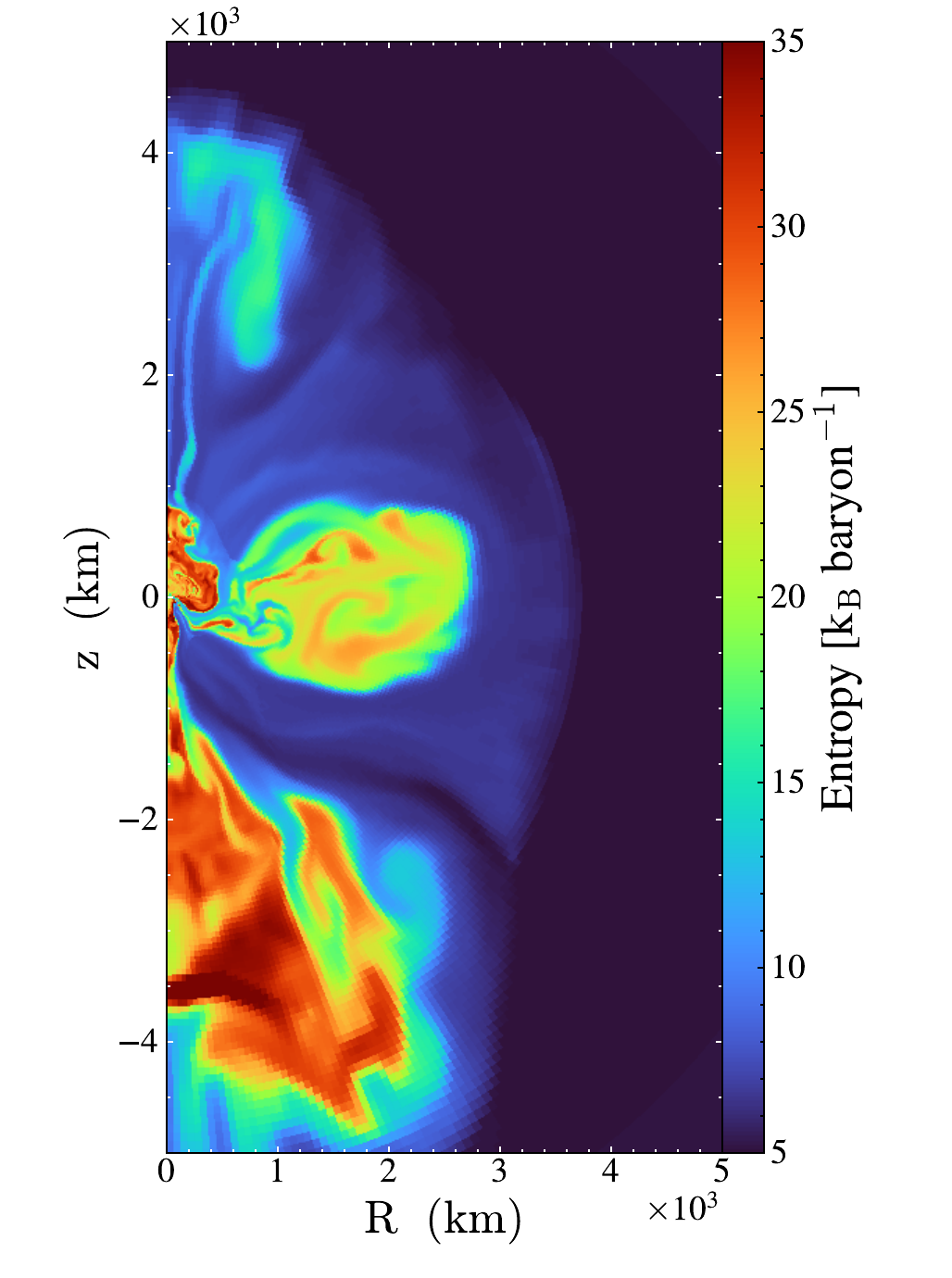}{0.32\textwidth}{(e) M5}
    \fig{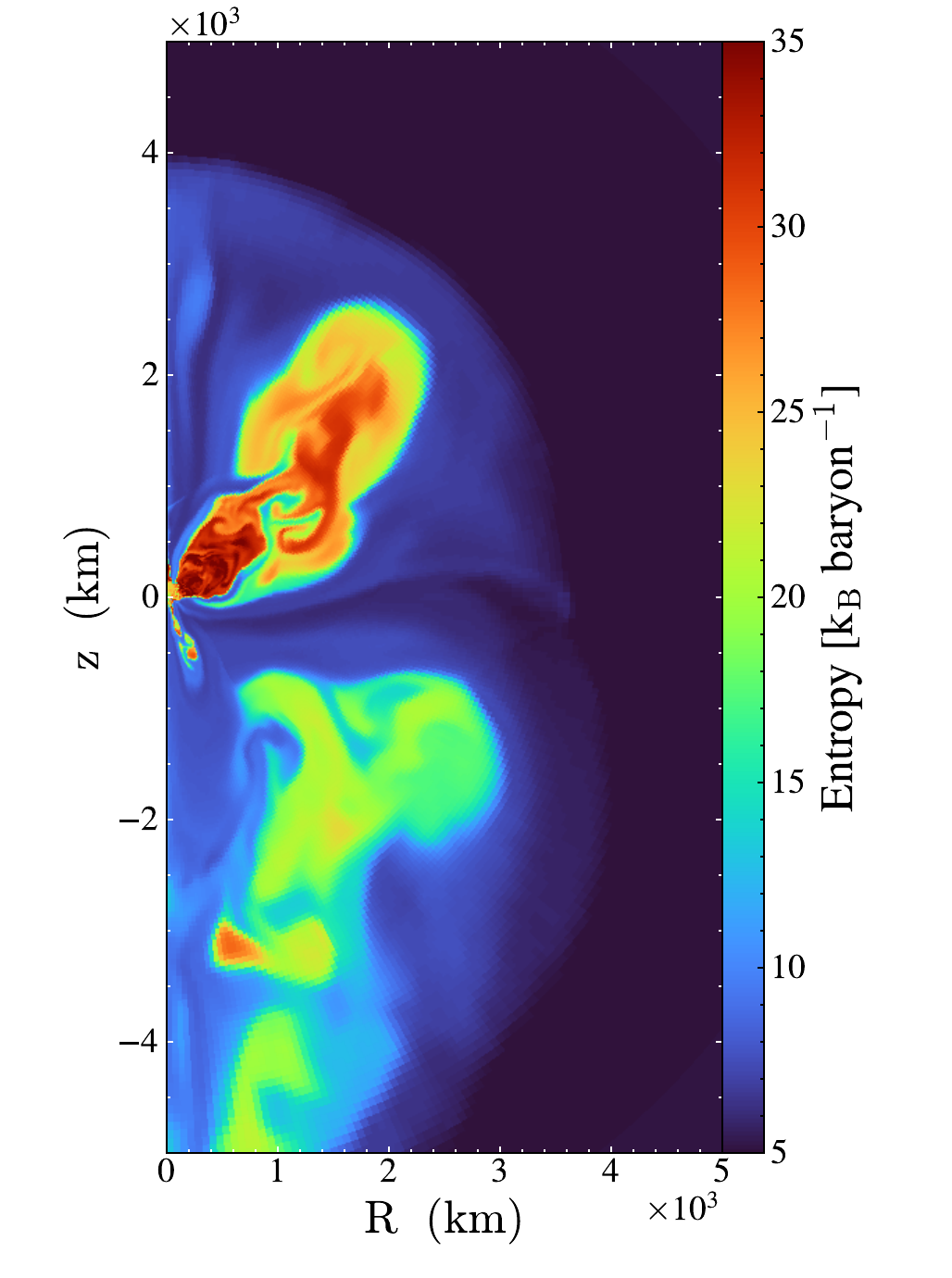}{0.32\textwidth}{(f) M6}
  }
  \caption{Entropy pseudocolor plot to radius 5000 km at 600 ms after bounce for (a) M1 to (f) M6 on the same 5--35~\kbbar\ scale. Explosion is sufficiently mature and explosion morphology is mostly determined at this time.}
  \label{fig:entropy-ms600}
\end{figure*}

After the onset of explosion, the general ejecta morphologies for our models and their changing process are characterized by the shock dipole shown in Figure \ref{fig:dipole-time}(b). The shocks in M1 and M4 are relatively symmetric with respect to their equatorial planes, $\ZoverR \approx 0$, for their entire runs.  M2 and M3 are clearly asymmetric that $\ZoverR$ tends to large negative and positive values respectively, indicating the strong and consistently developing unipolar ejecta in these two models. M5 and M6, though both show strong asymmetry in their earliest explosion stages, are affected by individual strong accretion events and thus change their overall morphology. These features are shown in Figure \ref{fig:entropy-ms600}, the entropy slice plots at $\tpb=600$ ms.

\paragraph{M1}
The mean shock radius of M1 deviates from the other five models after $\tpb=300$ ms (Figure \ref{fig:mean-shock}(c)). 
We deduce this deviation is not caused by its weak early SASI-like motions; otherwise M5 and M6 would behave similarly; but rather by the lower \Ediag\ of M1 in its early explosion phase ($\approx$200--500 ms after bounce). 
This indicates a more slowly developing explosion, resulting in a smaller mean shock radius. 
Notably, M1 exhibits a comparable or even higher neutrino heating rate, accretion rate, and neutrino luminosity than M5 and M6. 
The explosion geometry of M1 at 600 ms after bounce, as shown in Figure \ref{fig:entropy-ms600}(a), is nearly symmetric about the equator with an equatorial accretion stream. 
We therefore believe its smaller \Ediag\ is likely a result of the more spherical shock morphology of M1, 
which inhibits the formation of a sustained, unidirectional accretion stream onto the PNS and disperses the energy deposited into the post-shock materials over a wider spatial distribution. 
During the early explosion phase, no single strong, steady accretion stream dominates; instead, 2--3 moderate streams from different directions compete and interact with rising plumes. 
This competition suppresses the development of stable ejecta and initially favors a bipolar configuration. 
It first accretes onto the PNS from its northern side near the equator, and is gradually pushed to the north pole as the southern ejecta wins until $\tpb \approx 550$ ms. 
Between 550--560 ms, the upper portion of the south ejecta collides with the accretion stream and splits off a small plume ejecting approximately 50$^\circ$ from the north pole, which pushes the accretion stream toward the south and gradually truncates the southern plume. 
Eventually, a steady equatorial accretion stream accreting onto the PNS south pole and a northern ejecta plume with an opening angle of $\sim 40^\circ$ are set at the end of the simulation. 
Together with the early propagating southern ejecta, the explosion morphology of M1 is prolonged to an elliptical shape. 
We also suggest that concentrating ejecta into a narrower angular range in one hemisphere contributes to the rising \Ediag\ observed in M1 after 690 ms after bounce.

\paragraph{M2 and -M3} 
The explosion morphology of M2 is dominated by an unipolar ejecta toward the south with accretion onto the PNS from its north pole (Figure~\ref{fig:entropy-ms600}(b)). At a later time (after $\tpb \approx 700$ ms), it develops a new minor lobe along the +$z$-axis (Section \ref{sec:exploE}) without changing its unipolarity. 
M3 has a strong unipolar ejecta toward the north with a weaker southern plume and accretion onto the PNS from its south hemisphere near the pole (Figure~\ref{fig:entropy-ms600}(c)). These two models with strong unipolarity have relatively simpler interactions and development histories between their accretion streams and main ejecta plumes compared to the other four models. 

\paragraph{M4}
The bipolar explosion morphology of M4 has a strong equatorial accretion stream onto the PNS, as shown in Figure \ref{fig:entropy-ms600}(d), and such a feature forms before the onset of explosion. 
The large north and south plumes force the accretion streams to converge at the PNS equator, but which pole they accrete onto the PNS is shaped by the interaction between them and the plumes. 
This unimpeded flow of the equatorial stream to the PNS without interruption from minor plumes likely enables the growth of the explosion energy, as is demonstrated in Section~\ref{sec:m2b}.
Although more energy is deposited into the south plume during the early explosion stage, the accretion event (see above Section \ref{sec:accretion}) restores the north plume and makes it the dominant site of energy deposition. However, this change does not completely cut off the deposition into the south plume, and the overall explosion morphology of M4 remains bipolar to the end of the simulation. 
This plume switching is reflected in the dipole change shown in Figure \ref{fig:dipole-time}(b) with a delayed-time feature.

\paragraph{M5}
Similar to M1, M5 and M6 have more symmetric geometries as plotted in Figure \ref{fig:entropy-ms600}(e) and (f), but their dipoles (Figure \ref{fig:dipole-time}) indicate that the geometry of M5 is more dominated by its southern ejecta, while M6 is balanced by different plume orientations. In both models their relatively symmetric geometries form from periodic switching of dominant plume pole during explosion. In the earliest explosion phase of M5, three plumes develop in the north, equatorial, and south, divided by two accretion streams, and the southern plume has the largest radius. The accretion stream in the northern hemisphere cuts off continuous material ejection in the north plume, as the accretion stream in the southern hemisphere blends with the material near the PNS south pole, weakening the material supplement to the south plume. Consequently, the equatorial plume dominates and gradually grows into the southern hemisphere and the the dipole of M5 tends towards $\ZoverR \approx 0$. However, this plume is soon divided into two plumes again at the equatorial and south by the accretion, which comes from the southern hemisphere and accretes onto the PNS north pole at around 420 ms after bounce, setting the continuous material supplement to the south plume and growing negative dipole. 
At a later time, about 500 ms post-bounce, the attempt to develop a plume at the PNS north pole pushes accretion streams in both hemispheres and the equatorial lobe, causing the weakening of the southern ejecta. 
Eventually, at the end of simulation, M5 has most of the materials ejecting out from its PNS north pole with an angle of approximately 10$^\circ$, as well as a weaker equatorial plume. 

\paragraph{M6}
Unlike the other five models, the overall symmetric explosion morphology of M6 shown in Figure \ref{fig:entropy-ms600}(f) is not roughly established at its onset of explosion. Its preliminary geometry is dominated by a large southern lobe and a minor northern lobe, but this unipolarity trend is altered soon after the onset of the explosion. At 250 ms after bounce, the accretion stream comes from near equatorial accreting onto the PNS south pole, pushing the upper part of the lobe to the northern hemisphere, and gradually cuts off the material supplement to the southern ejecta. The split plume stably develops in the northern hemisphere with a angle of $\sim 35^\circ$. As the result, this early accretion event changes the overall explosion morphology of M6, setting its dipole recovering towards a more symmetric form. At the end of the simulation, we observed nearly all the ejecta go into the north plume. 
As a result, the shock radius in the northern hemisphere gradually catches up to that in the south, leading to a symmetric shock geometry and $\ZoverR \approx 0$.  

\subsubsection{Morphology and Explosion Energy}
\label{sec:morph&E}

As we try to build a general pattern between the explosion morphology and the explosion diagnostic energy, we find that the relevance between them is weak. Figure \ref{fig:explo-energy} shows that M4 has the largest \Ediag\ and M3 has the lowest \Ediag\ among our models. If the preferential heating into one hemisphere stands, as deduced by \cite{VaCoBu22}, we should see that M2 and M3 have higher heating rates and diagnostic energy than the other four models; however, this is not what we observe from our models. Compared to their 3D models, our 2D models inherently force the ejecta to be axisymmetric, and thence preferential heating is more prominent in 2D than in 3D. 

The diagnostic energy is more tied to the mass accretion into the the emission region on top of the PNS. 
As shown in Figure \ref{fig:lumin-accrete}, the higher accretion rates in M2 and M4 (panel b) and the larger total accreted masses (panel c) correspond to the higher diagnostic energies seen in Figure \ref{fig:explo-energy}, while models with lower accretion rates and accumulated masses result in lower diagnostic energies. 
Comparing Figure \ref{fig:lumin-accrete}(c) with Figure \ref{fig:explo-energy}, we find that accretion at 100 km correlates strongly with the PNS mass increase and also links the cumulative mass accretion to \Ediag. 
Instead of strict order mapping, grouping the models with similar trends is more reasonable. M2 and M4 have higher accretion rates and cumulative inward mass accretion, which indicate more kinetic energies carried by the downflow are converted to thermal energies to supply the explosion; thus higher \Ediag. 
The higher accretion in M2 is due to its unique structure (Section~\ref{sec:inicondi}), while in M4 it results from unimpeded equatorial accretion downstream (Section~\ref{sec:morphM1toM6}). 
The other four models, M1, M3, M5, and M6, have similar yet lower total accreted masses, as well as the resemblance to flattened \Ediag\ growth. The larger cumulative mass accretion slopes of M1 and M3 represent their continuously growing \Ediag, while the lower slopes on the cumulative mass accretion growth of M5 and M6 correspond to their more flattened \Ediag\ growth. 
This grouping by plume and accretion flow morphology seems to have a greater impact on outcomes than perturbations in the initial conditions emphasized by prior studies.
The impact of the stochasticity will be discussed further in Section \ref{sec:stochasticity}.

\subsection{Nucleosynthesis}
\label{sec:nucleosynthesis}

\begin{figure*}
    \includegraphics[width=\textwidth]{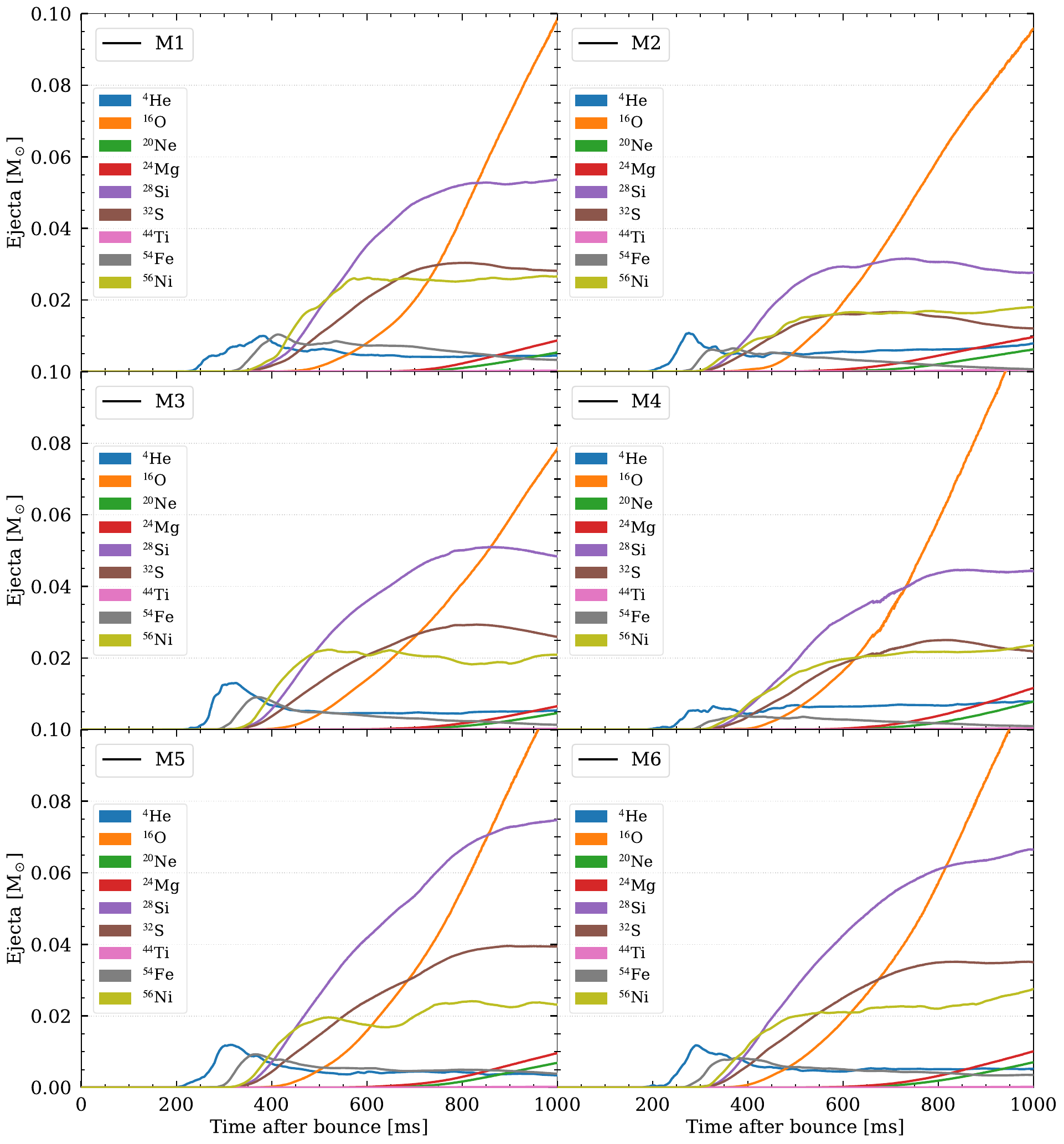}
    \caption{Ejecta (unbound materials with $v_r > 0$) masses for select nuclei for the first second after bounce. The sequence (a)--(f) follows a left-to-right, top-to-bottom arrangement, from M1 to M6.  }
    \label{fig:nucleosynthesis}
\end{figure*}

In this section, we focus on the nucleosynthetic yields obtained from our simulations. In CCSN ejecta, nuclear production encapsulates the thermodynamic histories of the ejecta.  They are determined by the interaction of explosive nuclear burning with progenitor compositions, and shaped by shock propagation, neutrino heating, and multi-D hydrodynamics. As we discussed in Section \ref{sec:inicondi}, the first three models, M1, M2, and M3, exhibit strong compositional differences in the silicon shell and modest compositional differences in the convective oxygen-burning shell at collapse.  The composition in the remainder of the oxygen shell is similar; however the radial coordinate of this transition is different by 1000 km, despite the mass coordinate being the same in the three models. In contrast, the other four models with the same \polaris-2D progenitor, M3 though M6, have the same initial composition; thus, differences in their ejecta relate solely to their different hydrodynamic and thermodynamic histories. 

In Figure \ref{fig:nucleosynthesis}, we track the most important nuclei in the ejecta, defined as unbound matter (see Equation \ref{eq:postive_energy_zone}) with positive radial velocity, included in our {\tt sn160} network to 1~s after bounce: \isotope{He}{4}, \isotope{O}{16}, \isotope{Ne}{20}, \isotope{Mg}{24}, \isotope{Si}{28}, \isotope{S}{32}, \isotope{Ti}{44}, \isotope{Fe}{54}, and \isotope{Ni}{56}. 
We ignore \isotope{C}{12} because the shock waves are still propagating in the lower non-convective oxygen shell at the end of our simulations (see Figure \ref{fig:abund-ms0}(a) and \ref{fig:mean-shock}(d)) and have not reached the carbon-rich material.
\isotope{He}{4} is the first species to be unbound, around 200 ms after bounce, matching the timing of shock revival, when the small amount of unbound material is still dominated by extremely hot and dense materials under NSE. 
This is followed by the appearance of \isotope{Fe}{54} at approximately 250--300 ms post-bounce, corresponding to a drop in \isotope{He}{4} as this NSE material expands and cools, experiencing \emph{$\alpha$-rich freezeout, which favors iron-group nuclei}.
\isotope{Ni}{56}, \isotope{Si}{28} and \isotope{S}{32} follow rapidly at $\tpb\approx 300$--350~ms. 
These timings are within roughly 50 ms after the shock enters the convective oxygen shell (see Table \ref{tab:Modelinfo}), indicating that iron-peak ejecta are products of both \emph{NSE freeze-out} and shock-introduced \emph{explosive silicon burning}. 
For $\Ye \approx 0.5$, \emph{incomplete silicon burning} favors the formation of \isotope{Fe}{54} (and two protons), which leads to the earlier appearance of \isotope{Fe}{54} than \isotope{Ni}{56}. As this material expands and cools, \isotope{Fe}{54} converts to \isotope{Ni}{56} \citep{TrCaGi66, BoClFo68}. 
The appearance of \isotope{Fe}{54} does not reflect the appearance of lower \Ye ($\approx 0.48$) material in the ejecta. 
The presence of \isotope{Si}{28} and \isotope{S}{32} represents both unburnt nuclei surviving from \emph{incomplete silicon burning} and the products of \emph{explosive oxygen burning}. 
As the shock moves outward, its peak temperature declines, allowing undestroyed \isotope{O}{16} to join the ejecta, generally around 400 ms post-bounce, while the shock is still in the convective oxygen-burning shell. 
This is delayed in M1 about 50 ms due to its slower shock at this time. 
\isotope{Mg}{24} and \isotope{Ne}{20} are the last species to join the ejecta when the shock reaches the non-convective oxygen shell, which has unburnt \isotope{Mg}{24} and \isotope{Ne}{20} in the progenitors. 
These two nuclei have nearly identical distributions across 1.7--2.6 \msun\ with \isotope{Mg}{24} owning a higher mass fraction of $\approx$0.02 (Figure \ref{fig:abund-ms0}(a)). 
However, the earlier appearance of \isotope{Mg}{24} in the ejecta indicates that a portion of \isotope{Mg}{24} is a product of \emph{explosive neon burning}. 
The \isotope{Ti}{44} yields remain between $10^{-4}$ and $10^{-3}$ \msun\ in all six models, so the lines are indistinguishable from zero in Figure~\ref{fig:nucleosynthesis}. 
This is likely to rise if the models were run further \citep{WaBu24c}. 

The shocks' ongoing propagation through the lower oxygen shells is responsible for the steep rise in yields for \isotope{O}{16}, \isotope{Ne}{20}, and \isotope{Mg}{24} that continues at 1~s for all models in Figure~\ref{fig:nucleosynthesis}, as the shock unbinds material transmuted during carbon shell burning.
At 1 s after bounce, the ejecta in models M4 to M6 contain at least 0.1 \msun\ of \isotope{O}{16}, while M1 (0.097 \msun), M2 (0.096 \msun), and M3 (0.079 \msun) are close to or below this value and exceed it by the end of simulations.
These values are reflective of the shocks' progress through the oxygen shells, less any \isotope{O}{16} destroyed by \emph{explosive oxygen burning}. 
These values, however, are not the final yields for \isotope{O}{16} as the shock is expected to, at a time beyond our simulations, unbind the remaining oxygen-shell mass. 

What does distinguish between progenitor models are \isotope{Si}{28}, \isotope{S}{32}, and \isotope{Ni}{56} yields. 
At 1 s after bounce, M1 has $\approx$0.054 \msun\ of \isotope{Si}{28}, 0.028 \msun\ of \isotope{S}{32}, and 0.027 \msun\ of \isotope{Ni}{56} in its ejecta; M3 has $\approx$0.048 \msun\ of \isotope{Si}{28}, 0.026 \msun\ of \isotope{S}{32}, and 0.021 \msun\ of \isotope{Ni}{56} in its ejecta. 
In contrast, M2 has only $\approx$0.028 \msun\ of \isotope{Si}{28}, 0.012 \msun\ of \isotope{S}{32}, and 0.018 \msun\ of \isotope{Ni}{56} in its ejecta, which is obviously lower than M1 and M3. 
Though the progenitor structure differences within the silicon shell have been destroyed, the significantly higher \isotope{Si}{28} and \isotope{S}{32} fractions in the silicon shell of \polaris-1D progenitor of M2 (Figure \ref{fig:abund-ms0}) offer more fuel for shock-induced \emph{explosive silicon burning}, which contributes to an earlier explosion and higher explosion energy in M2. 
The higher explosion energy also explains its lower \isotope{Si}{28} and \isotope{S}{32}, but we do not observe the corresponding higher amount of \isotope{Ni}{56}. 
Instead, other products of \emph{complete explosive silicon burning}, such as \isotope{Ti}{44} (if $\alpha$-rich), \isotope{Cr}{48}, and \isotope{Fe}{52}, have higher yields in M2 by small, but physical meaningful, amounts (a couple times of $10^{-4}$ \msun).

Models M3 to M6 share the same \polaris-2D progenitor but differ in the treatment of lateral velocity. 
Therefore, any differences in ejecta composition are attributed to variations in explosion development related to lateral velocity treatment rather than differences in the progenitor structure. 
At 1 s, M4 has $\approx$0.044 \msun\ of \isotope{Si}{28}, 0.022 \msun\ of \isotope{S}{32}, and 0.024 \msun\ of \isotope{Ni}{56} in its ejecta, which is similar to those values for the ejecta in M3. 
The faster decrease of \isotope{Fe}{54}, earlier appearance of \isotope{Mg}{24} and \isotope{Ne}{20}, and larger amount of \isotope{O}{16} match the faster shock progression in M4 (Figure \ref{fig:mean-shock}). 
The same reason leads to the larger amount of \isotope{O}{16} also observed in M5 and M6. Their \isotope{Ni}{56} yields at $\tpb = 1$~s are both 0.025 $\pm$ 0.03 \msun, similar to M4 and higher than M3. 
However, these two models have a notably higher \isotope{Si}{28} and \isotope{S}{32} yields in their ejecta: $\approx$0.075 \msun\ of \isotope{Si}{28} and 0.039 \msun\ of \isotope{S}{32} in M5, and $\approx$0.066 \msun\ of \isotope{Si}{28} and 0.035 \msun\ of \isotope{S}{32} in M6. 
We observe that models with non-zero lateral velocities (M3 and M4) have similar abundances in \isotope{Si}{28}, \isotope{S}{32}, and \isotope{Ni}{56}, and models without lateral velocities (M5 and M6) exhibit an even higher resemblance across our selected nuclei shown in Figure \ref{fig:nucleosynthesis}.

\begin{figure}
    \fig{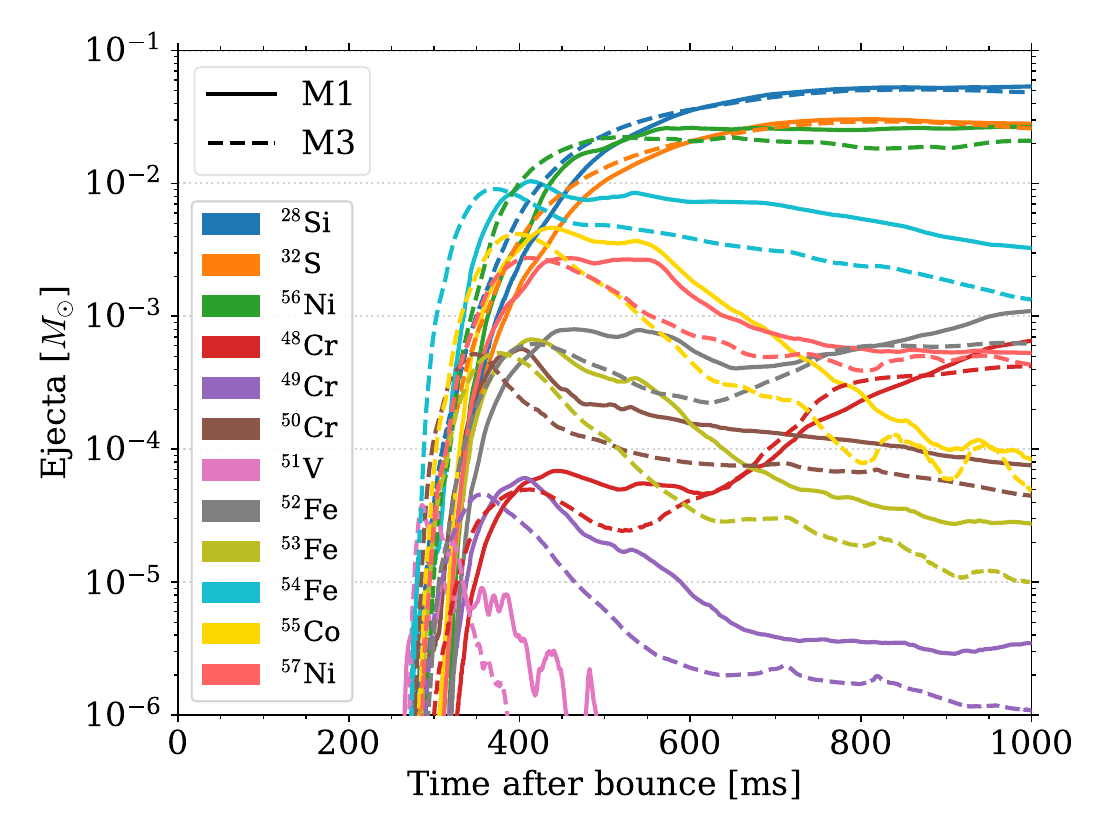}{\columnwidth}{(a)}
    \fig{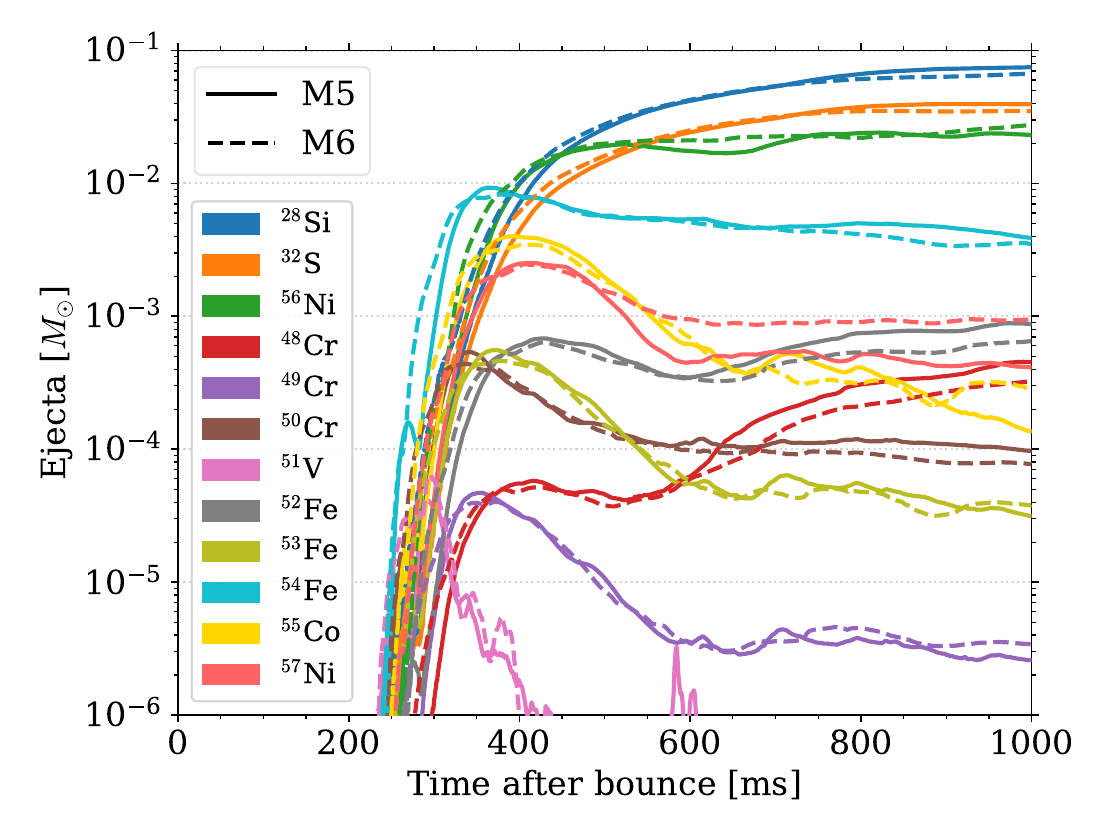}{\columnwidth}{(b)}
    \caption{Iron nuclei abundances in the ejecta (unbound materials with $v_r > 0$). These nuclei are produced through complete explosive silicon burning. (a) compares M1 and M3, and (b) compares M5 and M6. }
    \label{fig:abund-iron}
\end{figure}

We further analyze products from \emph{complete explosive silicon burning} in models M1, M3, M5, and M6. 
These models exhibit a similar flattening in their \Ediag\ (Figure~\ref{fig:explo-energy}). 
In addition to \isotope{Si}{28}, \isotope{S}{32}, and \isotope{Ni}{56} shown in Figure~\ref{fig:nucleosynthesis}, we track the abundances of iron nuclei, as suggested by \cite{WoHeWe02}, in Figure~\ref{fig:abund-iron}. 
The comparison between M1 and M3 shows that they have consistent differences in \isotope{Cr}{49}, \isotope{Cr}{50}, \isotope{Fe}{53}, and \isotope{Fe}{54} abundances, with M1 being $\approx$2--3 times more abundant than M3 at 1 s. 
M5 and M6 generally  have more similar iron abundances. 
We only find consistent abundance differences in \isotope{Cr}{48}, \isotope{Cr}{50}, \isotope{Fe}{52}, and \isotope{Ni}{57}. 
At 1 s, M5 is $\approx$1.25--1.4 times more abundant than M6 in \isotope{Cr}{48}, \isotope{Cr}{50}, and \isotope{Fe}{52}, whereas M6 contains $\approx$2.30 times \isotope{Ni}{57} than M5.

Against the backdrop at 1 s of steadily rising abundances of \isotope{O}{16}, \isotope{Ne}{20}, and \isotope{Mg}{24} and slightly rising abundance of \isotope{Ni}{56} and \isotope{Ti}{44}, the mixed behavior of \isotope{Si}{28} and \isotope{S}{32}, merits additional discussion.  
While there is steadily expanding Si+S-rich ejecta in all of the models, the accretion streams are also Si+S-rich with some of this matter rendered marginally unbound by interactions with the plumes.
This is the last matter in the supernova to have its fate determined, with much of it eventually accreting, contributing to the final growth of the explosion energy, but some joining the ejecta \citep{HaHiCh17}.  
As a result, the \isotope{Si}{28} and \isotope{S}{32} yields at 1 s are likely to decline over the next few seconds and are certainly subject to significant uncertainty.

\section{Impacts of Progenitor Asymmetry}
\label{sec:structure} 

In this section, we will discuss the impacts of asymmetry in the progenitor to the later explosion through comparison between M3 and its 1D counterparts, M5 and M6. 
In Section~\ref{sec:RD}, we have demonstrated the macroscopic similarities between these three models, as well as M4. 
However, we exclude M4 from the current discussion because it is not self-consistent. 
By spherically averaging the velocity in the \polaris-2D progenitor without zeroing its lateral velocity, M4 resembles 1D to multi-D mapping models from other studies with lateral features artificially implemented. 
Compared to the other three models with the same \polaris-2D progenitor, its outlier behaviors suggest that this self-consistency is salient when making a dimensionality transition. 

\begin{figure}
    \fig{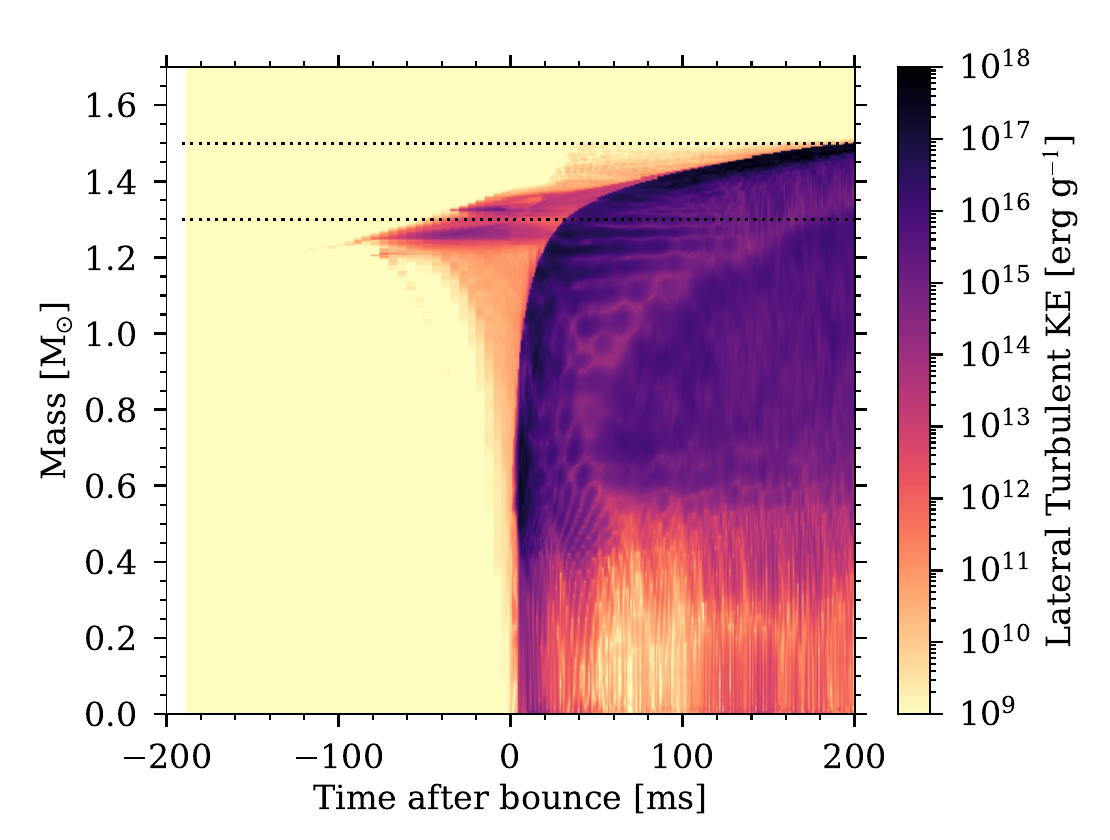}{.98\columnwidth}{(a)}
    \fig{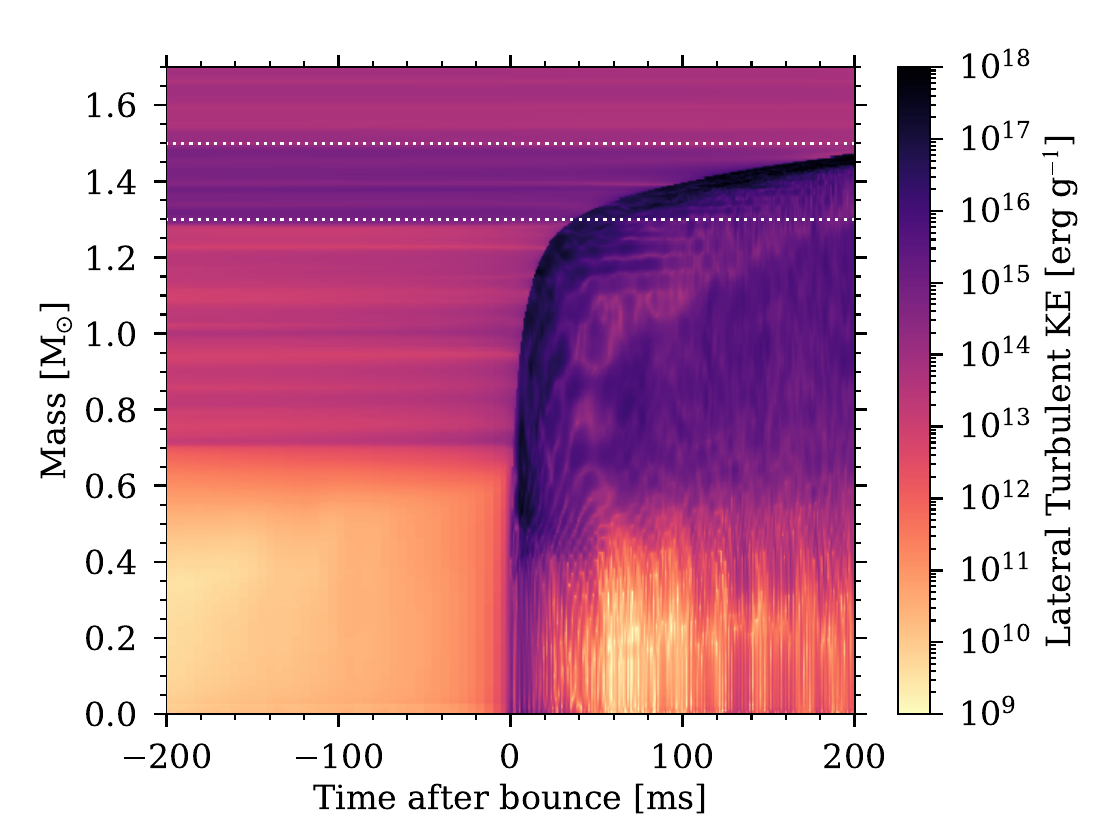}{.98\columnwidth}{(b)}
    \fig{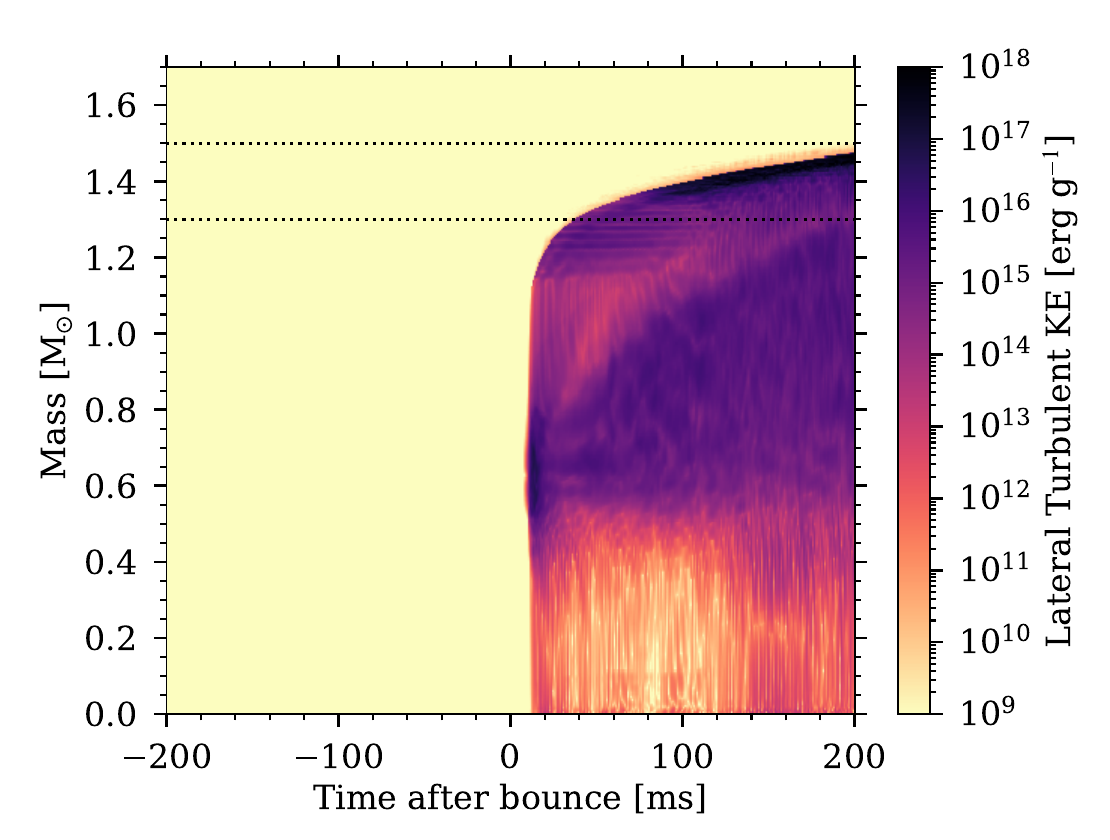}{.98\columnwidth}{(c)}
    \caption{Mass-time diagram between -200--200 ms post-bounce for lateral TKE for M2 (panel a), M3 (panel b), and M5 (panel c). The dotted lines indicate Fe/Si and Si/O boundaries at 1.3 and 1.5 \msun, respectively. }
    \label{fig:LTKE_spectrum}
\end{figure}

We plot the lateral TKE evolution in Figure~\ref{fig:LTKE_spectrum} for three of our models, which directly captures the impact of early asymmetry on the lateral TKE evolution. 
M3 (panel (b)) with the \polaris-2D progenitor has large lateral flows throughout the core before bounce. 
Its spherically averaged 1D counterpart M5 (panel (c)) with zero initial lateral velocity stays quiescent until $\approx$10 ms after bounce, when lateral TKE exceeding $10^{9}$ \ergg\ become visible in the plot.
M2, our anomalously noisy 1D model (panel (a)), starts from a `clean' pre-bounce configuration like other 1D models with spherical progenitors and no initial lateral velocities.
However, M2's lateral TKE reaches $10^{9}$ \ergg\ around 100 ms before bounce in its outer iron core where the network was active, unlike the rest of the models. 
The enhanced lateral TKE spreads both inward and outward, leading to a large turbulent region across the lower silicon-burning shell and outer iron core at bounce.
This results in post shock Lateral TKE in Figure~\ref{fig:LTKE_spectrum}(b) that is much more like Figure~\ref{fig:LTKE_spectrum}(a) than Figure~\ref{fig:LTKE_spectrum}(c).
We delve deeper by comparing M3, M5, and M6 in Section~\ref{sec:lateral}, and discuss M2 in Section~\ref{sec:M2_LTKE}. Lastly, we show a pre-shock density perturbation analysis in Section~\ref{sec:dens_mpole}, comparing our results to the literature. 

\subsection{Growth of Lateral Velocities}\label{sec:lateral}

\begin{figure}
    \fig{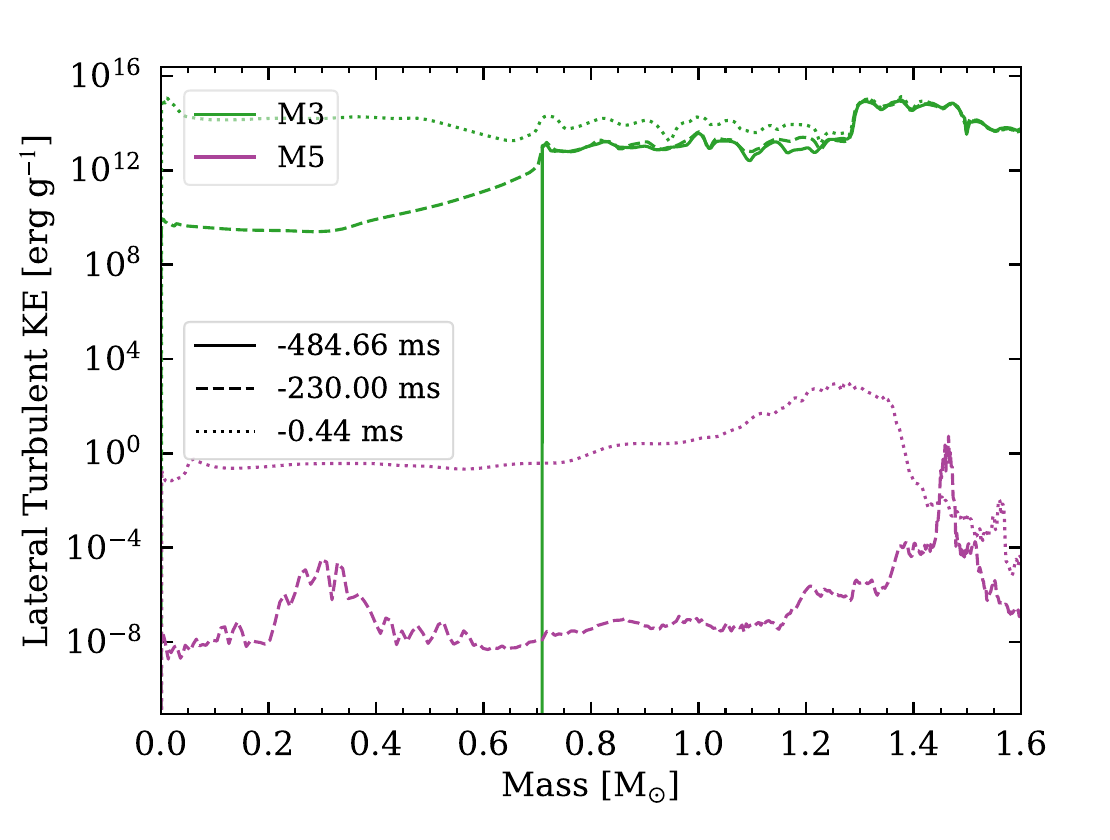}{\columnwidth}{(a)}
    \fig{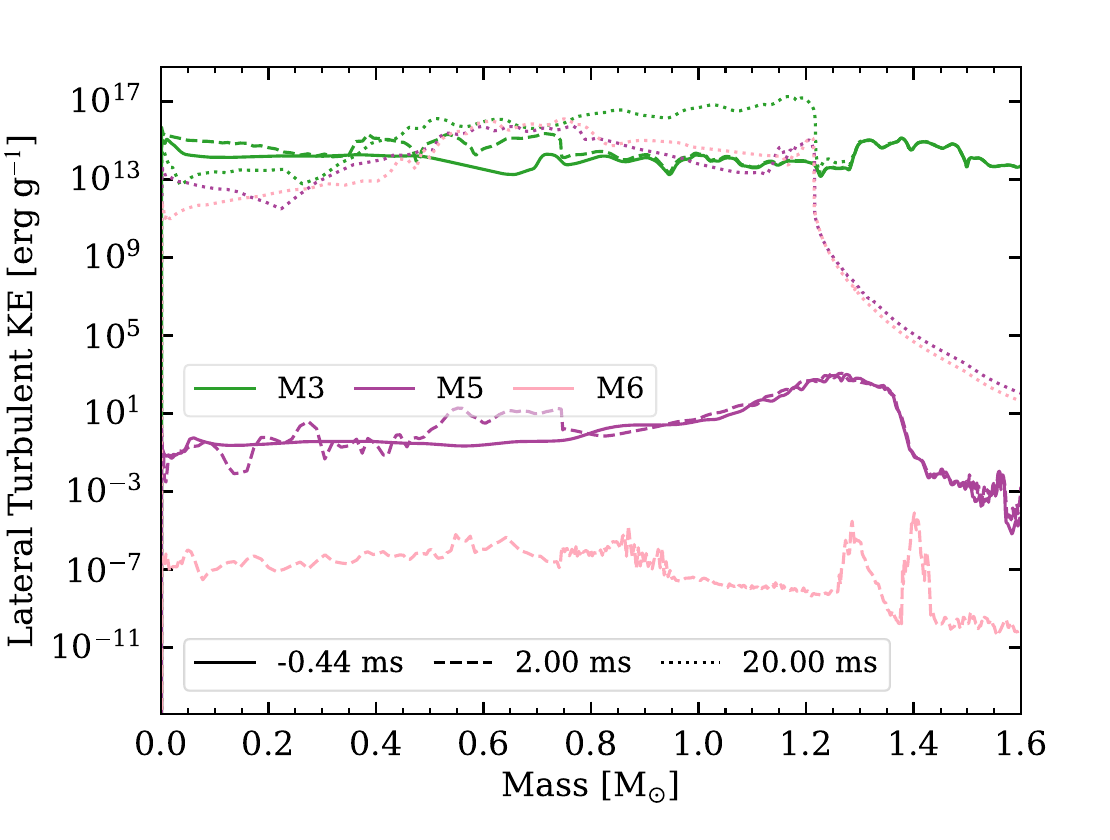}{\columnwidth}{(b)}
    \caption{Lateral TKE in the Fe+Si core region for the 2D progenitor model (M3, green) and spherically averaged version (M5, purple) (a) at the onset of collapse (solid lines), 230 ms prior to the core bounce (dashed lines), and just before the core bounce (dotted lines); and (b) just before core bounce (now solid lines), 2 ms (dashed lines) and 20 ms (dotted lines) after bounce. M6 (pink) is added in panel (b), and its 1D nature gives zero lateral TKE before core bounce. } 
    \label{fig:m3-m5-noise}
\end{figure}

First, we explore the effect of noise generation in the collapsing phase on our models, and thus the explosion dynamics, through measuring the variations in their radial and lateral TKE in the collapsing phase and immediately post-bounce, before shock asymmetry develops, roughly 20 ms after bounce. 

We begin by comparing the full multi-D progenitor model, M3, and its sphericalized counterpart, M5, through the early phases of collapse. 
In Figure \ref{fig:m3-m5-noise}(a), the lateral TKE in M5 is zero at the onset of collapse (thus not displayed in the plot), reaching values in the range $\sim 10^{-8}$--$10^{-4}$ \ergg\ by 230 ms prior to the bounce, with higher energies in the silicon-burning shell. 
Just before core bounce, the energies have increased further to $\sim 10^{-1}$--$10^{2}$ \ergg, peaking between 1.15--1.35 \msun\ at $\sim 10^{3}$ \ergg. 
In contrast, the lateral TKE of M3 is essentially unchanged within its intrinsically turbulent silicon shell (1.3--1.5 \msun) at a value close to $10^{15}$ \ergg.  It is amplified about ten times in its outer iron core, from $\sim 10^{13}$ \ergg\ to $\sim 10^{14}$ \ergg, while only the previously spherical inner iron core experiences dramatic changes in lateral TKE.

In Figure \ref{fig:m3-m5-noise}(b), we continue the analysis of the lateral TKE through bounce, adding M6.
The lateral TKE for M3 remains little changed through core bounce, however, it grows significantly in the region between 0.8-1.2 \msun\ by $\tpb=20$ ms.  
In M5 at 2 ms after bounce, the lateral TKE is also little changed from just before core bounce, but 18 ms later, it has increased to the level $\sim 10^{14}$ \ergg\ throughout the post-shock region, similar to the level of M3 near core bounce. 
Even in M6, whose enforced spherical symmetry during collapse kept the lateral TKE to zero until bounce (and hence missing from Figure \ref{fig:m3-m5-noise}(b) at $\tpb=-0.44$ ms), the lateral TKE rises to $\sim 10^{-7}$ \ergg\ within 2 ms and $\sim 10^{14}$ \ergg\ within 20 ms after bounce.

Shock deformation plays a key role in the growth of lateral kinetic energy by diverting radial kinetic energy, which grows large during collapse, into the lateral direction \citep{MuJa15}.  
The fully multi-D nature of M3 introduces shock asymmetry at $\tpb=0$ ms, measured as dipole \ZoverR\ (Figure \ref{fig:dipole-time}), leading to the growth of lateral TKE behind the shock (1.2 \msun\ at $\tpb=20$ ms).
The development of shock asymmetry of its sphericalized counterparts M5 and M6 is much slower until the shock stall phase, $\tpb\gtrsim75$ ms, when neutrino-driven convection develops and soon dominates the dynamics in the cavity. 
Shock deformation amplifies the lateral TKE of M5 from $\sim 10^{1}$ to $\sim 10^{14}$ \ergg, but for M3, its lateral TKE increases only from about $10^{14}$--$10^{15}$ to $10^{16}$--$10^{17}$~\ergg\ in the post-shock region, indicating that $\sim 10^{17}$ \ergg\ is the level at which the lateral TKE saturates.

\begin{figure}
    \fig{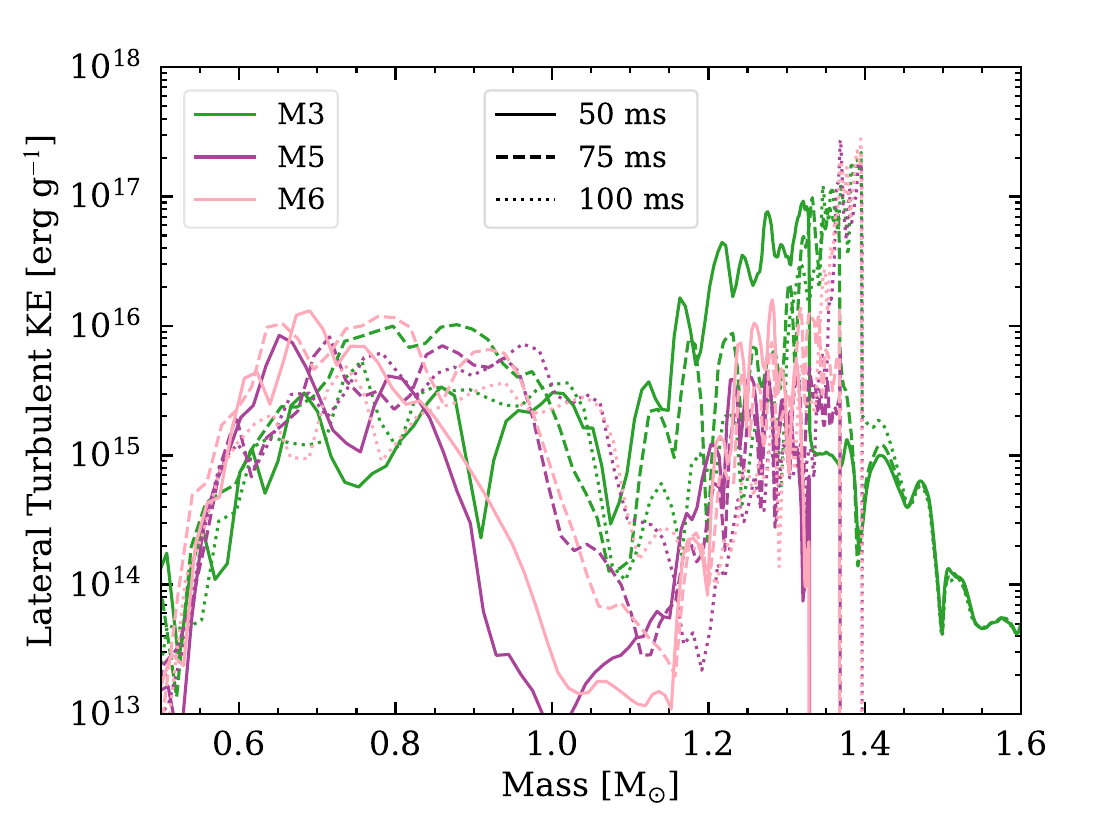}{\columnwidth}{(a)}
    \fig{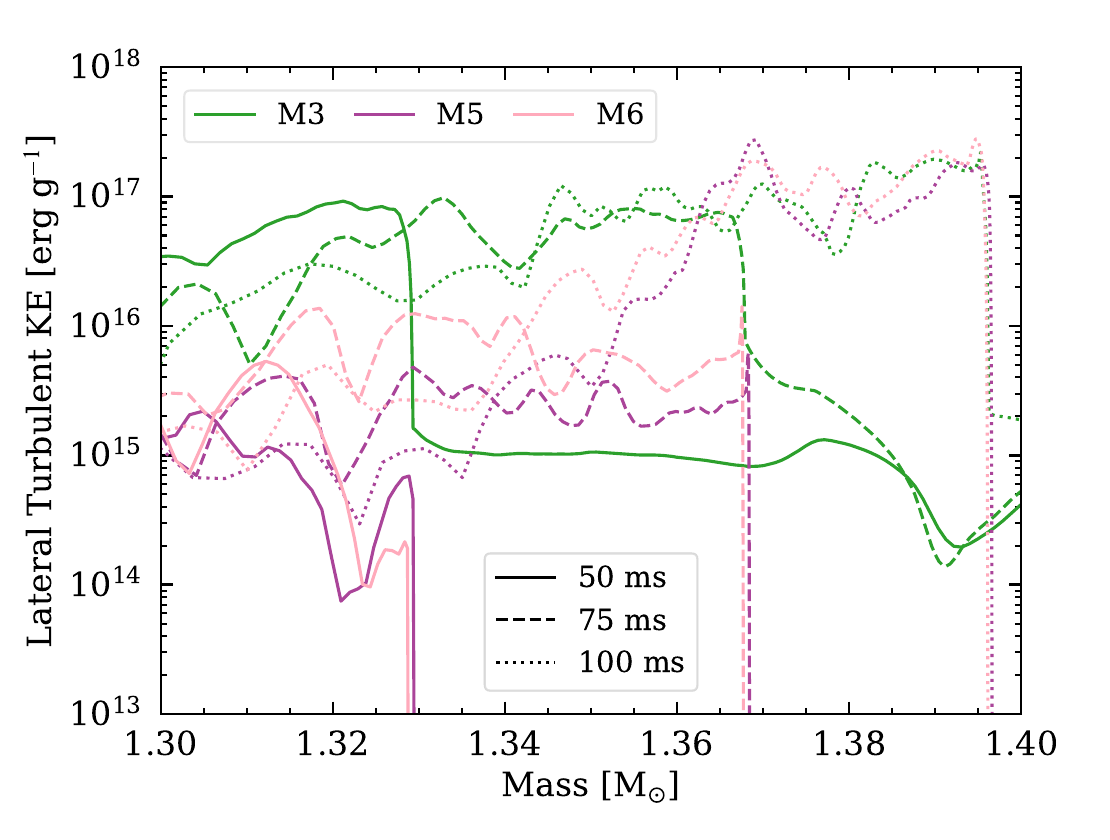}{\columnwidth}{(b)}
    \caption{Lateral TKE for M3, M5, and M6 at 50 ms (solid lines), 75 ms (dashed lines), and 100 ms (dotted lines) after bounce (a) in the outer Fe-core and Si shell (0.5--1.6~\msun) and (b) for 1.3--1.4~\msun\ to demonstrate details in the expanding shocked cavity.}
    \label{fig:ke-lat-m356}
\end{figure}

We next explore how the post-shock lateral TKE is impacted during shock stall phase ($\tpb=50$--100~ms). Figure \ref{fig:ke-lat-m356}(b) shows, in the cavity by $\tpb=75$ ms (dashed lines), the lateral turbulent kinetic energies of M5 and M6 has been amplified by the shock deformation to the level of $10^{15}$--$10^{16}$ \ergg, only 10--100 times smaller than the saturation energy level that M3 achieves early. This energy increase introduced by shock deformation proves the existence of SASI in our models. Between 75--100 ms post-bounce, the neutrino-driven convection develops in M5 and M6 and rapidly takes effect, further lifting up the energy level to saturation. Hence, we see these three models have quite similar levels of turbulent energy by 100 ms. With this final increase by the neutrino-driven convection between 75--100 ms post-bounce, the increase of lateral component of the kinetic energy of turbulent motion matches that measured in the gain region in Figure~\ref{fig:ke-lat}(b). 

\subsection{Noise Growth in M2}\label{sec:M2_LTKE}

\begin{figure}
    \fig{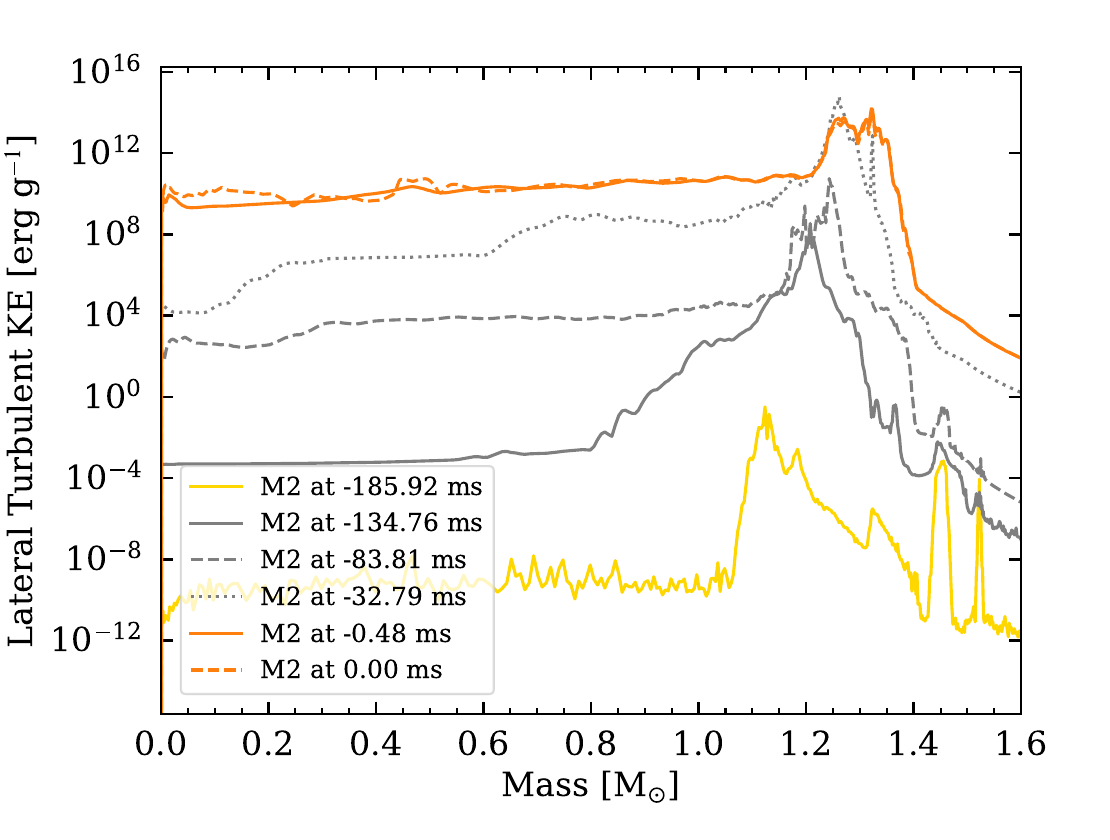}{\columnwidth}{(a)}
    \fig{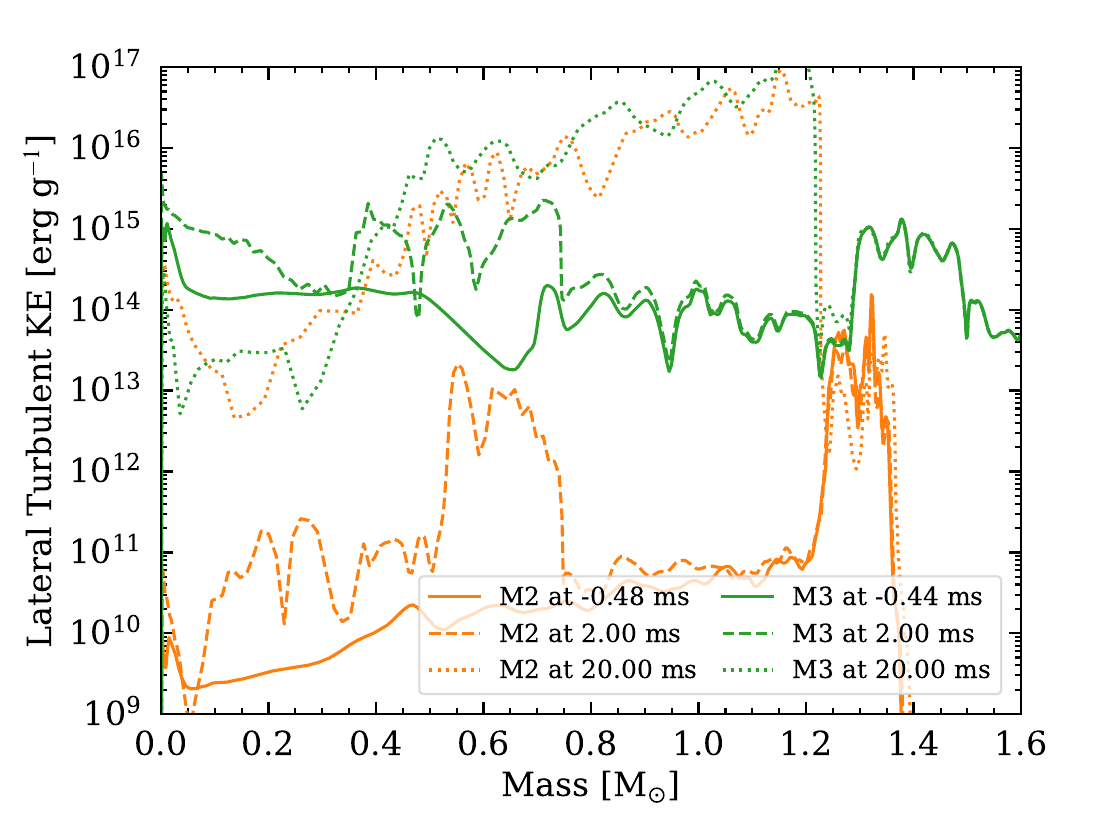}{\columnwidth}{(b)}
    \fig{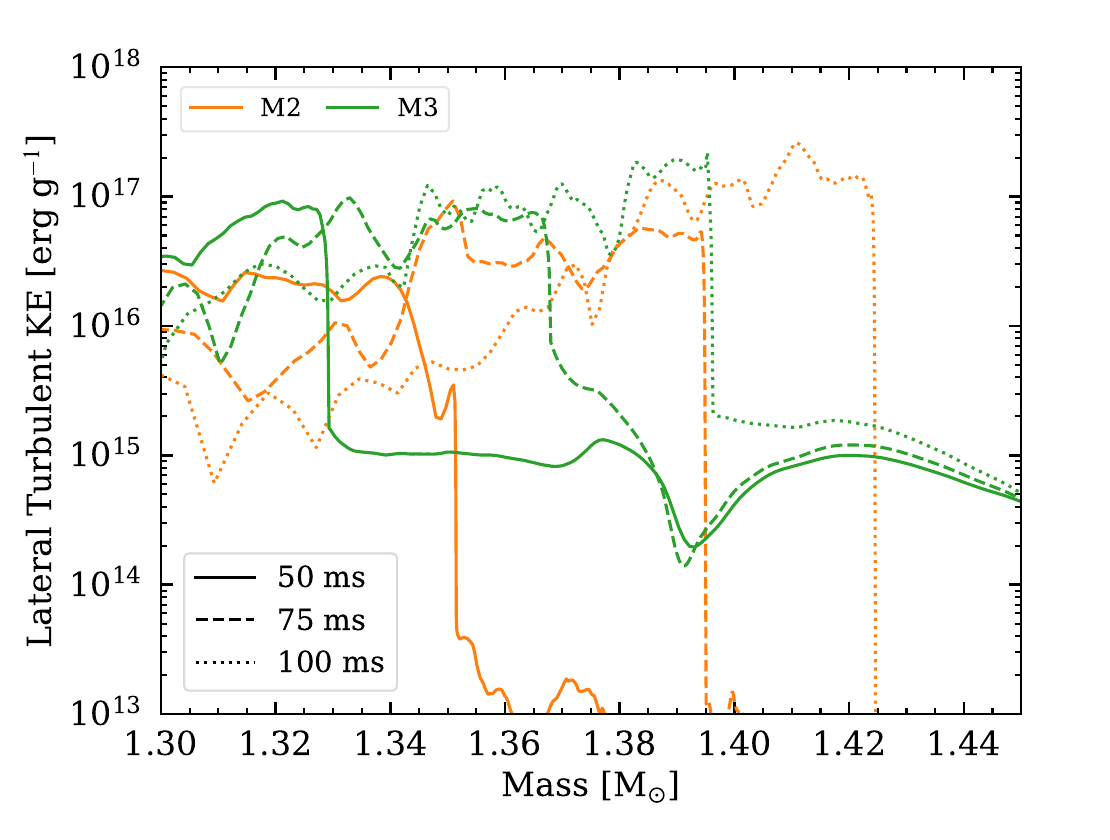}{\columnwidth}{(c)}
    \caption{Lateral TKE for (a) M2 in the collapsing phase; (b) M2 and M3 at the same times as in Figure~\ref{fig:m3-m5-noise}(b); and (c) M2 and M3 at the same times during shock stagnation as Figure~\ref{fig:ke-lat-m356} zoomed to 1.3--1.45~\msun. }
    \label{fig:m2-noise} 
\end{figure}

M2 is an anomaly among the models started from 1D progenitors. In the first 100 ms after bounce, it exhibits early development of shock asymmetry (Figure \ref{fig:dipole-time}(a)), leading to strong sloshing bulk fluid motions, and a higher lateral TKE in the gain region (Figure \ref{fig:ke-lat}(b)). These features resemble those found in our multi-D progenitor models M3 and M4. In addition, Figure \ref{fig:ke-lat}(a) shows M2 already has a considerable amount of lateral TKE everything at or interior to its Si-burning shell at bounce, while the values of other models with 1D progenitor stay low. 
Further, as shown in Figure \ref{fig:m2-noise}(a), while there is zero lateral TKE in M2 at the onset of collapse, within 5~ms ($\tpb=-186$ ms), the  lateral TKE has reached as high as $10^{-2}$ \ergg. As the collapse evolves, more and more noise is generated, eventually lifting the overall lateral TKE level to $10^{9}-10^{11}$ \ergg, with $\sim 10^{13}$ \ergg\ between 1.25--1.35 \msun: its upper PNS mantle and silicon shell bottom. This is approaching the energy level of M3.

Unlike the other models, the outer iron core in this model was inadvertently not treated in NSE during core collapse, which is computational anomaly relative to standard \chimera\ practice. Thus the NSE composition was maintained by the nuclear reaction network as the temperature and density increased during collapse.  Due to the exponential temperature dependence of nuclear reactions, the nuclear energy release from the nuclear network near NSE has a tendency to strongly amplify noise, as small temperature variations lead to large variations in energy release or absorption. This seems to be the cause of the noise in this model growing rapidly. 

After bounce, shown in Figure \ref{fig:m2-noise}(b), we witness the newly formed shock (located at 0.75 \msun) rapidly amplify the post-shock turbulent energy 100-fold, from $\sim 10^{11}$ at $\tpb=-0.5$ ms to to $\sim 10^{13}$ \ergg\ at $\tpb=2$~ms. Note that the shock deformation of M2 begins approximately 5 ms after bounce (Figure \ref{fig:dipole-time}(a)). Hence, the post-shock lateral TKE of M2 reaching $\sim 10^{16}$ \ergg\ at 20 ms after bounce, approaching the saturation level, can be attributed to SASI-like fluid motions (Section \ref{sec:SASI}). In Figure \ref{fig:m2-noise}(c), we examine the lateral TKE in the shock stall phase at 50, 75, and 100 ms after bounce, zoomed in to the cavity as Figure \ref{fig:ke-lat-m356}(b). At $\tpb=50$ ms, the post-shock lateral TKE is lower than M3 because its pre-shock matter is less turbulent, but notably higher than that for the other spherical progenitor models M5 or M6 in Figure \ref{fig:ke-lat-m356}(b). The neutrino-driven convection in M2 starts $\approx$60~ms after bounce, lifting the post-shock energy up to $\sim 10^{16}$ \ergg\ by $\tpb=75$ ms. By $\tpb=100$ ms, the lateral TKE reaches a similar $\sim 10^{17}$~\ergg\ saturation level as M3. This result is identical to our comparison in Section~\ref{sec:lateral}.

The comparison in the relative growth of lateral TKE in M2 compared to the full multi-D model M3 and its 1D counterparts, M5 and M6, near bounce (Figure \ref{fig:m2-noise}(b) relative to Figure \ref{fig:m3-m5-noise}) and during the development of neutrino driven convection (Figure \ref{fig:m2-noise}(c) relative to Figure~\ref{fig:ke-lat-m356}) provides a useful, if inadvertent, insight. 
We find that a discernible amount of turbulent energy is generated during the collapsing phase for 1D progenitor models due to nuclear burning. This is particularly true for M2, because the network is active throughout the outer Fe core.
This is the numerical result of noise amplification in \chimera\ by the nuclear network, but seems consistent with the physics of thermonuclear burning, small temperature inhomogeneities get amplified by nuclear burning, leading to small scale velocity inhomogeneities. 
In addition, regardless of the initial structure, the shock deformation generates most of the post-shock lateral TKE that saturates after reaching a certain level. 
Our results indicate that sufficient noise in the heating region, from any source, allows strong neutrino-driven convection that drives our models toward explosion. 
We therefore do not see strong connections between the pre-collapse structure and the explosion product, as indicated in the literature \citep{HaMaMu12, CoOc14, CoOt15, MuMeHe17, BoYaKr21, VaCoBu22}.

\subsection{Fluctuation Analysis}\label{sec:dens_mpole}

\begin{figure}
    \fig{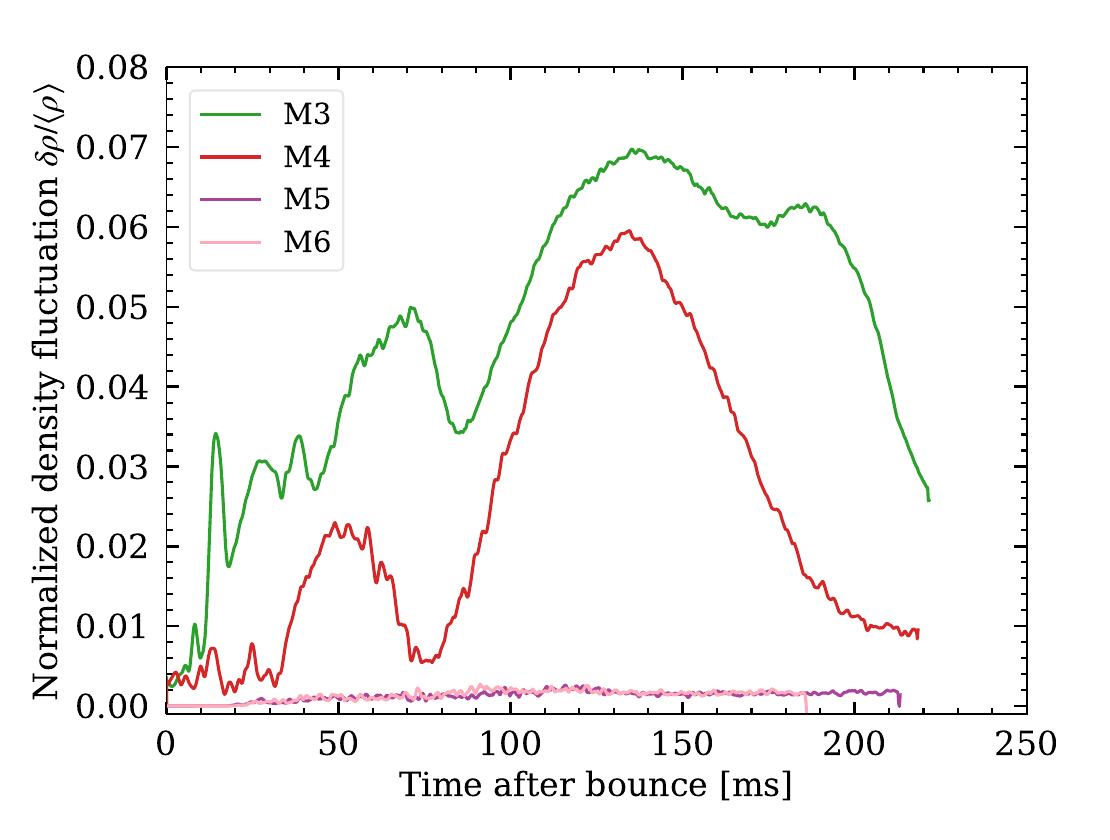}{\columnwidth}{(a)}
    \fig{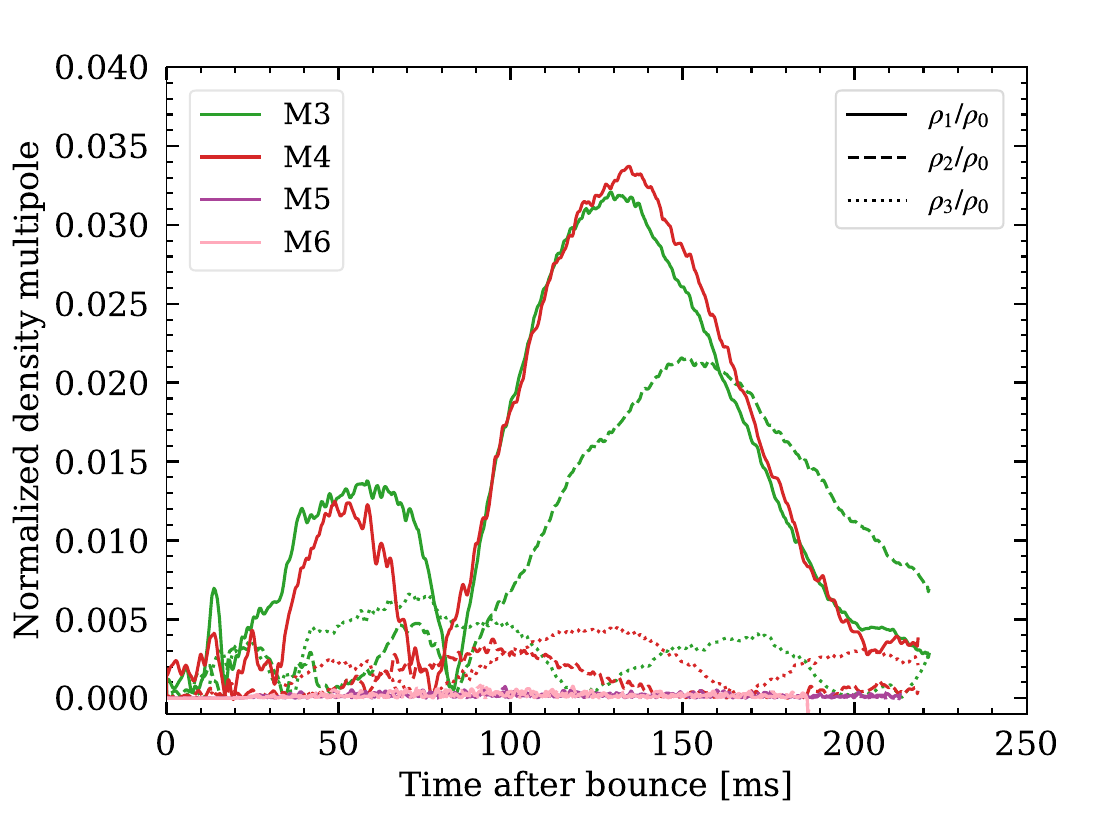}{\columnwidth}{(b)}
    \caption{Pre-shock density perturbation evaluated at radius 500~km by (a) normalized density fluctuations and (b) multipoles. No data are shown after the maximum shock reaches 500 km for each model. Lines for $\ell=1$, $\ell=2$, and $\ell=3$ modes are smoothed the data over 2 ms interval for clarity. $\langle\rho\rangle=\rho_{0}$ is the spherical average density at the analyzed radius. }
    \label{fig:dens-perturb}
\end{figure}

Prior work by others demonstrated the impact of pre-shock perturbations. \cite{MuJa15} showed large-scale modes, $\ell=1$ and $\ell=2$, of pre-shock perturbations facilitate shock revival as they distorted the shock and changed the matter infalling angles, redirecting more kinetic energy to tangential flows. This means more kinetic energy from the infalling matter is preserved and channels into the post-shock region, leading to more violent aspherical motions in post-shock region that aid the shock expansion. The same argument is demonstrated in \citet{MuMeHe17}. Their Figure 6 shows the infalling density perturbations in their s18-3D and s18-3Dr models are small, but greatly increase after the Si/O interface accretes through the shock. Their s18-3D model has larger pre-shock perturbations in its oxygen shell and explodes earlier than their reduced O burning model s18-3Dr after the oxygen shell pre-shock perturbations accretes to the shock. To show the interaction between the infalling perturbations with the shock, we follow the same prescription as \cite{MuMeHe17} (see their equation 2 and 3) to track the normalized root-mean-square (RMS) density fluctuation and the lowest three normalized density multipoles at certain radii in the pre-shock region. 

In Figure \ref{fig:dens-perturb}, we analyze pre-shock density perturbations at 500 km, and compare our models M3, M4, M5, and M6. We choose 500 km, though the shock reaches this radius after the explosion develops, because it shows the density fluctuations for almost the entire silicon shell. M5 and M6 have sphericalized progenitors, so naturally they do not have obvious pre-shock density perturbations but do show negligible small fluctuations due to wave propagation from the silicon shell, where noise is  generated in the collapsing phase. Both M3 and M4 show prominent pre-shock density perturbations as expected. The only difference is the normalized RMS density fluctuation of M4 is completely dominated by $\ell=1$ perturbations, while M3 also has a strong $\ell=2$ signal developing from $\tpb \approx$ 80 ms. The first rise of $\ell=1$ density perturbation in M3 and M4 at 30~ms after bounce corresponds to the timing when the Fe/Si interface accretes through the maximum shock. At this time the maximum, mean, and minimum shock radii are nearly the same, but this time is slightly earlier than the time of the mean shock reaching the Fe/Si interface as recorded in Table~\ref{tab:Modelinfo}.  The contribution for the $\ell=2$ mode makes the normalized RMS density fluctuation of M3 ($v_{\theta}$ not averaged  at the onset of collapse) larger than M4 (averaged $v_{\theta}$ but $v_{\theta} \neq 0$). To this point, our results agree with  \cite{MuMeHe17} (their Figure 6). However, we disagree with their conclusion that the level of seed perturbations impacts the shock revival. Despite no large scale (low $\ell$) pre-shock density perturbation accreting onto and interacting with the shock, our sphericalized models without pre-shock perturbations in the Si-burning shell, M5 and M6, still explode at a time close to the models with perturbations, either reduced, M3, or not M4. Again, we do not see the impacts of the large scale pre-shock perturbations on the explosion.

\begin{figure}
    \includegraphics[width=\columnwidth]{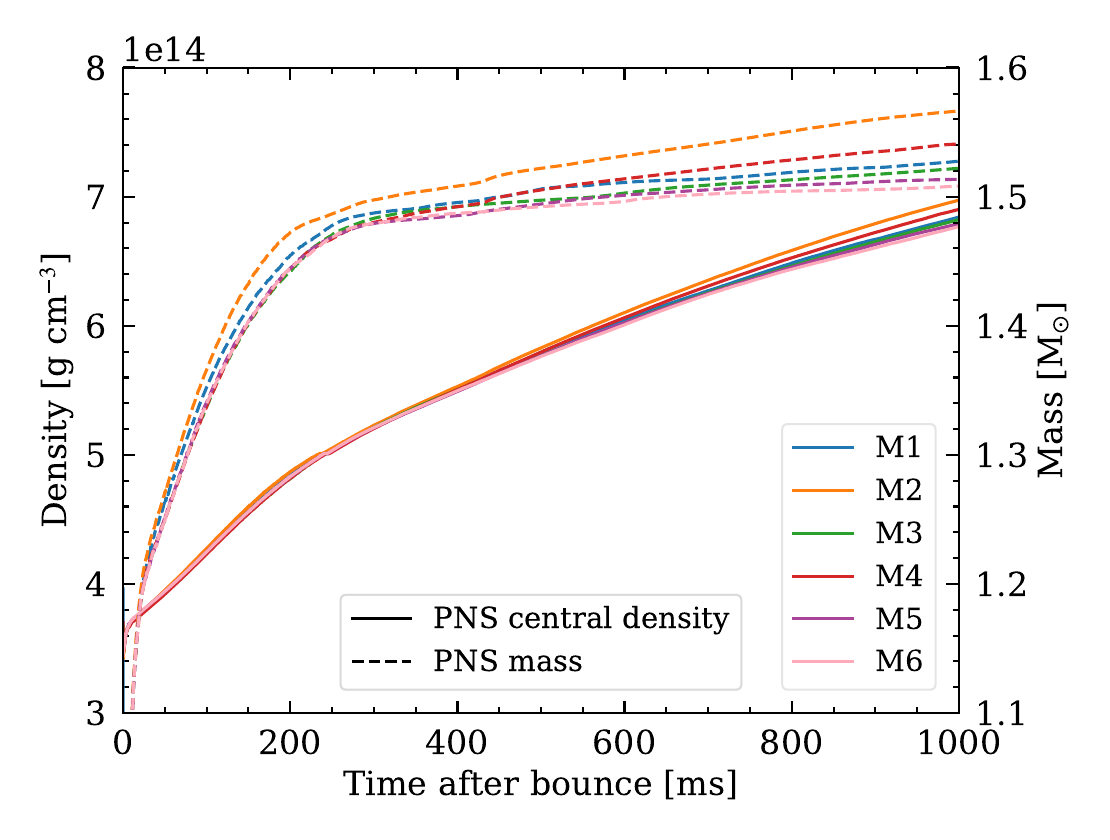}
    \caption{Mass and central density of PNS between 0--1000~ms after bounce for all models. }
    \label{fig:PNS-all}
\end{figure}

\begin{figure}
    \fig{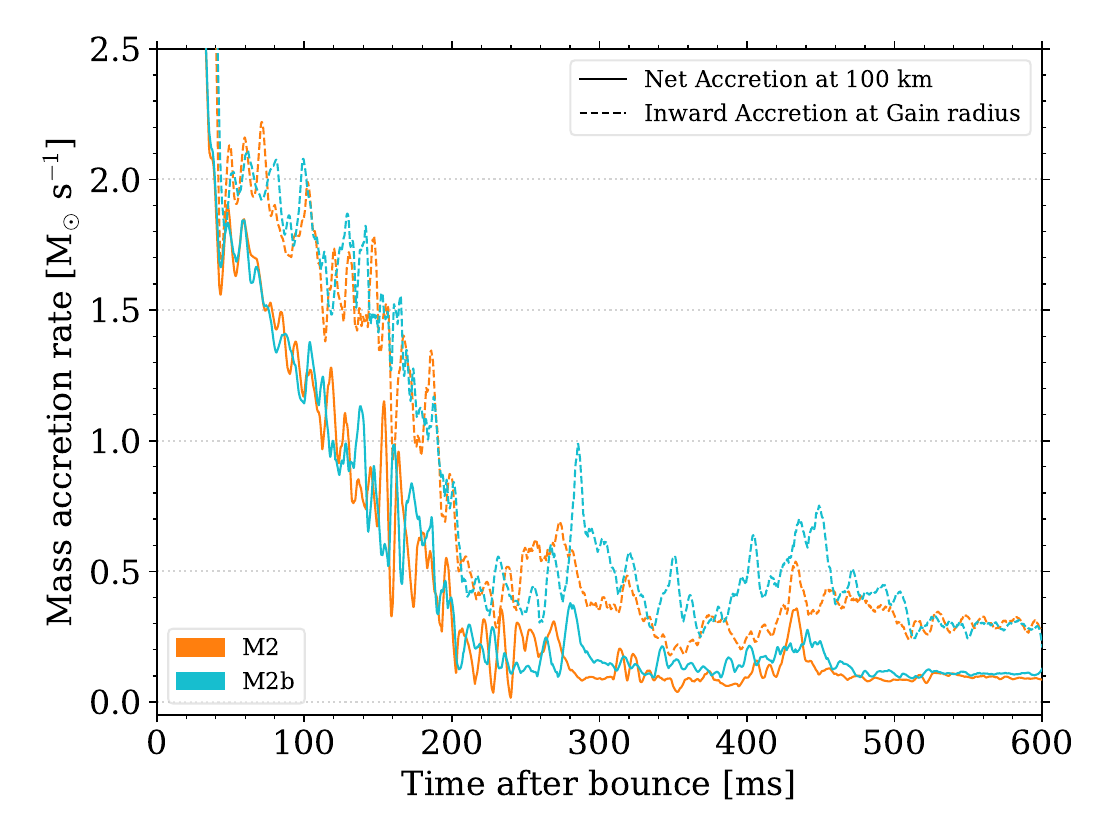}{\columnwidth}{(a)}
    \fig{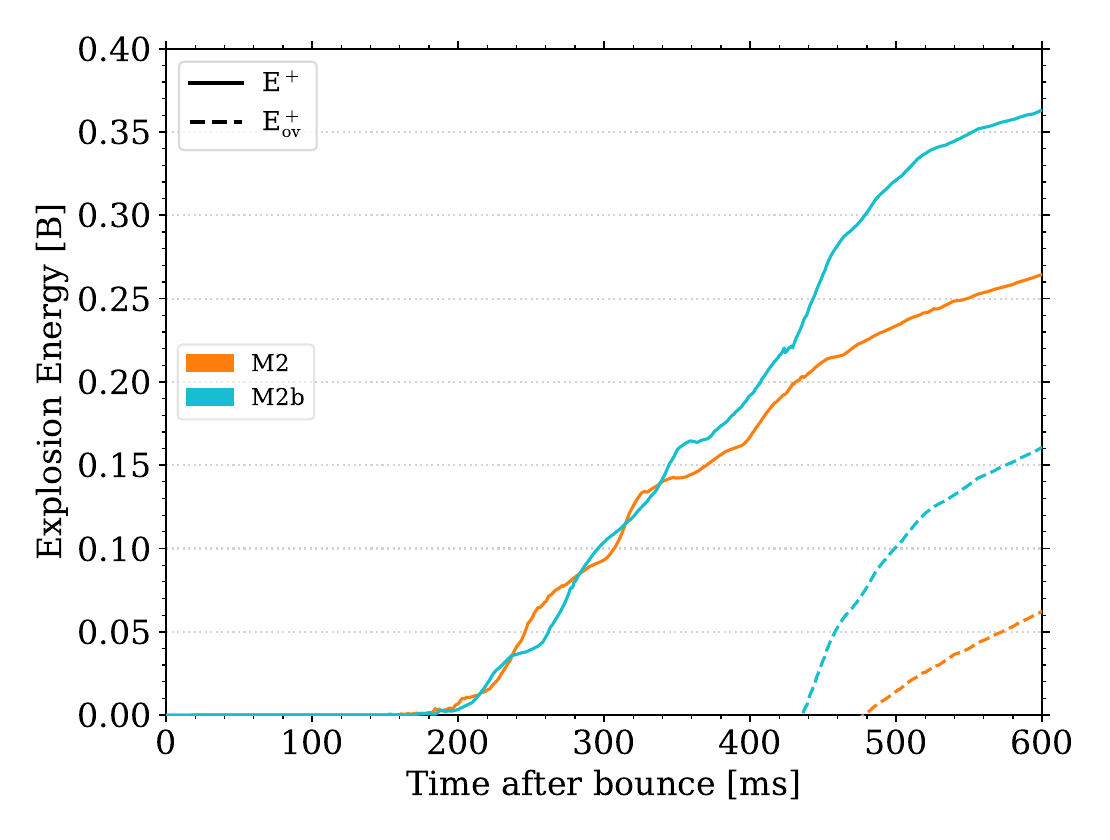}{\columnwidth}{(b)}
    \caption{Comparison of (a) mass accretion rates and (b) diagnostic energies of M2 and M2b for the first 600 ms after core bounce. (a) is smoothed over a 10 ms interval. }
    \label{fig:m2m2b-accrete-energy}
\end{figure}

\section{Stochasticity}
\label{sec:stochasticity}

Despite different progenitor structures at the onset of collapse, our models show very similar behaviors and have similar PNS as shown in Figure \ref{fig:PNS-all}, especially for M1, M3, M5, and M6. Two most energetic models M2 and M4 have heavier PNS and slightly higher PNS central density because they have more cumulative mass accretion onto their PNS (Figure \ref{fig:lumin-accrete}(c)). The subtle differences between models from changes in initial conditions can be lost in stochastic variations in explosion geometry and accretion. We construct two experiments to examine the impact of stochastic variation on our results.

\subsection{M2b}
\label{sec:m2b}

To isolate the role of stochasticity, we take one of our models, M2, and re-evolve it from core bounce to 600 ms as M2b. For this model, in addition to the standard numerical sources of variation (parallel summation order, recompiled executable, etc.) the updated angular momentum conservation in M2b will also trigger tiny differences the seed perturbations.

Based on the features and evolution shown in Figure \ref{fig:m2m2b-accrete-energy}, we consider three stage comparisons between M2 and M2b: the first 200 ms after bounce, 200--400 ms after bounce, and 400--600 ms after bounce. In the first 200 ms after core bounce, the ``pre-explosion" phase, the accretion rates change drastically and the diagnostic energies remain zero. Between 200--400 ms after bounce, two models have similar behaviors and dynamic features and between 400--600 ms they deviate from each other but maintain a similar trend, which will be discussed below.

\begin{figure}
    \begin{center}
    \gridline{
       \fig{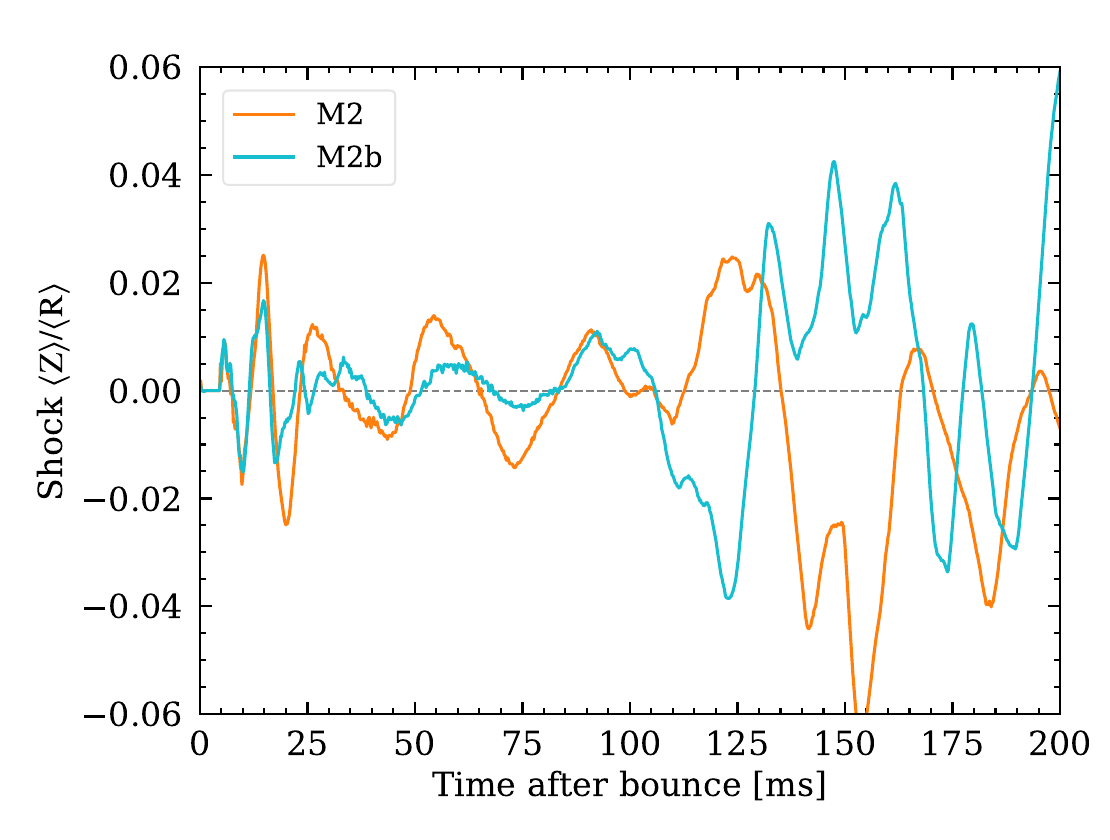}{0.95\columnwidth}{(a)}
    }
    \gridline{
       \fig{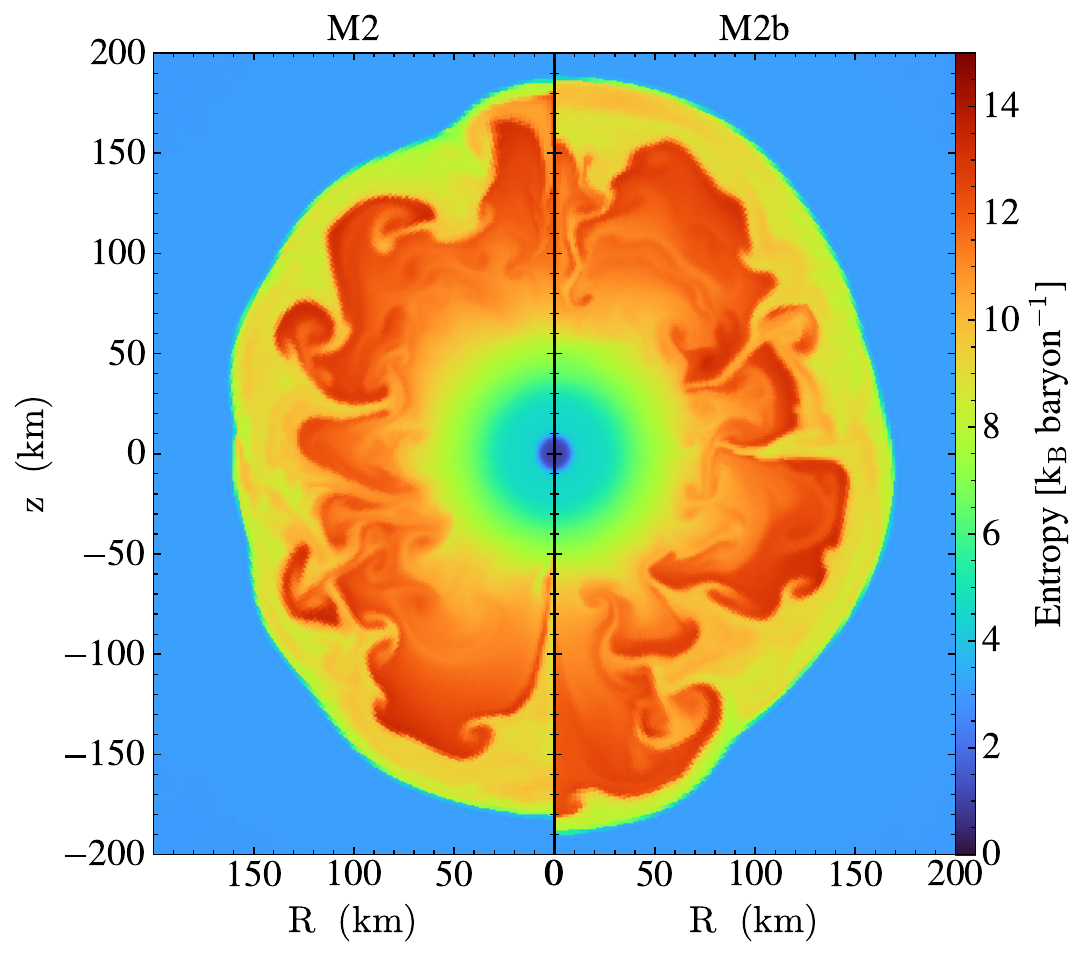}{0.95\columnwidth}{(b)}
    }
    \gridline{
      \fig{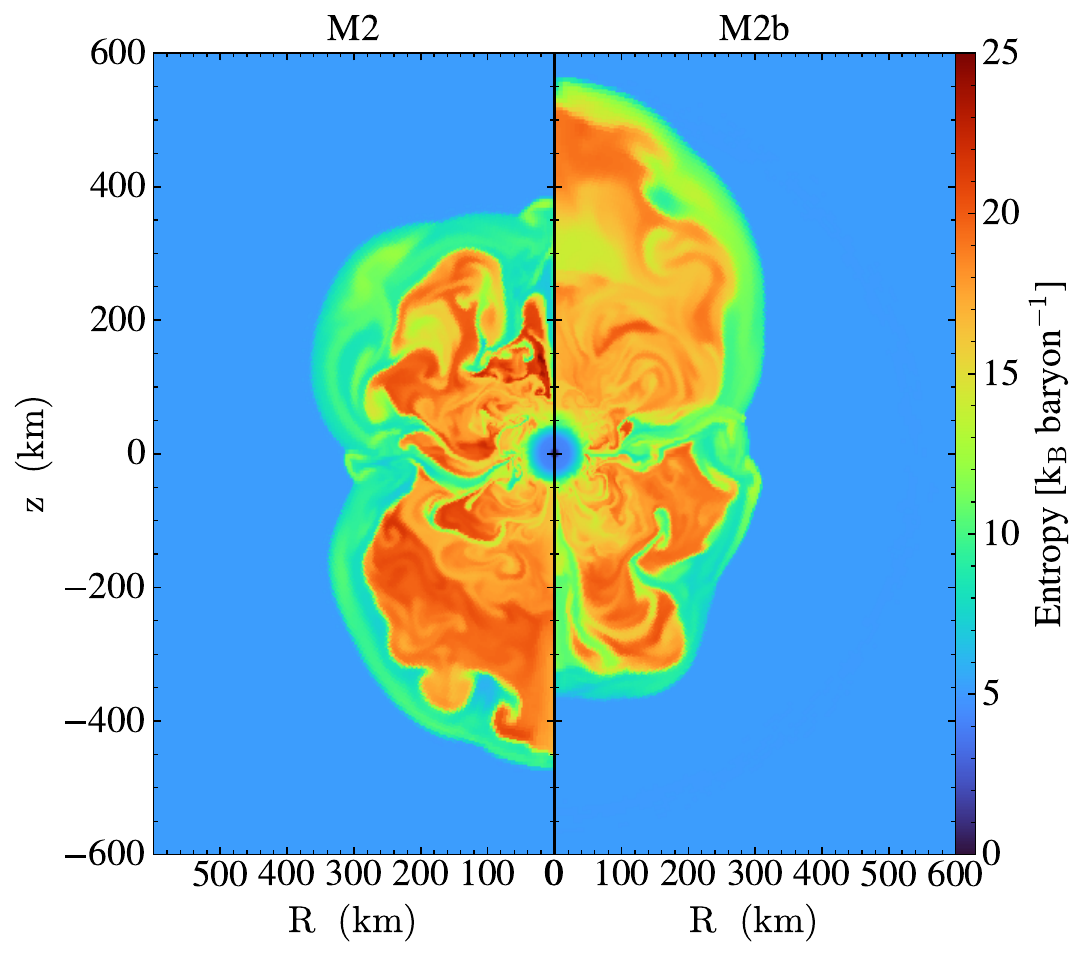}{0.95\columnwidth}{(c)}
    }
    \end{center}
  \caption{Dipole deformation of the shock, \ZoverR, (a); and entropy pseudocolor plot at (b) 100 ms and (c) 200 ms after bounce.} 
  \label{fig:m2m2b-ms0-ms200}
\end{figure}

The explosion geometry is impacted by stochasticity in the first 200 ms after bounce. From the shock dipole deformation in Figure \ref{fig:m2m2b-ms0-ms200}(a), we observe the first obvious difference between M2 and M2b emerging at $\tpb \approx 15$ ms, with variations limited to amplitude, but not phase, until $\tpb \approx$ 110 ms. As shown in Figure~\ref{fig:m2m2b-ms0-ms200}(b), the macroscopic features,  the geometry and shock radius, are approximately the same at $\tpb = 100$ ms, but we can notice the smaller features are different --- they have distinct convective plume patterns. Both models explode between 180--190 ms after bounce (Table~\ref{tab:Modelinfo}), and the explosion geometry is roughly determined at the onset of explosion. The explosion geometries of M2 and M2b at $\tpb = 200$ ms are plotted in Figure \ref{fig:m2m2b-ms0-ms200}(c), and we can see that M2 has a stronger south plume and M2b has a stronger north plume. Their explosion geometries should sustain unless they encounter a sufficiently strong accretion event that changes the dominating plume orientation.

There is not a reliable method to quantify the contribution or relative weight of individual convective plumes to the overall explosion dynamics. 
What we can state with reasonable confidence is the importance of equatorial accretion downflows. 
Although 2D slice plots suggest that accretion downflows at different latitudes occupy similar solid angles, a full 3D perspective, including the azimuthal direction, reveals that equatorial downflows have larger solid angles and therefore carry the greatest weight and therefore have the strongest influence on the overall explosion geometry. 
Combining with the nature of 2D axis-symmetric simulation, the position where the equatorial downflows reaches the PNS forces the ejecta at the oppose direction. 
The changes in downflow orientation and stronger plumes are recoverable in the post-bounce accretion phase, but gradually become difficult as the explosion is launched and the role of the dominant plume is amplified. 
This is particularly obvious in models with unipolar explosion morphologies. 

\begin{figure}
     \gridline{
        \fig{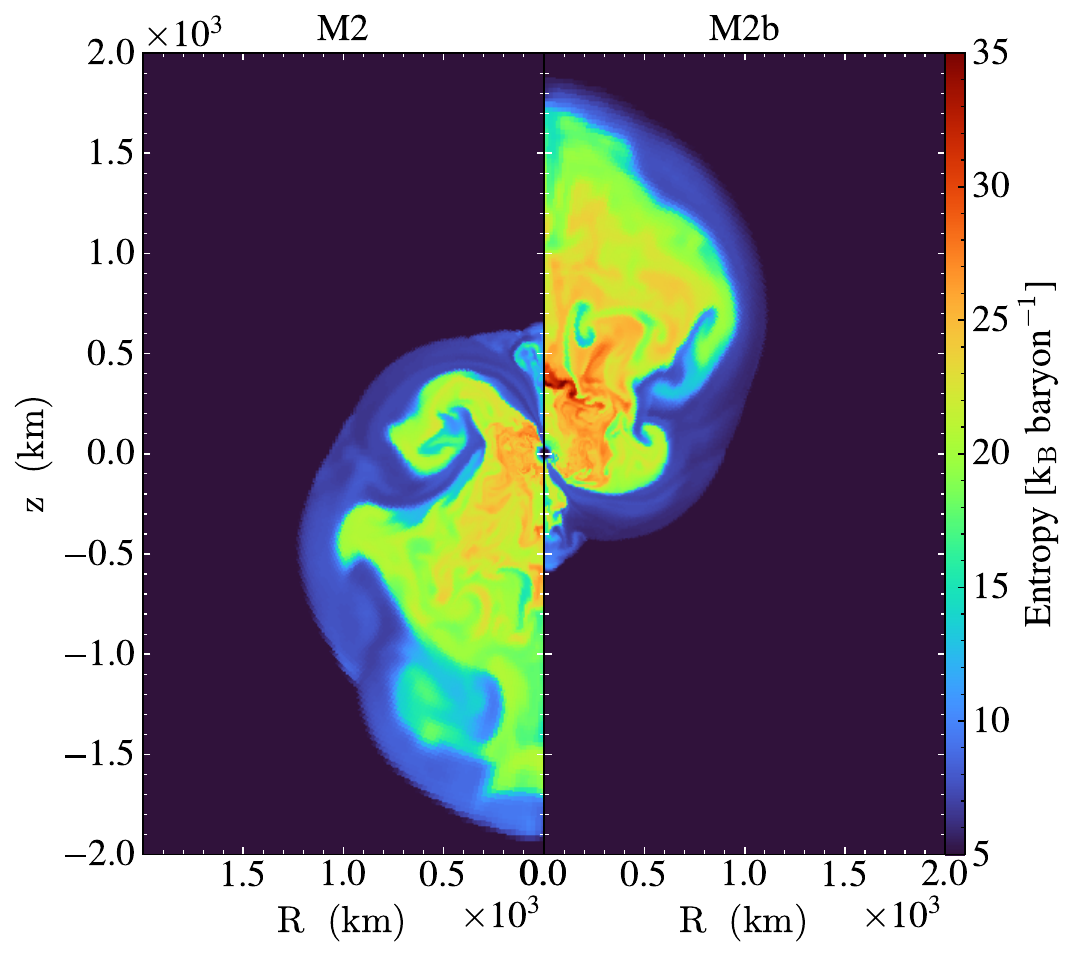}{.85\columnwidth}{(a)}
        }
    \gridline{
        \fig{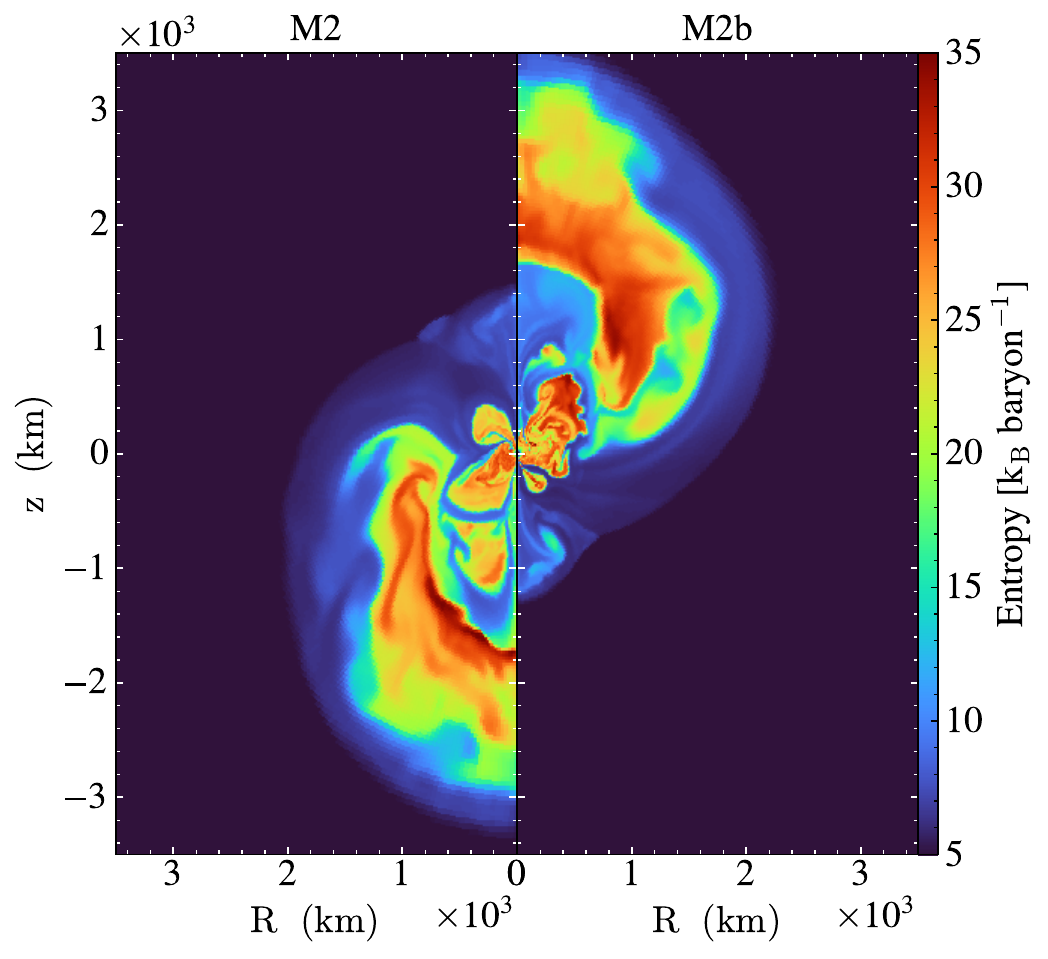}{.85\columnwidth}{(b)}
        }
    \gridline{
        \fig{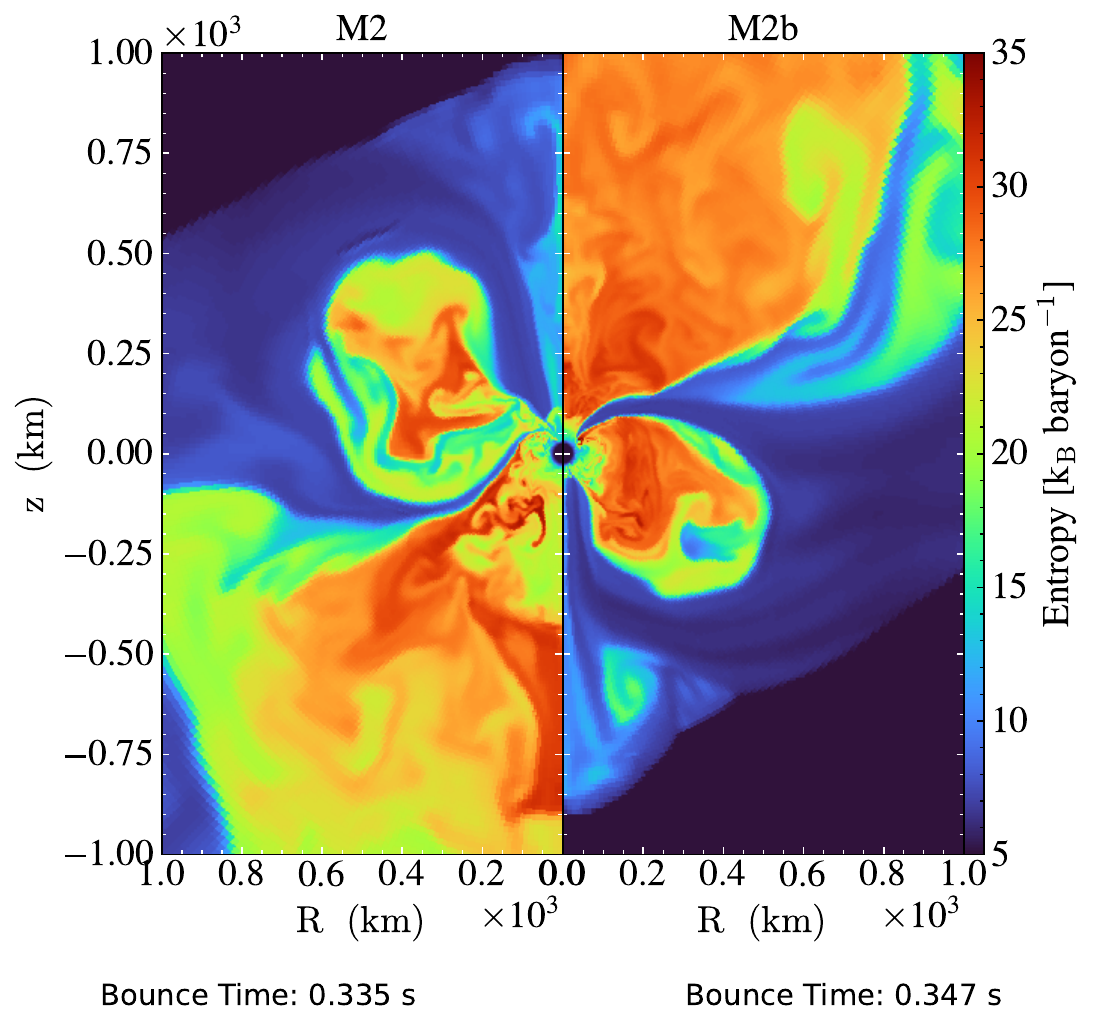}{.85\columnwidth}{(c)}
        }
    \caption{Pseudocolor plots of entropy for M2 and M2b at (a) 300 ms, (b) 400 ms, and (c) $\approx$340~ms (zoomed to show minor lobes) after bounce. }
    \label{fig:m2m2b-ms200-ms400}
\end{figure}

Between 200--400 ms after bounce, the explosions stabilize and mature, and the two models behave similarly. Their features start to deviate around $\tpb = 400$ ms due to strong local accretion events, where the accretion streams inhibit the formations of minor lobes as discussed in Section~\ref{sec:accretion}. Figure \ref{fig:m2m2b-ms200-ms400}(a) and (b) depict entropy slices of M2 and M2b at 300 ms and 400 ms after bounce, respectively. It is interesting that their explosion geometries are almost identical but mirrored. This suggests the resemblance between two models in macroscopic view. In Figure \ref{fig:m2m2b-ms200-ms400}(b) we notice the existence of minor plumes in both models, whose formations trace back to $\tpb=335$ ms for M2 and $\tpb=347$ ms for M2b, as plotted in Figure \ref{fig:m2m2b-ms200-ms400}(c) and zoomed in to a radius of 1000 km for a better view of the minor lobes. 

\begin{figure*}[htbp]
  \gridline{
    \fig{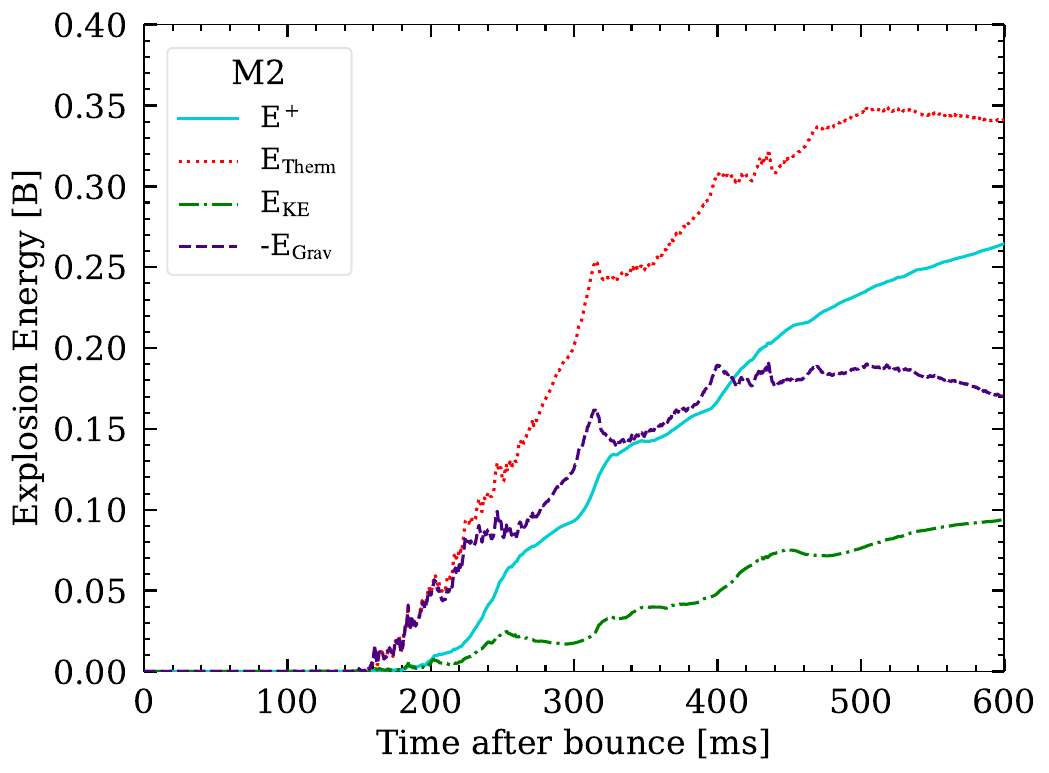}{0.48\textwidth}{(a)}
    \fig{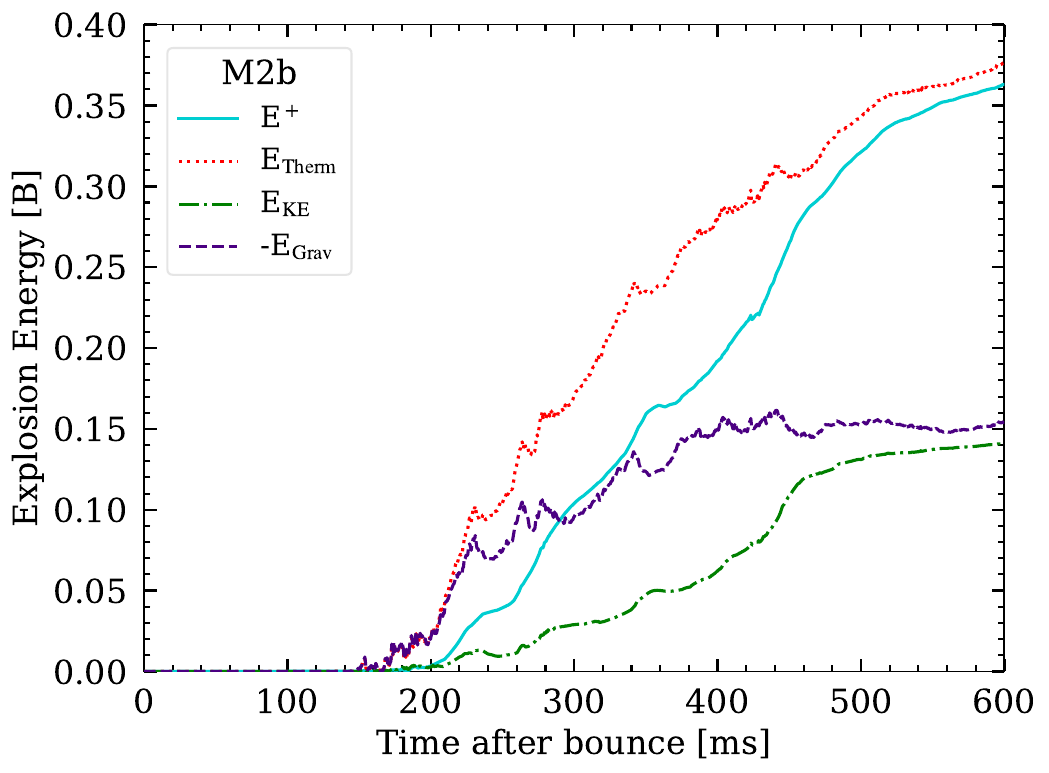}{0.48\textwidth}{(b)}
  }
  \gridline{
    \fig{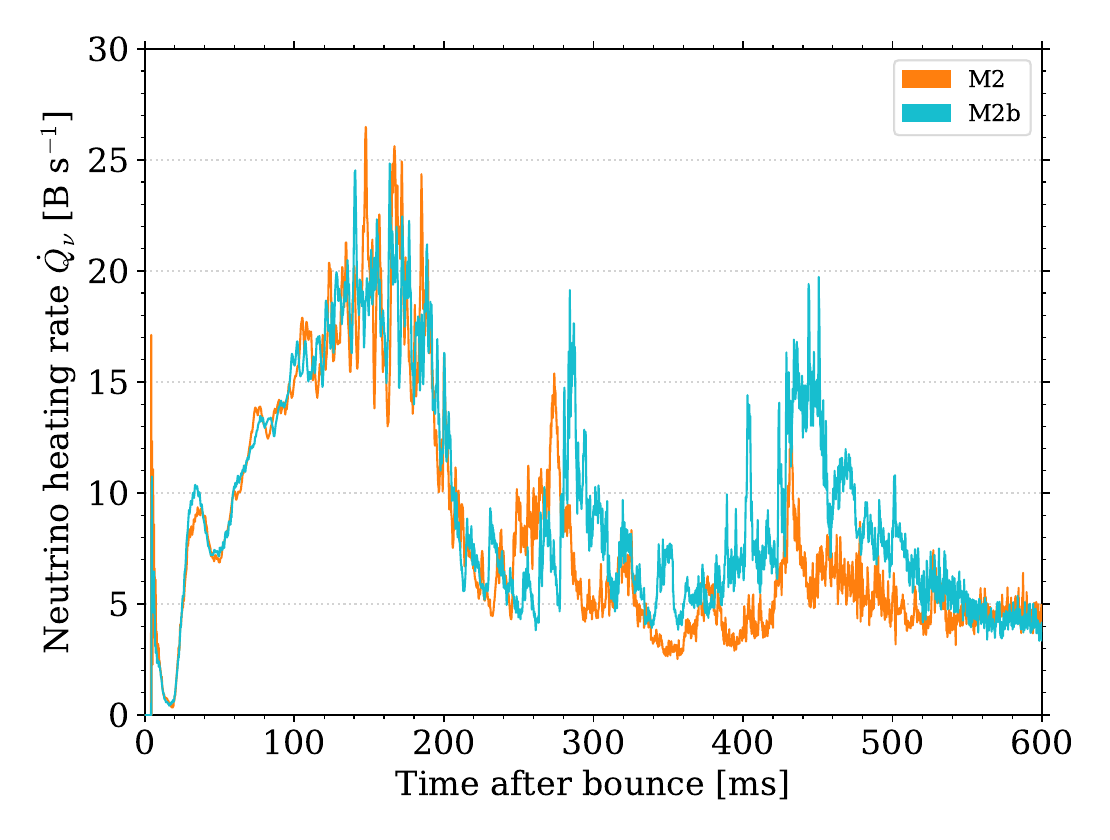}{0.48\textwidth}{(c)}
    \fig{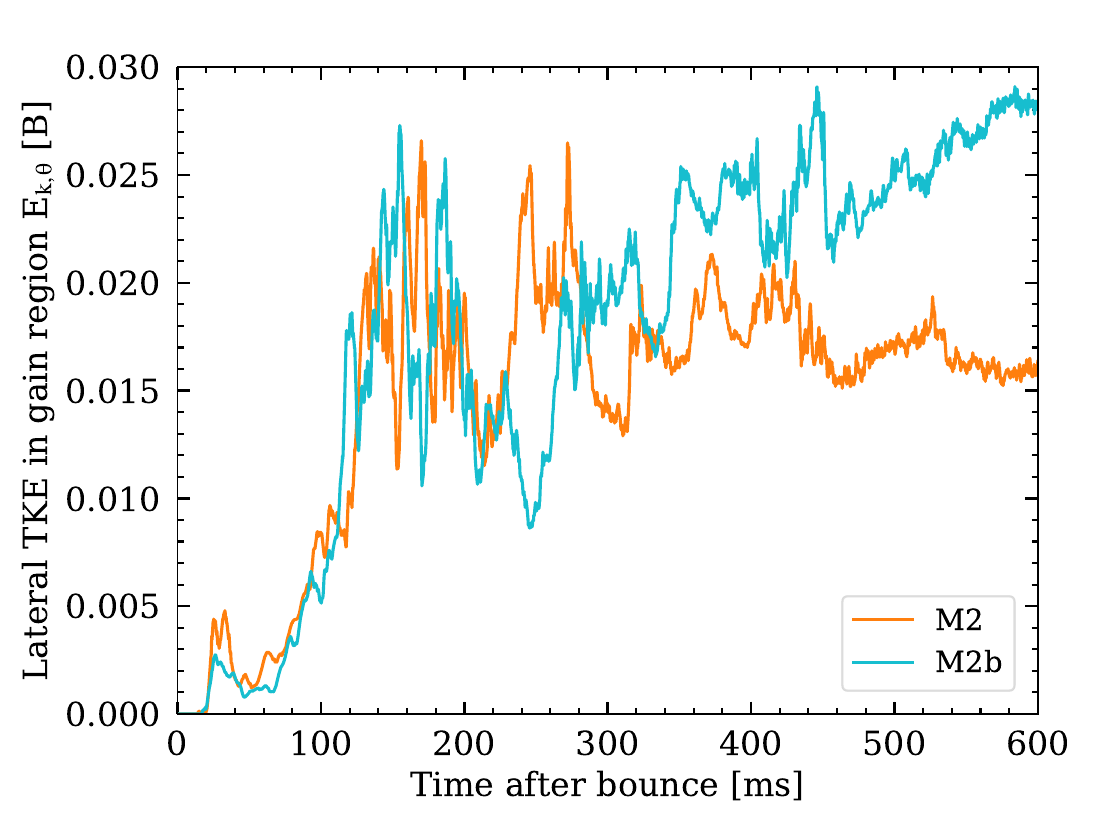}{0.48\textwidth}{(d)}
  }
  \caption{\Ediag\ decomposition plot for (a) M2 and (b) M2b as in Figure \ref{fig:energy-split}; neutrino heating rate (c) and lateral TKE (d) in the gain region for same models.}
  \label{fig:m2m2b-accrete-event}
\end{figure*}

The competition between the minor lobe development and accretion downflows is the driver making these two models different in this phase. The minor lobe in M2 exists for a longer period between $\tpb=335$--433~ms, while the actual inhibition happens between $\tpb=400$--433~ms as stated in Section~\ref{sec:accretion}. Between $\tpb=335$--400~ms, this minor lobe gets divided twice before the accretion stream entirely pushes it onto the PNS. 
The minor lobe in M2b exists between $\tpb=347$--429~ms; however, the minor lobe development is suppressed immediately after its formation. Most of this lobe is soon pushed toward and merges with the main ejecta plume at the north, with the divided remaining small portion of it being forced accreting onto the PNS between $\tpb=388$--429~ms. 
Hence, the smaller energy waste in the competition and more efficient energy conversion from gravitational to heat lead to a higher \Ediag\ in M2b. From Figure~\ref{fig:m2m2b-accrete-event} we can see the increasing heat component of \Ediag, the heating rate, and gain region lateral TKE all support the influence of these accretion events we observed.

\begin{figure*}[htbp]
  \gridline{
    \fig{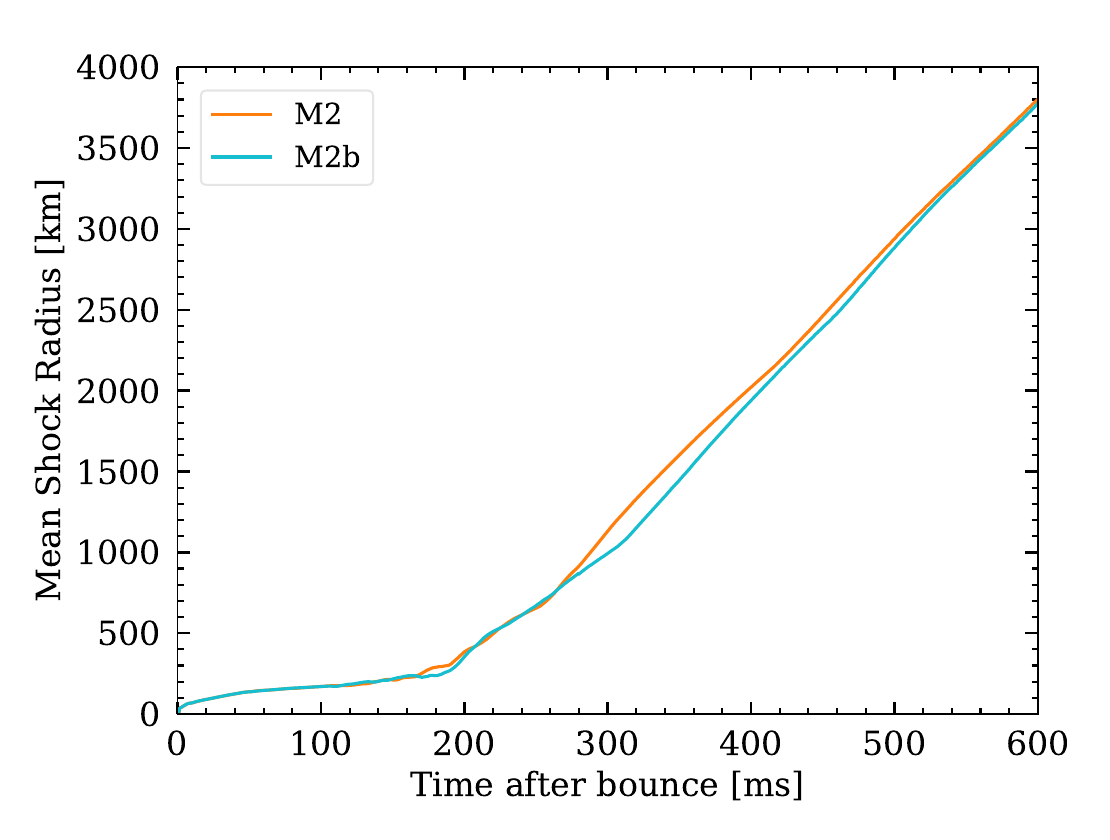}{0.48\textwidth}{(a)}
    \fig{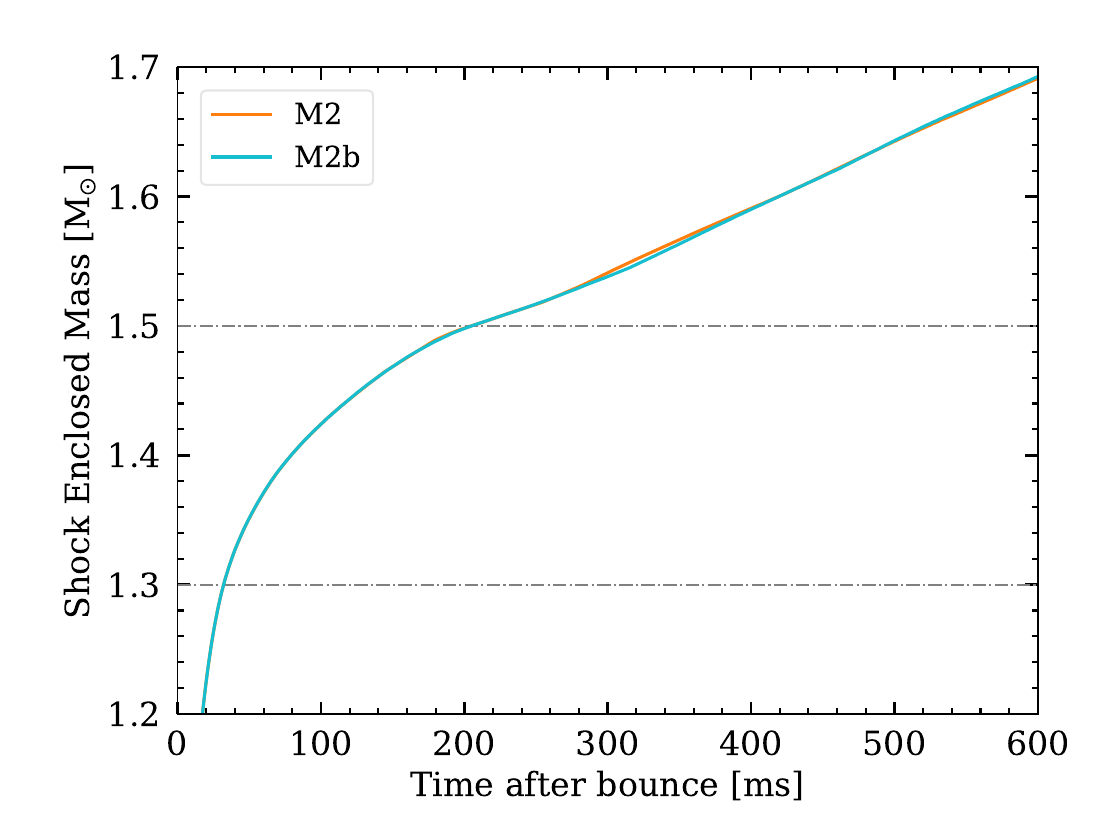}{0.48\textwidth}{(b)}
  }
  \gridline{
    \fig{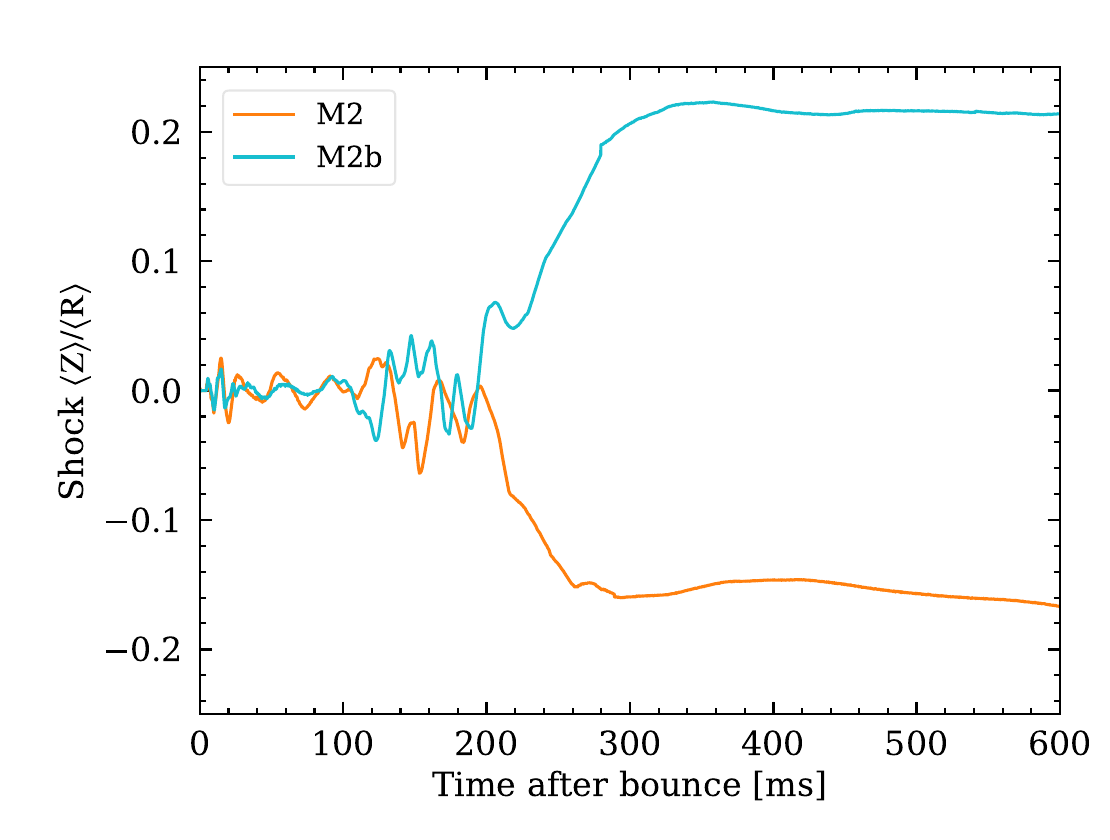}{0.48\textwidth}{(c)}
    \fig{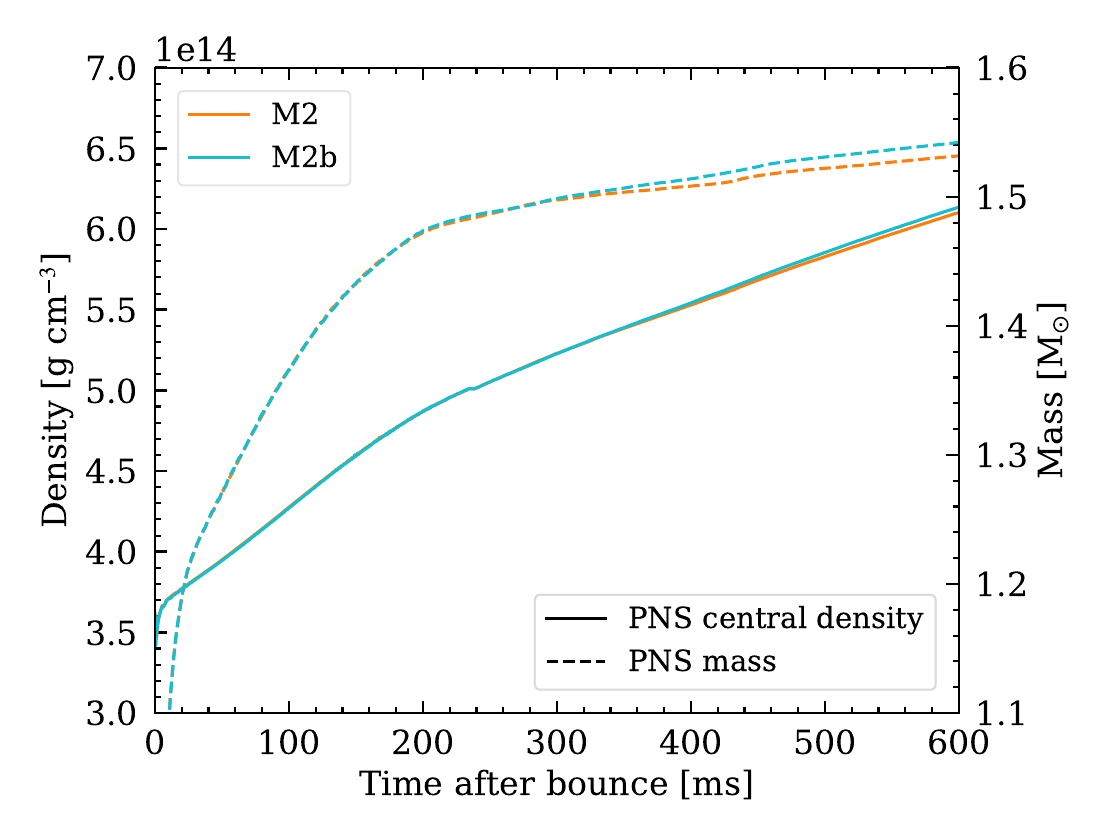}{0.48\textwidth}{(d)}
  }
  \caption{Comparison of M2 and M2b (a) mean shock radii, (b) shock enclosed mass, \Mshock, (c) shock dipole deformation, and (d) PNS properties for $\tpb=0$--600~ms. }
  \label{fig:m2m2b-similarity}
\end{figure*}

Between $\tpb=400$--500~ms, the influence of the accretion event remains, but the behaviors of two models are similar for $\tpb=500$--600~ms as the impacts from the accretion event fade away. Therefore, we see their \Ediag\ growths are parallel (Figure~\ref{fig:m2m2b-accrete-energy}(b)). When we examine the entire 600 ms (Figure \ref{fig:m2m2b-similarity}) we find M2 and M2b are nearly identical but with mirrored explosion geometry, same mean shock radius and \Mshock\ growth, and similar PNS evolution with M2b being slightly more massive and denser PNS from the additional accretion event. 

\subsection{M5 vs M6}
\label{sec:M5vsM6}

Aside from the comparison between M2 and M2b, we use another pair of models, M5 and M6, to show how stochasticity affects simulation outcomes. These two models have nearly identical initial conditions, differing only by the numerical noise accumulated during collapse in 2D of M5 as M6 is collapsed in 1D and spherically symmetric at bounce. 
We observe that M5 and M6 have nearly identical shock progressions both in Eulerian and Lagrangian space (Figure~\ref{fig:mean-shock}(c) and (d)), \Ediag\ and corresponding components (Figure \ref{fig:explo-energy} and \ref{fig:energy-split}) and nucleosynthetic yields for signature nuclei at 1 s after bounce (Figure \ref{fig:nucleosynthesis} and \ref{fig:abund-iron}(b)) are similar. 
Both models in the pair maintain spherical shocks in the first 100 ms after bounce (Figure \ref{fig:dipole-time}). 
Though their shock dipoles and explosion morphologies diverge after that due to the position randomness of developed convection plumes, M5 and M6 still have very similar pre-explosion lateral turbulent kinetic energies in their heating layers (Figure \ref{fig:ke-lat}(b)). 
Analysis of the lateral kinetic energy (Figure \ref{fig:m3-m5-noise}(b) and \ref{fig:ke-lat-m356}) shows that while M5 has built up lateral kinetic energy from multi-D collapse in its Fe/Si core region at bounce and M6 has $v_\theta\equiv 0$, by 20 ms after bounce both have similar $\sim 10^{14}$ \ergg\ lateral TKE in the post-shock region from prompt convection and remain in good agreement afterward with no delay in the onset of neutrino-driven convection or explosion.

Based on these two pairs of models, we find that stochastic variation plays an important role in the post-bounce accretion phase when explosion geometry changes the nature of accretion events, such as seen in the M2 and M2b pair, but when such accretion events are not present, as in 
M5 and M6, stochastic variation has minimal impact on the outcomes.

\section{Summary}
\label{sec:summary}

We have examined the impact of models with a variety of treatments of pre-collapse silicon burning and related perturbations through seven non-rotating axisymmetric 2D CCSN simulations with the \chimera\ code \citep{BrBlHi20} evolved from core collapse, through collapse, core bounce, and explosion phases. 
The simulations were initialized from 15 \msun\ progenitors with variations only in the final minutes of silicon-shell burning.  The six principle models G15-SiSB-M1 through G15-SiSB-M6 ran to at least 1~s after bounce, with one model, G15-SiSB-M2b, rerun for 600 ms after bounce.  

By analyzing the structure of different progenitors at the onset of collapse, especially their non-radial velocities, we demonstrate the impact of a multi-D pre-collapse burning environment (\polaris-2D) during silicon-shell burning and the subsequent transition to core collapse.
Among these insights, our results reveal clear structural differences within the convective oxygen-burning and silicon-burning shells among MESA, \polaris-1D, and \polaris-2D progenitors at the onset of core collapse. 
In the \polaris-2D progenitor, the silicon-burning shell has nearly exhausted its Si/S fuel, converting most of it into iron-group nuclei. 
In contrast, the MESA progenitor retains a modest amount of Si/S, primarily concentrated near the Si/O interface, while the \polaris-1D progenitor leaves a large reservoir of unburned Si/S in its upper silicon-burning shell. 
In the density structure, we observe sharp density declines at both the Fe/Si and Si/O interfaces, with the latter consistent with the pronounced Si/O density drop widely reported in the literature \citep{VaLaRe21, VaCoBu22, BrSiLe23}. 
We confirm these pre-collapse features are preserved and amplified in the collapsing stage, in agreement with the suggestion by several groups \citep{CoChAr15, MuMeHe17, BoYaKr21, VaCoBu22}. We find that perturbations introduced by multi-D stellar evolution can be maintained and amplified. 

However, while this earlier work suggested these features shorten the time to rejuvenate the stalled shock and also lead to more powerful explosions, our simulation based on the 2D progenitor model, G15-SiSB-M3, does not exhibit earlier shock revival or a more energetic explosion. 
All our models transition to explosion approximately 200 ms after bounce, and their diagnostic energies, \Ediag, range between 0.25--0.42 B at 1000 ms after bounce, with G15-SiSB-M3 at the lower bound of this range. 
Robust `early' ($\tpb\approx 200$~ms) explosions are characteristic of \chimera\ models. 
This may make \chimera\ models less sensitive to pre-collapse perturbations. 
Specific to this study, the `base' MESA progenitor model, G15-SiSB-M1, explodes promptly without perturbations in the typical \chimera\ fashion.
In addition, we observe G15-SiSB-M3 has similar dynamic features to its 1D, spherically averaged counterparts, G15-SiSB-M5 and G15-SiSB-M6. 
By analyzing the lateral turbulent energy across these three models and contrasting G15-SiSB-M3 with the model that generates significant noise during collapse, G15-SiSB-M2, we arrive at three results from this study.

The first result is that, regardless of the initial structure, turbulence generated by prompt convection reaches at least $\sim 10^{14}$ \ergg\ as the shock progresses through the iron core as exemplified by G15-SiSB-M5.
Prompt convection is delayed in the `quiet' models (M1, M5, and M6) by a few ms relative to the `noisier' models (M2, M3, and M4) that have larger lateral velocities at bounce, but in all cases signs of prompt convection are visible within 10~ms of bounce. 

Our second result is that a discernible amount of turbulent energy can be generated during the collapse phase from 1D progenitor models when nuclear burning is included in the outer iron core as in G15-SiSB-M2. Lateral TKE of $\sim 10^{13}$~\ergg\ are generated in the upper iron core where the network was active during collapse and are $\sim 10^{10}$~\ergg\ throughout the rest of the iron core at bounce.
This is an area where \chimera\ is more sophisticated than the codes used in prior studies of the impact of perturbations.
This arises from the numerical amplification of noise in \chimera\ by the nuclear network, and the simulation outcomes indicate that this noise has a non-negligible impact, seeding convection in a fashion similar to large scale anisotropies.

Our third result is that the `noisy' models with high lateral velocities at bounce ($>1$~\kmps) exhibit a strong dipole shock oscillation that is not seen in `quiet' models that have little or no noise at bounce. The shock asphericity is somewhat oblique to the infalling material of the iron core and diverts some of the radial infall momentum into lateral motion and accelerates the growth of post-shock lateral TKE to a characteristic $\sim 10^{17}$ \ergg\ level, well above that of just prompt convection. Until we added the final two, low-noise, models (M5 and M6) which did not exhibit this behavior, we were concerned that it was an artifact of the cell merging, which allows non-radial motion deep in the core unlike all prior \chimera\ models. By 100~ms after bounce, neutrino-driven convection has built the lateral TKE to the same level in the gain region behind the shock in all models, but the `noisy' models with the shock oscillations during prompt convection have larger lateral velocities in the outer layers of the PNS.

These results indicate that the environment in the gain region is similarly turbulent in all models when neutrino-driven convection is fully developed at 100~ms after bounce regardless of the impact of perturbations prior to that point and our models do not need additional boosting from accretion of external turbulence to explode. 
We therefore do not find a strong connection between perturbations to the pre-collapse structure and the explosion outcome, which differs from trends reported in previous studies \citep{HaMaMu12, CoOc14, CoOt15, MuMeHe17, BoYaKr21, VaCoBu22}.

Aside from lateral turbulent energy analysis, we also replicate the density analysis in \cite{MuMeHe17}. 
Our comparison between four models with \polaris-2D progenitor at a radius of 500 km confirms the existence of pre-shock RMS density fluctuations and the dominance of $\ell=1$ and $\ell=2$ modes in full 2D progenitor model G15-SiSB-M3 and reduced 2D progenitor model G15-SiSB-M4. 
This aligns with the results of \cite{MuJa15, MuMeHe17} indicating that the pre-shock perturbation in their models are similarly in ours. 
However, our results disagree with their conclusion that the strength of seed perturbations affects shock revival. 
Our sphericalized models without pre-shock density perturbations in the silicon-burning shell, G15-SiSB-M5 and G15-SiSB-M6, explode at nearly the same time as the perturbed models, G15-SiSB-M3 and G15-SiSB-M4, regardless of perturbation amplitude. 
Thus, we find no clear impact of pre-shock perturbations on the explosion. 

Small differences between models G15-SiSB-M1, G15-SiSB-M3 and its spherically averaged counterparts, G15-SiSB-M5 and G15-SiSB-M6, are attributed to stochasticity, which mildly affects the explosion geometry and accretion pattern. 
We also find that unipolar ejecta are less affected by individual accretion events in the explosion phase. 
Following this line, we compare G15-SiSB-M2, the model with the most unipolar ejecta, and its recomputed version, G15-SiSB-M2b, to 600 ms after bounce. 
From this comparison, we find that they share similar dynamic features and behaviors, but with mirrored morphologies. 
Their mirrored explosion geometries and small differences caused by strong accretion events both trace back to stochasticity. 
The former are decided in the post-bounce accretion phase, while the latter occur as the explosions mature. 

The second pairwise comparison between G15-SiSB-M5 and G15-SiSB-M6 demonstrates high consistency except for their greatly divergent explosion morphologies. 
Hence, we conclude that stochasticity plays a crucial role in the post-bounce accretion phase and determines explosion geometry. 
More symmetric explosion morphologies may be influenced to some extent by stochasticity when explosions develop, such as changes of dominant ejecta plumes and overall geometries as discussed in Section \ref{sec:morph}. 
However, the close similarities between G15-SiSB-M5 and G15-SiSB-M6 suggest that random accretion events have limited effects on the macroscopic features and outcomes of the simulations.  
Adding the standard 1D progenitor (G15-SiSB-M1) and 2D convective progenitor (G15-SiSB-M3) models to the comparison suggests that variation due to progenitor initiation (2D versus MLT convection, perturbations related to convection) are relatively small and fall within the nominal range of stochastic variation in explosion energetics as seen in parameterized 3D models \citep{WoJaMu13,TaKoSu14,HaPlOd14,IwNaYa14,CaBu15}. To do so, we must exclude the two more energetic explosions, G15-SiSB-M2 and G15-SiSB-M4, as being outside nominal stochastic variation.
Both of these models show accretion events at approximately 450~ms that boost accretion at the point where their \Ediag\ diverges from the other models.  G15-SiSB-M2 also shows larger sustained accretion in the latter phases of explosion. This larger accretion is likely connected to the difference in the structure of the \polaris-1D progenitor and is reflected in the more rapid growth of the PNS mass and mass enclosed by the shock.
Excluding G15-SiSB-M4 from the other models is more difficult, but through careful examination of the simulation, we find that the big difference in the growth of G15-SiSB-M4's \Ediag\ correlates with the burst in accretion, which is similar to the accretion-related difference between G15-SiSB-M2 and G15-SiSB-M2b.
This suggests that 2D \chimera\ models, and likely other 2D simulations with self-consistent neutrino heating, may experience large, random, accretion episodes that can dominate subsequent differences in explosion energetics, but these are tractable.
For G15-SiSB-M4, this indicates that the accretion event is the proximal cause for the difference in explosion energy, and we are unable to determine if the unusual $v_\theta \ne 0$, spherically averaged progenitor had any impact or was the cause of the accretion event. 
For the rest of the models, those with an MLT or 2D treatment of silicon-burning convection with self-consistent non-radial perturbations do not have any impact beyond the level of nominal stochastic variation. 

Supplementary movies illustrating the temporal evolution are provided, and all data used for line plots in this paper are publicly available on Zenodo at \dataset[doi:10.5281/zenodo.20550532]{https://doi.org/10.5281/zenodo.20550532}.

\section{Conclusions}
\label{sec:conclusions}

When we examined the role of perturbations on the revival of the stalled supernova shock, we found no evidence that they played any role in the timing, strength, or likelihood of shock revival and explosion in our most basic comparison tests. We were able to demonstrate that the perturbations advected on the the stalled shock in our multi-D progenitor model were similar in character to those used in prior studies and that the total lateral turbulent kinetic energy inside the shocked cavity was similar in all of our models. Perhaps this should have been expected. The model using the standard 1D stellar evolution input (M1) explodes without difficulty as is characteristic of \chimera\ models. The difference in developed explosion energies among our models can be traced to differences in accretion, primarily in the those caused by stochastic accretion events. These highlight the hazard of relying on individual CCSN models to assess the difference in pieces of supernova physics, particularly subtle ones.

Whether, or not, dramatic differences in outcome due to perturbations are suppressed because of some unidentified difference in the effectiveness of \chimera's neutrino heating that is outside the scope of this study, we have highlighted an important reality of iron-core stars --- they are inherently noisy environments that can impact how modeling responds to instabilities. The first example of this is the response of our models to the unstable gradients created by the bounce shock known as prompt convection. Even the tiny amount of noise from the PPMLR hydrodynamics solver is enough to trigger prompt convection in our `low-noise' models. Prompt convection then floods the post-shock cavity with turbulent noise that seeds neutrino-driven convection once neutrino heating has generated an unstable gradient. The artificial noise of those models is likely much smaller than the natural noise coming from overlying convective regions, or from the capacity of the steep temperature dependence in nuclear reactions to amplify passing fluctuations. `High-noise' models from these physical noise sources develop a dipole oscillation of the shock as it propagates through the iron core that injects non-radial motion behind the non-spherical shock. Any shock asphericity will generate significant lateral velocities from infall velocities of up to 50,000~\kmps. 
The connection of shock asphericity during prompt convection to the level of noise in the iron core at bounce makes it more likely that this is a real phenomenon that should be seen in the noisy iron cores of real stars and simulations that reflect that environment, though verification is needed. A better understanding of how noise grows and influences the earliest phases of CCSN development, particularly from core bounce to fully developed neutrino-driven convection, and how that is realized in simulation codes is an important question to be explored. This insufficiently examined and subtle aspect of CCSN models could be responsible for some of the differences in numerical realizations of CCSN explosions.

\begin{acknowledgements}

This research was supported by the U.S. Department of Energy, Offices of Nuclear Physics and Advanced Scientific Computing Research.
Past development of \chimera\ has also been supported by the NASA Astrophysics Theory Program (grant NNH11AQ72I); 
and the National Science Foundation PetaApps Program (grants  OCI-0749242, OCI-0749204, and OCI-0749248), Nuclear Theory Program (PHY-2309988, PHY-1913531, PHY-1516197) and Stellar Astronomy and Astrophysics program AST-1716134, AST-0653376).
This research used resources of the National Energy Research Scientific Computing Center (NERSC), a U.S. Department of Energy Office of Science User Facility located at Lawrence Berkeley National Laboratory, operated under Contract No. DE-AC02-05CH11231.
This research used resources of the Oak Ridge Leadership Computing Facility at Oak Ridge National Laboratory.
Research at Oak Ridge National Laboratory is supported under contract DE-AC05-00OR22725  from the Office of Science of the U.S. Department of Energy to UT-Battelle, LLC.

\end{acknowledgements}

\facility{NERSC, OLCF}

\software{\chimera\ \citep{BrBlHi20}, \texttt{matplotlib} \citep{hunter2007} }

\end{document}